\documentclass[pdflatex,sn-nature]{sn-jnl}%
\usepackage{graphicx}%
\usepackage{float} %

\usepackage{microtype} %
\usepackage{newfloat} %
\usepackage{multirow}%
\usepackage{amsmath,amssymb,amsfonts}%
\usepackage{amsthm}%
\usepackage{mathrsfs}%
\usepackage[title]{appendix}%
\usepackage{xcolor}%
\usepackage{textcomp}%
\usepackage[utf8]{inputenc}
\usepackage{textgreek}
\usepackage{xspace} %
\usepackage{manyfoot}%
\usepackage{booktabs}%
\usepackage{algorithm}%
\usepackage{algorithmicx}%
\usepackage{algpseudocode}%
\usepackage{listings}%
\usepackage{caption}
\usepackage{subcaption}
\usepackage[capitalize,noabbrev]{cleveref}
\usepackage{hyperref} %
\usepackage{fancyhdr} %
\usepackage{etoolbox} %

\AtBeginEnvironment{thebibliography}{%
  \renewcommand{\href}[2]{#2}%
  \renewcommand{\url}[1]{}%
}

\usepackage[section]{placeins} %

\usepackage[nolists,nomarkers]{endfloat}

\usepackage{array}
\usepackage{tabularx}

\definecolor{peptidepurple}{RGB}{153,0,255}

\DeclareFloatingEnvironment[name=Scheme]{scheme}
\crefname{scheme}{Scheme}{Schemes}
\Crefname{scheme}{Scheme}{Schemes}
\DeclareDelayedFloatFlavor{scheme}{figure}
\theoremstyle{thmstyleone}%

\theoremstyle{thmstyletwo}%

\theoremstyle{thmstylethree}%

\usepackage{amsmath,amsfonts,bm}

\def\eqref#1{equation~\ref{#1}}

\def\1{\bm{1}}

\def\rmX{{\mathbf{X}}}

\DeclareMathAlphabet{\mathsfit}{\encodingdefault}{\sfdefault}{m}{sl}
\SetMathAlphabet{\mathsfit}{bold}{\encodingdefault}{\sfdefault}{bx}{n}

\AtBeginDocument{%
  \pagestyle{fancy}%
  \fancyhf{}%
  \fancyfoot[C]{%
    \ifnum\value{page}>43\relax\ifnum\value{page}<47\relax
    \else
      \thepage
    \fi\fi
  }%
  \fancypagestyle{plain}{\fancyhf{}\fancyfoot[C]{\ifnum\value{page}>43\relax\ifnum\value{page}<47\relax\else\thepage\fi\fi}}%
}%

\usepackage{xr-hyper}
\makeatletter
\DeclareUnicodeCharacter{2009}{\,}
\DeclareRobustCommand\onedot{\futurelet\@let@token\@onedot}
\def\@onedot{\ifx\@let@token.\else.\null\fi\xspace}

\makeatother

\definecolor{uclablue}{rgb}{0.15, 0.45, 0.68}
\definecolor{aliceblue}{RGB}{178, 217, 245}
\definecolor{babyblue}{RGB}{217, 239, 251}
\definecolor{babypink}{RGB}{251, 231, 230}

\makeatletter
\patchcmd{\@maketitle}{\vskip21pt}{\vskip8pt}{}{}
\patchcmd{\@maketitle}{\removelastskip\vskip20pt}{\removelastskip\vskip10pt}{}{}
\makeatother

\begin{document}

\title[An Integrative Computational Approach to Predict Viral
Epitopes]{An Integrative Computational Approach to Predict Viral
Epitopes by Targeting the MHC-TCR Complexation}

\author[1]{\fnm{Jaya Vasavi} \sur{Pamidimukkala}}
\author[1,2]{\fnm{Roshan} \sur{Balaji}}
\author[1,2]{\fnm{Nirav Pravinbhai} \sur{Bhatt}}
\author*[1]{\fnm{Sanjib} \sur{Senapati}}\email{sanjibs@iitm.ac.in}
\affil[1]{\orgdiv{Department of Biotechnology},
  \orgname{BJM School of Biosciences, Indian Institute of Technology Madras},
  \orgaddress{\city{Chennai}, \country{India}}}
\affil[2]{\orgname{Wadhwani School of Data Science and AI, Indian Institute of Technology Madras},
  \orgaddress{\city{Chennai}, \country{India}}}

\abstract{T-cell immunity acts as a major defense system against
  controlling viral infections in vertebrates. During viral entry,
  innate immune cells degrade the viral proteins (antigens) and present
  them on their surface via Major Histocompatibility (MHC) proteins.
  T-cell receptors (TCRs) recognize these antigens/peptides presented
  by MHC (pMHC), initiating a T-cell-mediated immune response. Despite
  its significance, the mechanism by which pMHC-TCR binding triggers
  T-cell activation remains unclear. In this study, we employed an
  integrative computational approach combining Bioinformatics,
  Molecular Dynamics (MD) simulations, and Machine Learning (ML) to
  identify viral epitopes as potential vaccine candidates. We performed
  large-scale all-atom and coarse-grained MD simulations on
  MHC-peptide-TCR complexes embedded into dendritic and T-cells, for
  which experimental immunogenicity data is available. One hundred
  fifty such systems are simulated for 1 $\mu$s each to capture the
  conformational and dynamical changes that underlie T-cell activation.
  Our ML model (DynamiT), trained on simulation-derived structural and dynamical
  features extracted from 2500 time points, revealed key determinants
  responsible for T-cell activation with an accuracy of
  73.3\%. Notably, we have identified the bending of the TCR transmembrane region, major dynamic motions of the TCR$\alpha$ constant region and the buried surface area at the pMHC and TCR interface as
  critical factors influencing immune response initiation. Our approach
  unravels the mechanism of T-cell mediated immune response and helps
ML-guided screening of viral epitopes for vaccine development.}

\keywords{Antiviral Therapeutics, Epitope Prediction, Vaccine Design, Antigen Presentation, MHC-TCR Interactions, TCR Dynamics, Molecular Dynamics Simulations, Coarse-Grained Simulations, Membrane Protein Simulations, Machine Learning, Time-Series Classification}

\maketitle

\section{Introduction}

Viral infections continue to impose a substantial global health
burden, causing millions of deaths annually and
affecting populations worldwide. According to the Global Burden of
Disease 2021 study, viral infections caused an
estimated 8.7 million deaths and 259.2 million disability‑adjusted
life years (DALYs), accounting for nearly 13\% of global mortality and
highlighting their profound impact on public healthcare~\cite{li_global_2024}. Individual viral pathogens have
caused some of the most devastating outbreaks in history, including
HIV/AIDS, which has led to over 40 million deaths since the early
1980s, influenza pandemics with millions of fatalities, and the
recent COVID-19 pandemic with hundreds of millions of confirmed cases
and over 7 million deaths~\cite{who_health_stats}. More
recent estimates for 2024–2025 continue to underscore the ongoing
impact of viral diseases. Chronic viral hepatitis (B and C) accounted
for approximately 1.3 million deaths in 2024, while dengue virus caused around 14.1
million cases and roughly 9,500 deaths
globally~\cite{who_health_stats}\cite{haider_global_2025}. %
Other emerging viruses, such as Nipah, Ebola, Zika and the 2022-2025
Mpox epidemic, continue to pose localized health challenges.

Viruses present unique therapeutic challenges due to their rapid
mutation rates, antigenic variability and ability to evade host
immune responses, often bypassing both natural immunity and
existing antiviral therapies. For many clinically significant
viruses, effective licensed therapies or vaccines are unavailable or
limited, making it difficult in achieving long-term viral control.
This highlights the urgent need for novel preventive strategies.
Peptide-based vaccines have emerged as a promising alternative,
utilizing short, well-defined peptide fragments to elicit highly
targeted immune responses while minimizing allergenicity. Antiviral
peptides (AVPs) offer broad-spectrum activity, low cytotoxicity, and
multi-stage inhibition of viral replication, often combined with
immune-modulatory
effects~\cite{kuroki_broad_spectrum_2021}, thus
opening avenues for next-generation antiviral therapies.

Within the host immune response, T-cell immunity plays an important role in controlling viral infections, enabling recognition and elimination of
infected cells while establishing long-lived immunological memory.
CD8$^+$ cytotoxic T lymphocytes (CTLs) recognize viral peptides
presented on major histocompatibility complex class I (MHC I)
molecules in dendritic or infected cells, triggering targeted
cytolysis and limiting viral dissemination. Croft et al. found that
in vaccinia virus-infected mice, more than 80\% of MHC I presented
viral peptides were immunogenic, highlighting the broad capacity of
CD8$^+$ T cells to detect and respond to viral
antigens~\cite{croft_most_2019}. Thus, T-cell epitope based vaccines
are designed to generate long-lived memory T-cells capable of rapidly
activating effector functions upon re-exposure, thereby providing
durable protection even against viral variants that evade
antibody-mediated immunity, as observed with emerging pathogens such
as SARS-CoV-2~\cite{heitmann_covid_19_2022}. For example,
peptide-based vaccines such as CoVac-1,
comprising six HLA-DR-restricted SARS-CoV-2 peptides, induced robust
multifunctional CD4$^+$ and CD8$^+$ T-cell responses in vaccinated
individuals, demonstrating the feasibility of eliciting strong
cellular immunity in humans~\cite{heitmann_covid_19_2022}. Similarly,
the hepatitis C virus peptide vaccine IC41 having five
synthetic T-cell epitopes, elicited virus-specific T-cell responses, except  for patients receiving interferon therapy ~\cite{firbas_immunogenicity_2006}. These findings
emphasize the critical role of TCR-mediated immunity in antiviral
defense, contributing to viral clearance, memory formation, and
resilience against viral mutation, making T-cell based peptide
vaccine strategies a promising avenue for next-generation vaccines
and antiviral therapies.

Current TCR-mediated vaccine design strategies increasingly rely on integrated \textit{in-silico} pipelines that combine viral epitope identification, structural validation and MHC-epitope binding affinity predictions to prioritize candidates for downstream \textit{in-vitro} and \textit{in-vivo} evaluation. Recent advances in machine learning (ML) have fundamentally reshaped Immunoinformatics by enabling data-driven modeling of complex sequence-structure-function relationships from large-scale immune datasets. Tools such as NetMHCpan\cite{nilsson2025netmhcpan}, NetCTLpan\cite{stranzl2010netctlpan} etc., identify peptides likely to be processed and presented by integrating sequence-derived features, proteasomal cleavage, TAP transport efficiency and peptide-MHC binding affinity predictions, drawing on databases such as IEDB\cite{vita2025immune}, VDJdb\cite{shugay2018vdjdb} and SYFPEITHI\cite{rammensee1999syfpeithi}. More recently, predictive frameworks have expanded beyond antigen presentation to estimate peptide immunogenicity. For example, BigMHC\cite{bigmhc2023} employs an ensemble of pan-allelic deep neural networks trained on mass spectrometry-derived MHC-eluted ligands followed by transfer learning on experimentally validated immunogenicity datasets. Whereas ImmunoStruct\cite{immunostruct2025} integrates peptide sequence, peptide-MHC structural information and biochemical features within a multimodal deep learning framework to improve immunogenicity prediction and interpretability. However, peptide-MHC binding affinity alone does not reliably predict cytotoxic T lymphocyte (CTL) responses, as immunogenicity does not strictly correlate with MHC binding strength \cite{ochoa1997ability}. 

TCRs continuously scan pMHC complexes on antigen-presenting cells, yet only specific interactions initiate signaling. Effective immune activation ultimately requires recognition of the peptide-MHC complex by a cognate TCR, underscoring the need to model pMHC-TCR interactions. Accordingly, TCR-specific frameworks such as POPISK\cite{tung2011popisk}, NetTCR2.2\cite{jensen2024elife93934}, PRIME2.1\cite{schmidt2023prime2}, T-SCAPE\cite{morrone2024tscape} have leveraged experimentally validated TCR-epitope interaction datasets curated in publicly available immunological databases \cite{vita2025immune,shugay2018vdjdb,rammensee1999syfpeithi}  to learn statistical relationships between complementarity-determining regions (CDRs), particularly the hypervariable CDR3 loops, and epitope recognition. For instance, PRIME2.1 combines antigen presentation predictions with experimentally validated neoepitope datasets to estimate the likelihood of productive T-cell recognition \cite{schmidt2023prime2}, whereas T-SCAPE employs adversarial domain adaptation to integrate heterogeneous immunological information, including MHC presentation, peptide-MHC binding, pMHC-TCR interactions, source organism information and T-cell activation \cite{morrone2024tscape}. Despite these advances, current models predominantly learn statistical relationships from static sequence or structural datasets, providing limited mechanistic insight into the dynamic molecular determinants governing selective TCR activation. Consequently, the conformational changes associated with TCR signaling remain poorly understood.

In this context, molecular dynamics (MD) simulations provide a
complementary and mechanistic framework for understanding how dynamic
structural fluctuations triggers T-cell activation. Early MD studies
by Knapp et al. explored conformational
variations in the
LC13 TCR when bound to immunogenic and non-immunogenic peptides using
all-atom simulations of multiple pMHC-TCR complexes, but reported no
striking structural differences between the two classes \cite{knapp2014large}. Subsequent
simulations of different TCRs in bound and unbound states similarly
failed to reveal clear discriminative features \cite{knapp2019mhc}.
More recent studies have incorporated membrane environments to better mimic physiological
conditions. Alba et al. performed
membrane-bound all-atom
simulations of pMHC alone and pMHC-TCR complexes and demonstrated
that TCR binding restricts the conformational space of pMHC and that
fluctuations in the TCR$\beta$ constant region correlate with immune
response \cite{alba2020molecular}. Extending this work, the same group reported increased
solvent exposure of CD3$\varepsilon$2 and CD3$\delta$ regions upon
pMHC binding in membrane-bound pMHC-TCR-CD3 complexes, suggesting a
mechanistic link between extracellular recognition and intracellular
signaling. Similarly, Floris et al. observed that pMHC binding
induces TCR rigidification and decoupling of the CD3 complex,
facilitating immunoreceptor tyrosine-based activation motif (ITAM)
phosphorylation and signal propagation \cite{van2023tcr} showed. However, it is important to
note that most existing MD studies have primarily compared bound and
unbound states of pMHC-TCR or pMHC-TCR-CD3 complexes, whereas in
physiological settings TCRs continuously sample and bind a broad
spectrum of peptides presented by MHC molecules, including
self-antigens with weak affinity (\cref{sch:overflow}). Discriminating immunogenic from
non-immunogenic interactions therefore requires understanding how TCR
conformational dynamics differ across functionally distinct bound
states, rather than simply between bound and unbound configurations.
Capturing these subtle yet functionally relevant dynamic signatures
is essential for elucidating the mechanistic basis of TCR-mediated
immune activation.
\begin{scheme}[!t]
  \centering
  \includegraphics[width=\linewidth]{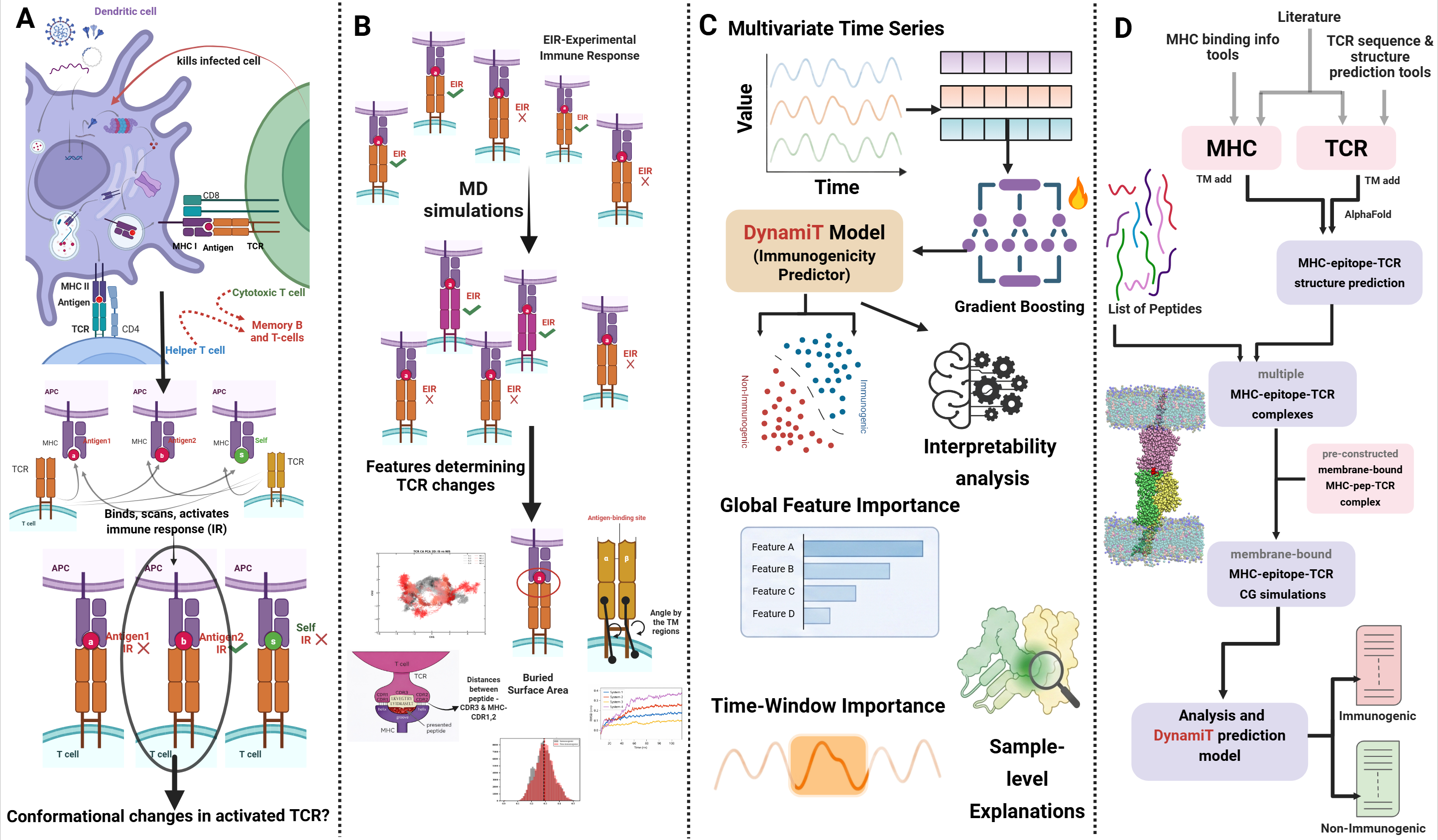}
  \caption{\textbf{Overview of the proposed epitope-prediction pipeline (DynamiT) integrating molecular dynamics (MD), and machine learning. }(A) T-cell activation is initiated when a T-cell receptor (TCR) recognizes a peptide--MHC (pMHC) complex and only a subset of pMHC--TCR interactions induces the conformational changes required for downstream signaling. (B) MD simulations of multiple immunogenic-response (IR) and non-immunogenic-response (NIR) pMHC--TCR complexes to capture conformational changes and generate structural/dynamical descriptors. (C) Time-series features extracted from the simulations are used to build the DynamiT framework. DynamiT classifies peptide immunogenicity while identifying the key conformational features that drive T-cell activation. (D) End-to-end application of the pipeline from candidate peptides to immunogenicity predictions.}
  \label{sch:overflow}
\end{scheme}

In this study, we adopt an integrative computational framework
combining bioinformatics tools, molecular dynamics (MD) simulations,
and machine learning (ML) algorithms to identify viral epitopes as
promising candidates for peptide vaccine design. Bioinformatics-based
methods are first employed to predict a list of antigenic epitopes of a given
virus based on sequence composition, antigenicity, and binding
affinity to major histocompatibility complex (MHC) molecules. For an
effective immune response to occur, the resulting peptide-MHC (pMHC)
complex must be recognized by a cognate T cell receptor (TCR),
thereby initiating downstream signaling cascades. To address this, we perform
large-scale all-atom and coarse-grained MD simulations to elucidate
the conformational and dynamical changes in TCRs upon binding to pMHC
complexes that drive T-cell activation. Subsequently, machine
learning algorithms are applied on the MD-derived trajectories data to
extract informative dynamic descriptors associated with T-cell
immunogenicity and to generate an Immunogenicity prediction model (DynamiT) to screen the most viable viral epitopes for
vaccine development (\cref{sch:overflow}).

\section{Results}
In this study, we focus on identifying conformational changes within the T cell receptor (TCR) that distinguish immunogenic from non-immunogenic pMHC complexes using MD simulations and time-series based ML classifiers. Although crystal structures of peptide-MHC-TCR (pMHC-TCR) complexes have provided valuable insights, they represent static snapshots and are often insufficient to differentiate immunogenic from non-immunogenic complexes \cite{ding_four_1999}, motivating the use of a dynamic, time-resolved framework. To achieve this, we performed 1~$\mu$s coarse-grained (CG) simulations of membrane-embedded pMHC-TCR complexes representing experimentally characterized immunogenic and non-immunogenic peptide variants to generate a large-scale dataset of TCR conformational dynamics. The resulting trajectories were analyzed to characterize TCR dynamics, develop an interpretable time-series machine learning classifier for immunogenicity prediction, and identify structural signatures associated with productive T-cell activation. Finally, the robustness and generalizability of these signatures were evaluated using independent pMHC-TCR complexes spanning multiple viral systems as an out-of-class test set.
\begin{figure}[!htbp]
\centering
\includegraphics[width=1\textwidth]{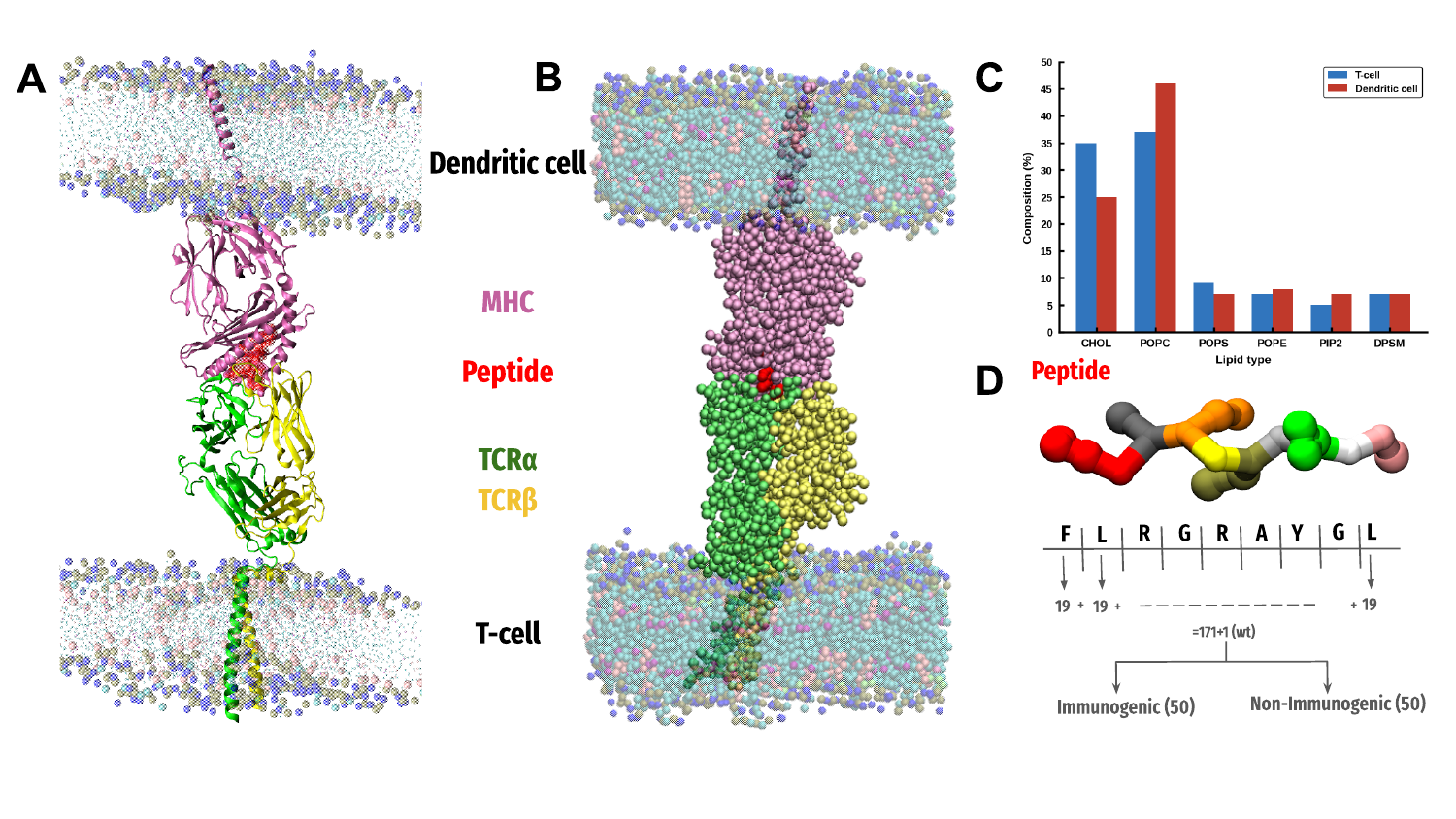}
\caption{\textbf{Construction of membrane-embedded pMHC-TCR systems and generation of the simulation dataset.} (A, B) Atomistic and coarse-grained (CG) representations of the MHC-peptide-TCR complex with transmembrane (TM) domains embedded in the dendritic cell and T-cell membranes. (C) Lipid composition of dendritic cell and T-cell membranes used in the simulations. (D) Representation of training set peptide composition and systematic peptide mutagenesis to produce immunogenic and non-immunogenic variants.}
\label{fig:fig1}
\end{figure}

 To establish the training dataset, we selected the well-characterized pMHC-TCR system reported by Kjer-Nielsen et al. \cite{kjer-nielsen_structural_2003}, in which the Epstein-Barr virus (EBV)-derived epitope FLRGRAYGL from the EBNA3A protein is presented by HLA-B08:01 and recognized by the LC13 TCR. The authors performed residue-wise mutations of the peptide and experimentally quantified the resulting T-cell responses, making it a well-suited system for linking conformational dynamics to functional activation. Using the crystal structure of the HLA-B*08:01–FLRGRAYGL–LC13 TCR complex (PDB ID: 1MI5), we generated both all-atom and Martini coarse-grained models of the membrane-embedded pMHC–TCR complex (Fig.~\ref{fig:fig1}A,B). The transmembrane regions of the MHC and TCR were first modeled and subsequently embedded into biologically relevant dendritic-cell and T-cell membrane compositions, respectively, using experimentally derived lipid compositions  \cite{luhr_maturation_2020,zech_accumulation_2009} (Fig.~\ref{fig:fig1}C; methodology detailed in Section~\ref{sec:methods}). Systematic single-residue mutations of the wild-type EBV peptide were  introduced to generate a dataset comprising 50 immunogenic systems (IS) and 50 non-immunogenic systems (NIS) having peptide variants based on experimentally measured T-cell response (Fig.~\ref{fig:fig1}D). Immunogenic variants correspond to the top 50 peptides have exhibiting high 50\% lysis at a peptide concentration of $10^{-7}$~M, whereas non-immunogenic variants failed to induce 50\% lysis at any tested peptide concentration \cite{kjer-nielsen_structural_2003}. Each membrane-embedded pMHC-TCR complex was subsequently subjected to 1~$\mu$s coarse-grained molecular dynamics simulations, as described in Section~\ref{sec:methods}.

\subsection{Membrane Dynamics Supports System Architecture}
Before performing large-scale simulations of the complete dataset, we first evaluated whether the CG simulation protocol accurately reproduces the structural differences observed experimentally between immunogenic and non-immunogenic pMHC-TCR complexes. For this purpose, we selected four experimentally characterized peptide variants sharing the same TCR and MHC but exhibiting distinct immunogenic outcomes (two immunogenic and two non-immunogenic complexes) with available crystal structures~\cite{ding_four_1999} (PDB IDs: 1AO7 (WT), 1QSE (V7R), 1QRN (P6A) and 1QSF (Y8A)). Starting from the wild-type structure (PDB ID: 1AO7), point mutations corresponding to V7R, P6A and Y8A were introduced, followed by modeling of the transmembrane regions and construction of full-length membrane-embedded pMHC-TCR complexes (Fig.~\ref{fig:fig1}A,B). All-atom (100 ns) and Martini coarse-grained (1~$\mu$s) simulations were performed in triplicates as discussed in Section~\ref{sec:methods}, allowing direct assessment of the ability of the CG framework to reproduce experimentally observed mutation-dependent structural differences.

\begin{figure}[!htbp]
\centering
\includegraphics[width=1\textwidth]{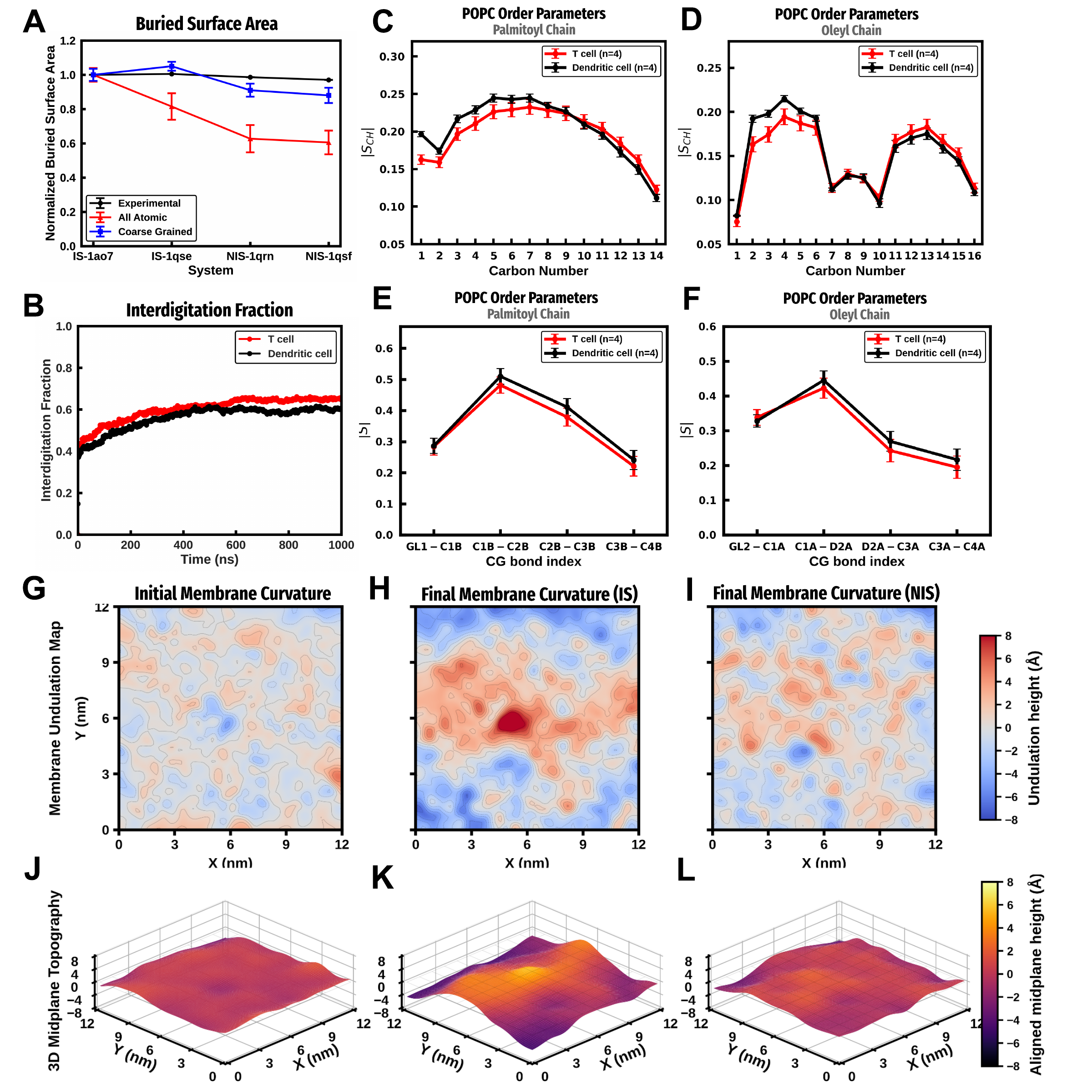}
\caption{\textbf{Validation of the coarse-grained simulation framework.} (A) Comparison of normalized buried surface area (BSA) for four pMHC-TCR complexes obtained from experimental structures, all-atom simulations, and coarse-grained (CG) simulations. All-atom and CG systems were generated from the wild-type structure by introducing the corresponding peptide mutations. (B) Leaflet interdigitation between the dendritic-cell and T-cell membranes, demonstrating stable overlap between opposing membrane leaflets throughout the simulations. Lipid order parameters ($S_{CD}$) of POPC for the palmitoyl and oleoyl acyl chains in the T-cell and dendritic-cell membranes obtained from all-atom (C,D) and CG (E,F) simulations. Membrane curvature of immunogenic (IS) and non-immunogenic (NIS) systems at the initial simulation frame (G,J) and during the production run (H,I,K,L). (G-I) Two-dimensional membrane undulation maps showing local deviations of the membrane midplane relative to the mean membrane plane. (J-L) Corresponding three-dimensional membrane surface reconstructions highlighting differences in membrane curvature between IS (K) and NIS (L) systems.}
\label{fig:fig2}
\end{figure}

Although the experimentally resolved crystal structures differ by only ~0.6 Å RMSD, they exhibit distinct buried surface areas (BSA) at the pMHC-TCR interface between immunogenic and non-immunogenic complexes \cite{ding_four_1999}. We therefore used interface BSA as an independent structural metric to evaluate the simulation protocols. Experimentally determined immunogenic complexes exhibited slightly larger interface BSA (20.73 and 20.85 nm$^2$) than the non-immunogenic complexes (20.46 and 20.12 nm$^2$)\cite{ding_four_1999}. To facilitate comparison of relative BSA trends, values in (\cref{fig:fig2}A) were normalized independently within each dataset (Experimental, AA and CG). The CG simulations closely reproduced the experimental trend, maintaining higher normalized BSA values for the immunogenic complexes than the non-immunogenic complexes. In contrast, the all-atom simulations showed a substantially larger reduction in normalized BSA, particularly for the non-immunogenic variants, resulting in greater deviation from experimental values (\cref{fig:fig2}A). This improved agreement of CG simualtions with the experimental structural details, likely due to the enhanced conformational sampling at longer timescales enabling mutation-induced structural rearrangements to relax towards conformations that are more closely to the experimentally observed structures. The inclusion of membrane environments, combined with extended sampling, further enables CG simulations to capture biologically relevant TCR dynamics more realistically. Given the substantial computational cost of all-atom simulations for large membrane-embedded systems at microsecond timescales, we therefore adopted the CG framework for simulating multiple pMHC-TCR complexes. 
Since TCR activation occurs within a highly heterogeneous cellular membrane, we next assessed whether the simulations accurately reproduce key membrane properties.

Leaflet interdigitation remained stable throughout the simulations, with an average normalized overlap fraction of approximately 0.6 (Fig.~\ref{fig:fig2}B), indicating stable coupling between opposing membrane leaflets and the absence of membrane packing defects \cite{venturoli2008mesoscopic}. This demonstrates that insertion of the pMHC-TCR complex does not perturb bilayer integrity, thereby preserving a realistic environment for transmembrane protein dynamics. Lipid order parameters ($S_{CD}$) were calculated for POPC, one of the major phospholipid components of both membrane models. As membrane order is strongly influenced by cholesterol content and lipid composition \cite{plesnar2011cholesterol}, the simulated order parameters were compared with previously reported experimental and computational values \cite{ferreira_cholesterol_popc_order_2013,piggot_acyl_chain_order_2017}. Both all-atom and CG simulations reproduced the expected qualitative ordering of lipid acyl chains (Fig.~\ref{fig:fig2}C-F), with minor deviations attributing to the more complex lipid compositions employed in the present cell-mimetic membrane models compared with simpler reference bilayers. Collectively, these results confirm that lipid packing is preserved throughout the simulations. We next examined membrane curvature, given its established role in regulating transmembrane protein organization and receptor activation \cite{jiang_membrane_curvature_2025}. At the beginning of the simulations, the membrane remained largely planar with minimal surface undulations (Fig.~\ref{fig:fig2}G,J). As the simulations progressed, localized membrane curvature emerged around the TCR transmembrane domain. In IS, these deformations became progressively more pronounced, producing membrane height differences approaching ~8~\AA{} (Fig.~\ref{fig:fig2}H,K), whereas NIS exhibited only slight curvature throughout the simulations (Fig.~\ref{fig:fig2}I,L). Analysis of multiple independent IS and NIS further confirmed that membrane curvature is consistently more pronounced in immunogenic complexes (Supplementary Fig.~S1). These observations suggest that productive TCR activation is accompanied by localized membrane remodeling, consistent with previous studies linking membrane curvature to lipid organization, receptor clustering and membrane mechanics during immune activation \cite{farrell_tcell_2020}.

Collectively, the agreement in interface BSA and membrane properties demonstrates that the coarse-grained simulation protocol captures the structural and biophysical characteristics of membrane-embedded pMHC-TCR complexes, supporting its application to the complete dataset for subsequent analyses of TCR dynamics and immunogenicity prediction.

\subsection{Data determining the complex TCR dynamics}

Following validation of the coarse-grained simulation protocol, we
performed 1~$\mu$s simulations for 100 pMHC-TCR complexes, comprising
50 immunogenic systems (IS) and 50 non-immunogenic systems (NIS), as
defined by experimentally measured immune responses
\cite{kjer-nielsen_structural_2003}. All comparative analyses were
restricted to the final 500~ns of each trajectory to minimize the
influence of initial structural relaxation.

We first examined commonly used structural descriptors, including the
distances between the peptide and the CDR3$\alpha$ and CDR3$\beta$
loops (as CDR3$\alpha$ and CDR3$\beta$ determine the peptide recognition), buried surface area (BSA) at the pMHC-TCR interface,
TCR-MHC distance, root-mean-square deviation (RMSD), and radius of
gyration ($R_g$) of different domains of the TCR. The distributions of these individual descriptors
showed substantial overlap between the IS and NIS data points, and
their average temporal profiles did not exhibit consistent
class-wise separation (Supplementary Fig.~S2). Considerable
system-to-system variation was also observed within each class.
Thus, immunogenicity could not be explained by a single structural
descriptor or a universal threshold in the raw feature space.

We therefore hypothesized that immunogenicity is encoded by
coordinated changes across multiple structural and dynamical
properties of the pMHC--TCR complex. To capture these collective
signatures, we extracted 25 time-dependent biophysical descriptors
spanning three interconnected aspects of TCR dynamics: structural
stability, interfacial organization, and domain-level conformational
motion (Supplementary Table.~1). Structural stability was characterized using the RMSD and
$R_g$ of the TCR variable ($V_\alpha$ and $V_\beta$), constant
($C_\alpha$ and $C_\beta$), and transmembrane (TM) regions.
Interfacial organization was described using the BSA at the
pMHC--TCR interface, distances between the peptide and the
CDR3$\alpha$ and CDR3$\beta$ loops, distances between the MHC and
the germline-encoded CDR1$\alpha$, CDR2$\alpha$, CDR1$\beta$, and
CDR2$\beta$ loops, and intermolecular contacts defined using a
0.6~nm cutoff (detailed in \cref{sec:methods}). These descriptors
provided complementary measures of global structural stability and
local recognition at the pMHC--TCR interface. To determine whether peptide recognition produces conformational
changes beyond the binding interface, we additionally characterized
domain-level motions of the $V_\alpha$, $V_\beta$, $C_\alpha$ and
$C_\beta$ regions. The orientations of the TCR$\alpha$ and
TCR$\beta$ TM helices relative to their corresponding constant
domains were also monitored. For each chain, the TM bending angle
was defined using three centres of mass: the constant domain and the
N- and C-terminal segments of the TM helix, represented by its first
and last four residues, respectively. The analysed TM spans
correspond to VIGFRILLLKVAGFNLLMTLRLW in TCR$\alpha$ and
TILYEILLGKATLYAVLVSALVLMAM in TCR$\beta$. Complete definitions and
calculation procedures for all descriptors (Supplementary Table.~1) are provided in
Section~\ref{sec:methods}. This representation therefore connected
local peptide recognition with conformational changes across the
complete TCR complex.

Each system contributed 2,500 time points from the final 500~ns of
its trajectory. The complete dataset was therefore represented as
$\rmX\in\mathbb{R}^{100\times2500\times25}$, where the dimensions
correspond to systems, time points, and biophysical features,
respectively. Preprocessing adjusted the temporal dimension to meet
the MANTIS input requirement, which produced
$\rmX_{\mathrm{MANTIS}}\in\mathbb{R}^{100\times2496\times25}$.
This adjustment retained all 25 features, the balanced class
composition, and the system-level unit of analysis. The resulting
representation preserved coordinated and time-dependent changes
across the pMHC--TCR interface, extracellular TCR domains, connecting
regions, and TM helices. It therefore supported system-level
classification based on the combined dynamics of the complete
complex.

\subsection{Performance on an independent external cohort}
\label{sec:model_performance}

\begin{figure}[!htbp]
\centering
\includegraphics[width=\linewidth]{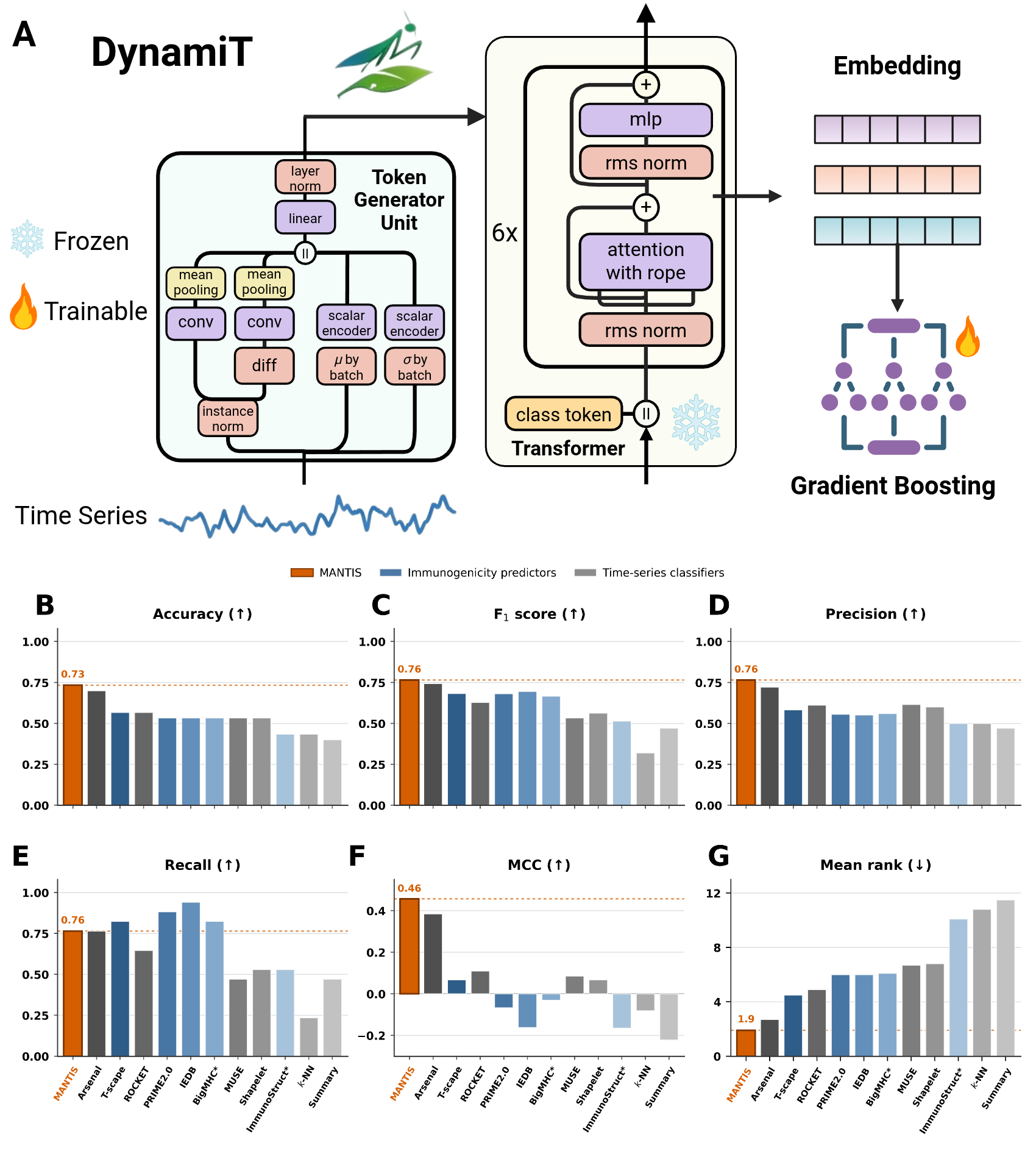}
\caption{\textbf{Performance of DynamiT and the baseline methods
on the independent external cohort ($n=30$).}
(A) DynamiT architecture. (B) Accuracy, (C) $F_1$ score,
(D) precision, (E) recall, and (F) Matthews correlation
coefficient (MCC). (G) Mean rank across the five performance
metrics. DynamiT achieved the highest accuracy, $F_1$ score,
precision, and MCC, and obtained the best mean rank. The asterisk
identifies the zero-shot evaluations of BigMHC and ImmunoStruct.
The arrows indicate whether higher or lower values are preferred.}
\label{fig:model_perf}
\end{figure}

DynamiT achieved the strongest overall performance on the independent
external cohort of 30 pMHC--TCR systems
(\cref{fig:model_perf}). This cohort contained 17 immunogenic and 13
non-immunogenic systems. Across the 12 evaluated methods, DynamiT
obtained the highest accuracy ($0.7333$), $F_1$ score ($0.7647$),
precision ($0.7647$), and Matthews correlation coefficient
(MCC; $0.4570$). Its recall was $0.7647$.

Arsenal was the strongest time-series baseline. DynamiT increased
accuracy from $0.7000$ to $0.7333$, $F_1$ score from $0.7429$ to
$0.7647$, precision from $0.7222$ to $0.7647$, and MCC from $0.3845$
to $0.4570$. Both models achieved a recall of $0.7647$. DynamiT also
outperformed ROCKET, MUSE, the Shapelet Transform, $k$-nearest
neighbours, and the summary-statistics classifier across the overall
set of metrics. The summary-statistics classifier achieved an
accuracy of $0.4000$ and an MCC of $-0.2217$. This result was
consistent with a loss of relevant temporal information when each
trajectory was reduced to global summary values.

The existing immunogenicity predictors showed a different performance
profile. T-SCAPE, PRIME2.1, the IEDB model, and zero-shot BigMHC
achieved recall values from $0.8235$ to $0.9412$. However, their
precision ranged from $0.5517$ to $0.5833$, and their MCC ranged from
$-0.1624$ to $0.0673$. Thus, their higher recall was accompanied by
more false-positive predictions and weaker discrimination between the
two classes. Zero-shot BigMHC achieved an accuracy of $0.5333$ and an
MCC of $-0.0301$. Zero-shot ImmunoStruct achieved an accuracy of
$0.4333$ and an MCC of $-0.1648$.

DynamiT obtained the best mean rank across the five metrics ($1.9$),
followed by Arsenal ($2.7$) and T-SCAPE ($4.5$). These results show
that DynamiT provided the most balanced performance among the methods
evaluated on this cohort. They also support the use of coordinated
pMHC--TCR dynamics as predictive information that complements
sequence- and structure-based immunogenicity scores. However, the
small external cohort limits the scope of this comparison, and larger
independent datasets are required to establish broader
generalizability.

\subsection{Key TCR dynamics associated with immunogenicity}
\label{sec:key_tcr_dynamics}

\begin{figure}[!htbp]
\centering
\includegraphics[width=\linewidth]{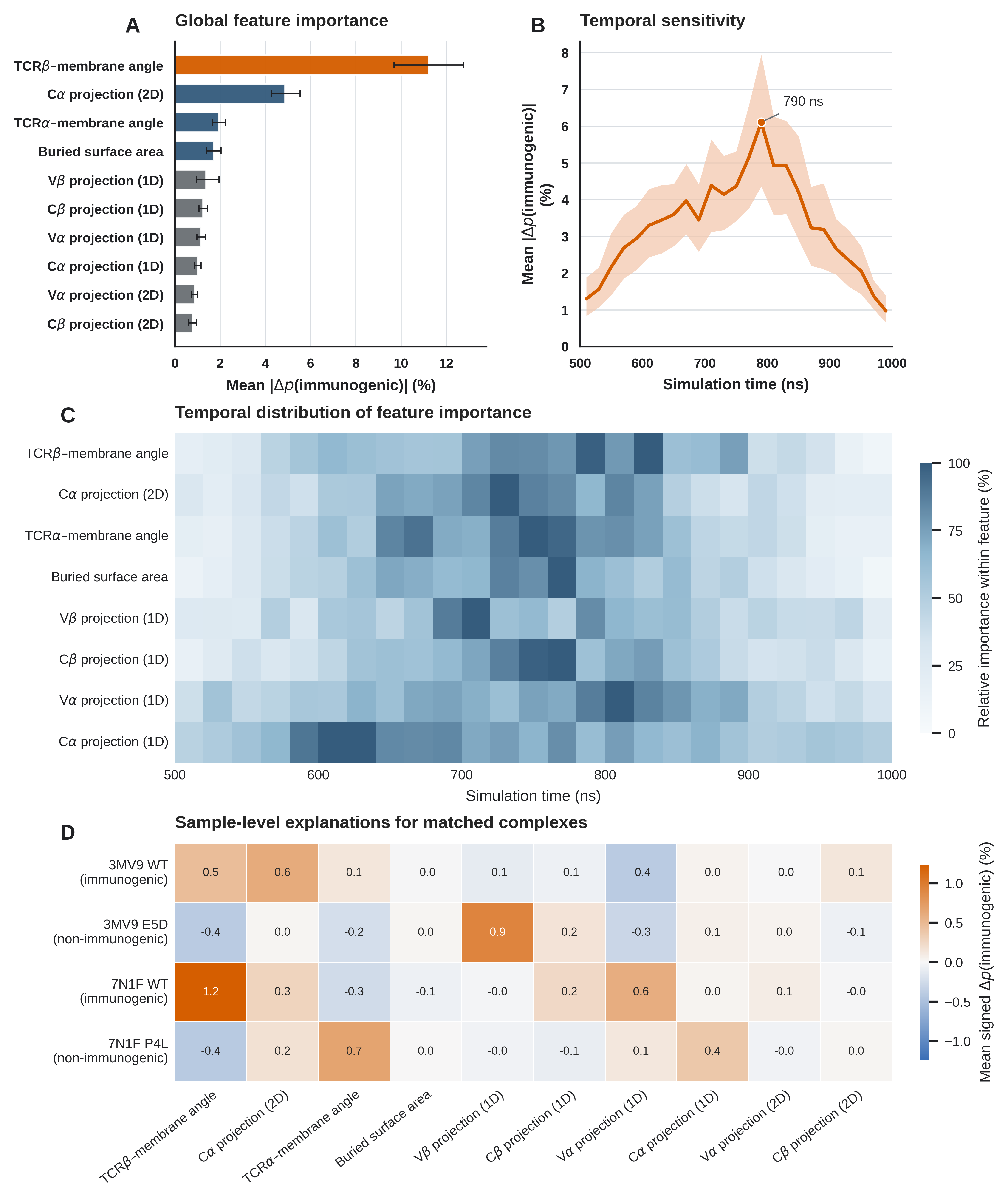}
\caption{\textbf{Feature- and time-resolved attributions associated
with predicted immunogenicity.}
(A) Grouped feature ablation identifies the TCR$\beta$ TM angle
as the leading global feature. (B) Temporal attribution reaches
its maximum at 790~ns. (C) Within-feature normalization shows the
most informative periods for each dynamical coordinate.
(D) Signed local attributions for matched wild-type and
substitution pairs show changes in the feature contributions
associated with the annotated class. Error bars in panel D show
95\% bootstrap confidence intervals. All attribution analyses
used the independent external test complexes.}
\label{fig:interpretability}
\end{figure}

The attribution analysis linked DynamiT predictions to coordinated
motions across the pMHC--TCR complex
(\cref{fig:interpretability}). Grouped feature ablation ranked the
TCR$\beta$ TM angle as the strongest contributor
(\cref{fig:interpretability}A). The dominant motion of the
TCR$\alpha$ constant domain ranked second. The TCR$\alpha$ TM angle
and the buried surface area (BSA) at the pMHC--TCR interface also
ranked among the leading features. These signals span the binding
interface, extracellular constant domains, and membrane-proximal
regions. DynamiT therefore used information distributed across the
receptor rather than one static interfacial descriptor.

Temporal importance varied across the analyzed trajectory and reached
its maximum at 790~ns (\cref{fig:interpretability}B). The strongest
classification signal therefore occurred during the later part of the
simulation. The channel-by-time map showed that different features
contributed during distinct periods
(\cref{fig:interpretability}C). Each row was normalized within its
corresponding feature. The colour scale therefore identifies the most
informative periods for each feature but does not compare absolute
importance between features. These results indicate that the model
used feature-specific temporal patterns rather than trajectory
averages alone.

Local attributions showed how these signals differed between matched
systems (\cref{fig:interpretability}D). The immunogenic wild-type
complexes 3MV9 and 7N1F were compared with their non-immunogenic
substitution variants, 3MV9 E5D and 7N1F P4L, respectively. Both
substitutions changed the features that supported or opposed the
experimentally assigned class. Thus, a single amino acid substitution
was associated with a change in the dynamic evidence used by the
model, despite the structural similarity of each matched pair. These
attributions describe model sensitivity and do not establish that
either substitution causes the observed immunogenicity outcome. The
95%
estimated contributions.

Together, the global, temporal, and local analyses associated
immunogenicity prediction with time-dependent motions across the
binding interface, TCR constant domains, connecting regions, and TM
helices. The following sections examine these leading features and
their possible structural roles.

\subsubsection{TCR TM Bending - Tunes Immunity?}
Interpretability analysis revealed the TCR$\beta$ transmembrane (TM) bending angle as the most prominent feature distinguishing immunogenic systems (IS) from non-immunogenic systems (NIS), with the TCR$\alpha$ TM bending angle emerging as the third most important feature. These features quantify the orientation of the TM helix relative to the extracellular constant domain and were designed to capture conformational changes that may be transmitted from the extracellular TCR domains toward the membrane-proximal signalling machinery. The TM bending angle was defined geometrically using three points: the centroid of the TCR constant domain, the start of the TM region, and the end of the TM region (detailed in Section~\ref{sec:methods}). The TCR$\alpha$ and TCR$\beta$ TM helices anchor the TCR within the T-cell membrane and are closely associated with the CD3 signaling complex. These CD3 chains contain immunoreceptor tyrosine-based activation motifs (ITAMs), which are phosphorylated upon TCR activation, initiating downstream signalling pathways including ZAP-70 recruitment, MAPK, NF-$\kappa$B, and calcium signalling~\cite{richard_staggered_2021}. Consequently, structural rearrangements within the TCR, particularly those propagating toward the TM region, are important for initiating intracellular signaling through the CD3 complex. We therefore examined whether TM bending dynamics differ systematically between IS and NIS.

\begin{figure}[!htbp]
  \centering
  \includegraphics[width=1\textwidth]{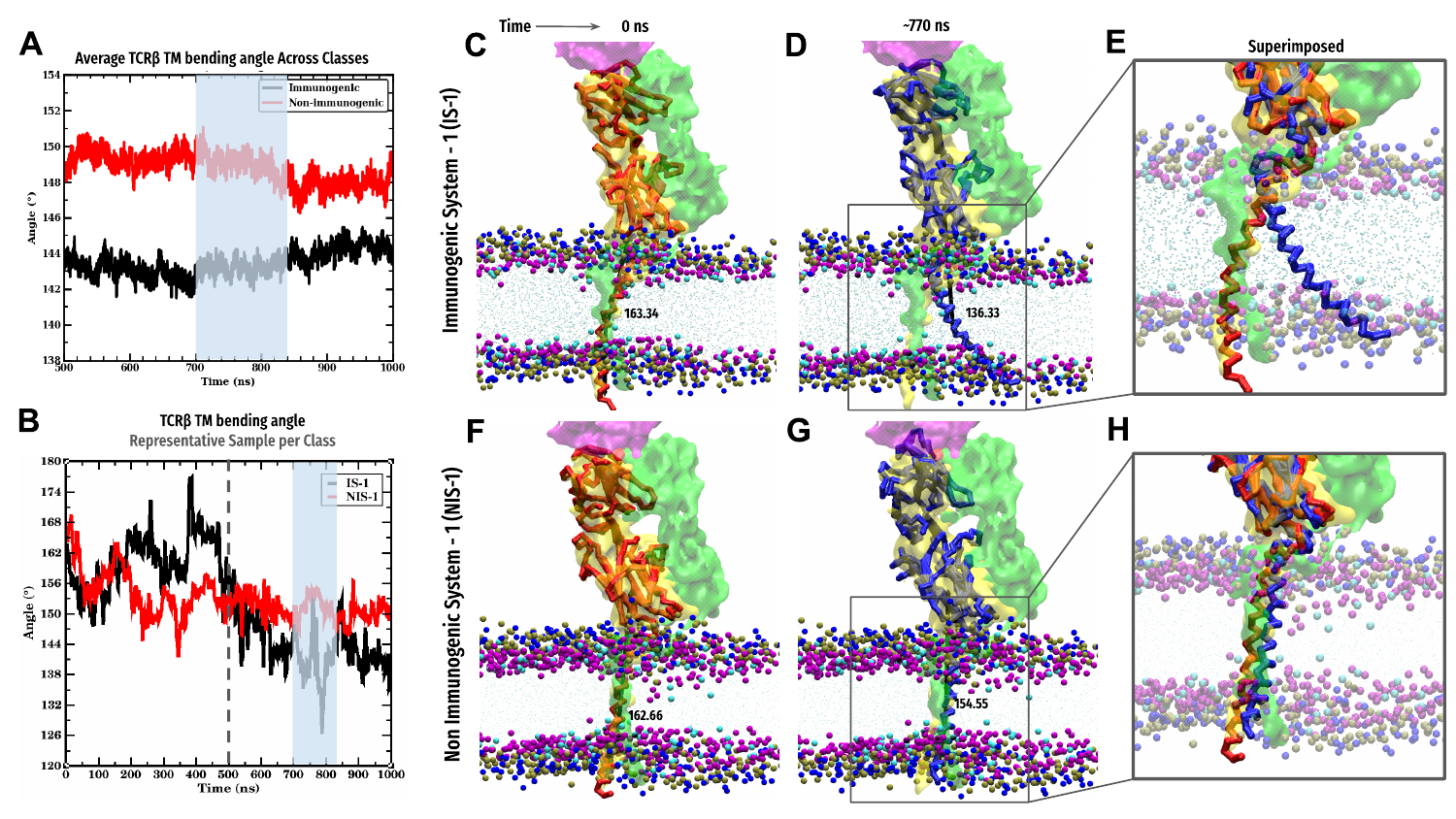}
  \caption{\textbf{TCR$\beta$ transmembrane (TM) bending dynamics in immunogenic and non-immunogenic systems. }(A) Average TCR$\beta$ TM bending angle for all immunogenic (IS) and non-immunogenic systems (NIS) during the final 500~ns of the simulations (500~ns--1~$\mu$s). (B) Time evolution of TCR$\beta$ TM bending angle for one representative IS (IS-1) and NIS system (NIS-1). The blue shaded region indicates the critical window identified by the interpretability analysis as contributing most strongly to the predictive importance of this feature. (C--H) Structural comparison of the TCR$\beta$ TM region in IS-1 and NIS-1, showing the initial structure (red), the average structure from the critical window (blue), and their superimposed orientations. Panels C--E correspond to IS-1 and illustrate a pronounced bending motion, whereas panels F--H correspond to NIS-1 and show comparatively limited bending. In all structural panels, the equilibriated MHC, TCR$\alpha$, and TCR$\beta$ are shown in magenta, green, and yellow, respectively.}
  \label{fig:tcrb_bend}
\end{figure}

\begin{figure}[!htbp]
  \centering
  \includegraphics[width=1\textwidth]{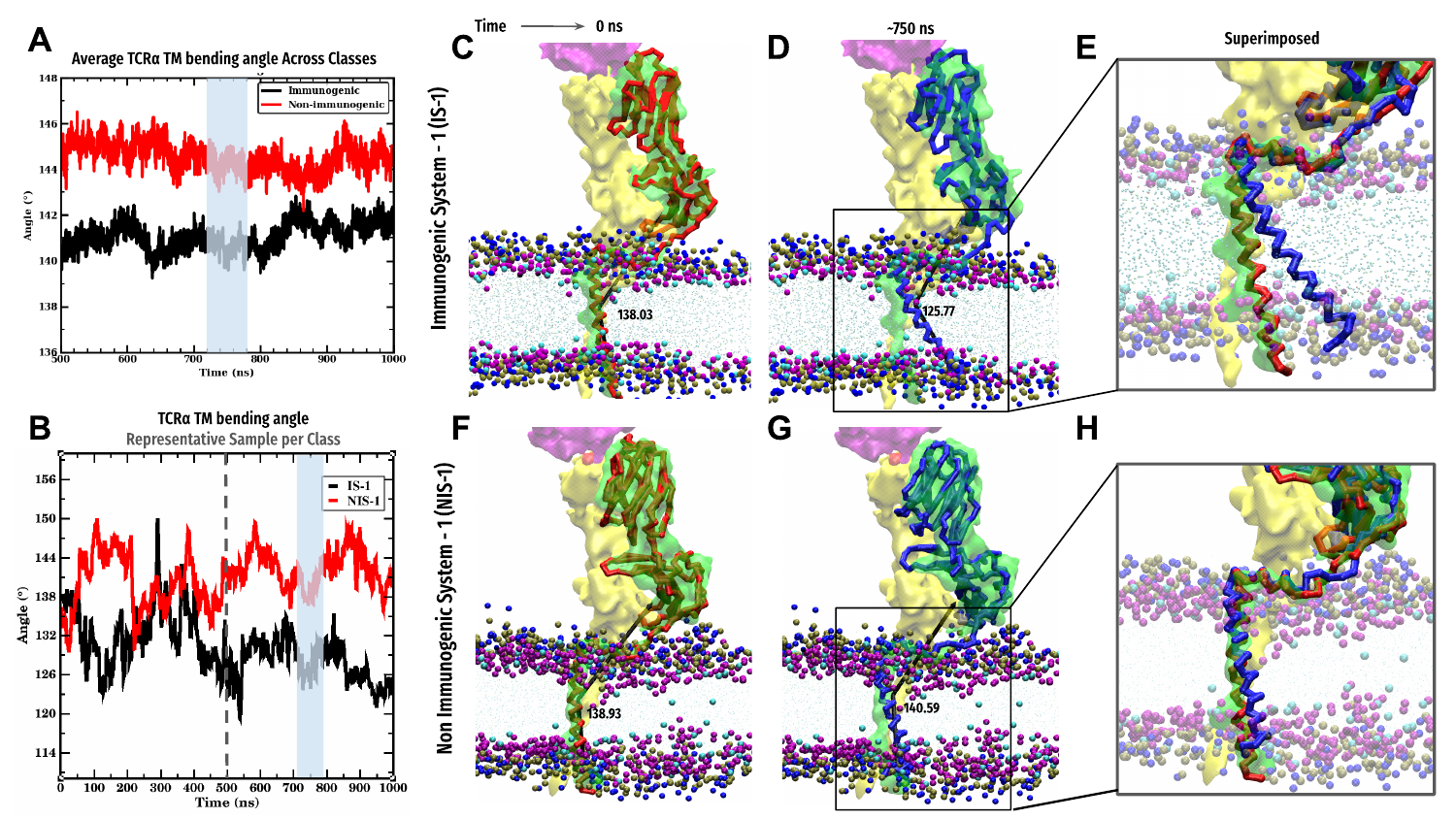}
  \caption{\textbf{TCR$\alpha$ transmembrane (TM) bending dynamics in immunogenic and non-immunogenic systems}. (A) Average TCR$\alpha$ TM bending angle for all immunogenic (IS) and non-immunogenic systems (NIS) during the final 500~ns of the simulations (500~ns--1~$\mu$s). (B) Time evolution of TCR$\alpha$ TM bending angle for one representative IS (IS-1) and NIS system (NIS-1). The blue shaded region indicates the critical window identified by the interpretability analysis as contributing most strongly to the predictive importance of this feature. (C--H) Structural comparison of the TCR$\alpha$ TM region in IS-1 and NIS-1, showing the initial structure (red), the average structure from the critical window (blue), and their superimposed orientations. Panels C--E correspond to IS-1 and illustrate a pronounced bending motion, whereas panels F--H correspond to NIS-1 and show comparatively limited bending. In all structural panels, the equilibriated MHC, TCR$\alpha$, and TCR$\beta$ are shown in magenta, green, and yellow, respectively.}
  \label{fig:tcra_bend}
\end{figure}

Class-averaged trajectories show a clear separation between IS and NIS throughout the final 500~ns of the simulations (\cref{fig:tcrb_bend}A), with interval between 700 and 840~ns contributing most strongly to the feature importance identified by our interpretability analysis (\cref{fig:interpretability}C). Though the average bending angle difference between the two classes remains nearly same, this temporal window was assigned the highest importance by the model. IS consistently adpot  lower TM bending angles ($\sim144^\circ$) than NIS ($\sim148^\circ$), indicating that the TCR$\beta$ TM helix acquires more tilted orientation relative to the extracellular domain and away from the membrane normal. Representative trajectories (IS-1, NIS-1) further illustrate this behaviour (\cref{fig:tcrb_bend}B). Following the initial equilibration phase (0-500 ns), both systems exhibit stable TM bending angles. Over time, the representative immunogenic system (IS-1) undergoes a pronounced reorientation of the TCR$\beta$ TM helix, with the TM bending angle decreasing by approximately $27^\circ$ ($\sim16.5\%$) (\cref{fig:tcrb_bend}D). In contrast, the representative non-immunogenic system (NIS-1) shows only a minor change in TM orientation (\cref{fig:tcrb_bend}G), indicating that large-scale TM reorientation is characteristic of the immune response. Similar divergence is observed across additional representative pairs (Supplementary Fig.~3), suggesting that increased bending of the TCR$\beta$ TM domain as a consistent structural signature of immunogenic triggering. 

Parallely, the TCR$\alpha$ TM bending with respect to its constant domain also emerges as an important feature (\cref{fig:tcra_bend}) with ~720 to 780 ns as the critical window (\cref{fig:interpretability}C). Across the simulations, IS display a lower average TCR$\alpha$ TM bending angle (~141°) compared to NIS systems (~144°), implying a more pronounced bending motion of TCR$\alpha$ away from the Z-axis over time (\cref{fig:tcra_bend}A). This difference is particularly evident in representative trajectories (IS-1, NIS-1). In \cref{fig:tcra_bend}B , both IS-1 and NIS-1 systems begin with similar initial bending angles, but by the end of the simulation, the IS-1 TCR$\alpha$ TM angle decreases to ~125°, representing a ~9\% reduction (\cref{fig:tcra_bend}C,D), whereas NIS-1 remains relatively higher at ~140° (\cref{fig:tcra_bend}F,G). A steady trend is observed in additional representative pairs (IS-2 and NIS-2; Supplementary Fig.~4), demonstrating that systematic reorientation of the TCR$\alpha$ TM domain is a recurring feature of immunogenic systems.

 Together, these results indicate that coordinated TM bending of both TCR$\alpha$ and TCR$\beta$ chains is a hallmark of functional activation in IS. Given that the TCR-CD3 complex forms a tightly coupled transmembrane signaling assembly, even subtle changes in TM orientation can alter CD3 interactions and downstream signaling~\cite{call2002organizing}. Notably, the absence of significant TM reorientation in NIS systems reflects a structurally constrained or inactive receptor state, consistent with previous reports of reduced conformational flexibility or altered TM organization in low-responsive systems~\cite{brazin_t_2018}. Having identified TM bending as the most informative predictor of immunogenicity, we next sought to determine which extracellular conformational changes give rise to these transmembrane rearrangements. Interpretability analysis identified the collective motions of the TCR$\alpha$ constant domain as the second-most informative feature, motivating a detailed analysis of its conformational landscape.

\subsubsection{Cryptic Dynamics in TCR$\alpha$ Constant Domain}
To investigate conformational dynamics across the various TCR domains and their influence on immunogenicity, we performed principal component analysis (PCA) on the constant (C$\alpha$, C$\beta$) and variable (V$\alpha$, V$\beta$) domains of TCR$\alpha$ and TCR$\beta$ chains. The first two principal components (PC1 and PC2) capture the dominant collective motions, together accounting for up to $\sim60\%$ of the total dynamic variance, with PC1 contributing $\sim45\%$ and PC2 $\sim15\%$. Such 1D or 2D projections are widely used to reveal functionally relevant conformational changes in TCR signaling systems. PC1 primarily represents large-scale conformational changes, while PC2 reflects secondary rotational or twisting motions. PC1 projections together with the combined PC1 and PC2 for C$\alpha$, C$\beta$, V$\alpha$ and V$\beta$, were included as input features for the machine learning classifier. Notably, model interpretability identified the TCR$\alpha$ constant domain (C$\alpha$) as the second most important feature in determining immunogenicity (\cref{fig:interpretability}C), highlighting the functional relevance of its collective motions.
\begin{figure}[!htbp]
  \centering
  \includegraphics[width=1\textwidth]{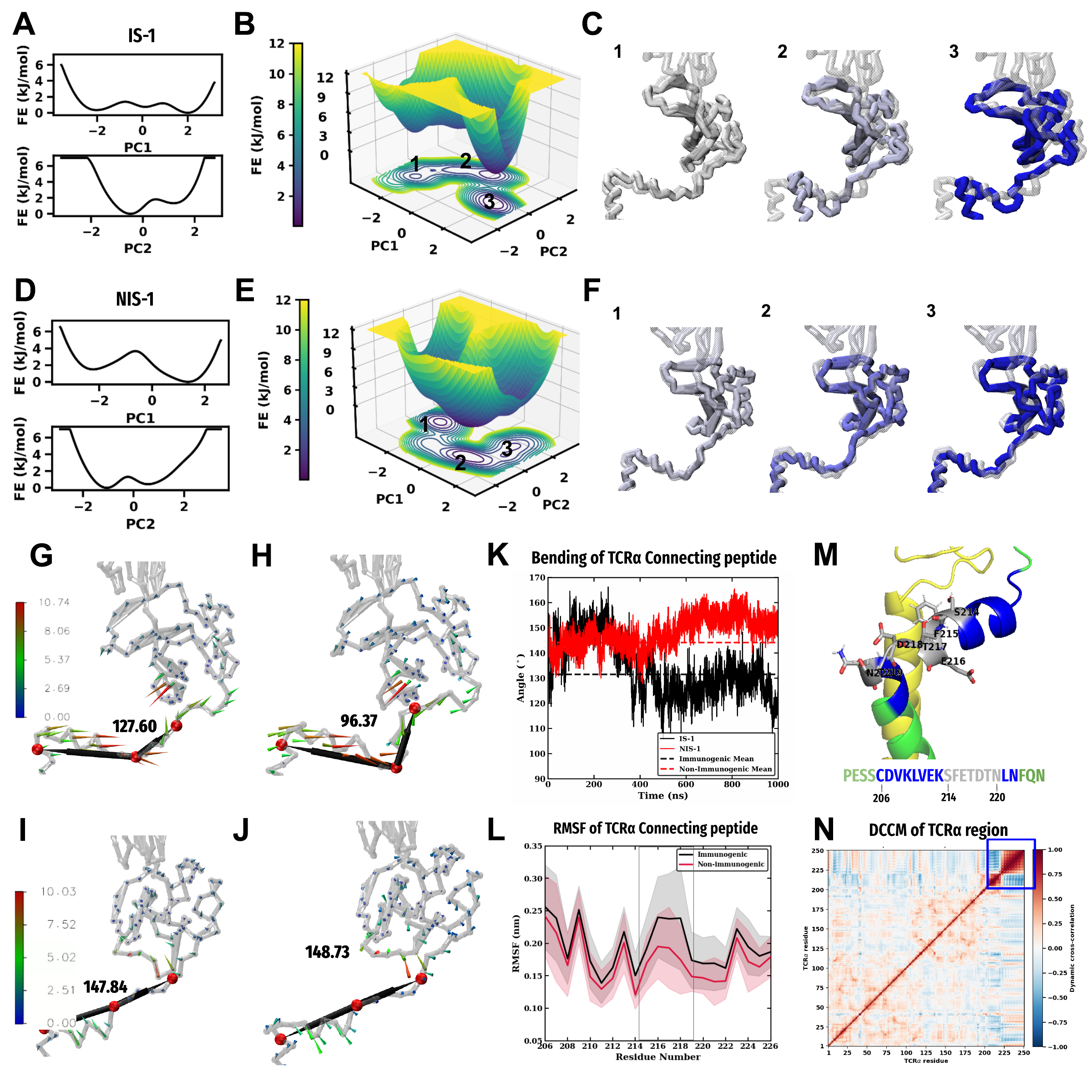}
  \thispagestyle{empty}
  \caption{\textbf{Principal component analysis and differential dynamics of the TCR$\alpha$ constant regions in immunogenic and non-immunogenic systems.} (A--C) Immunogenic system (IS) and (D--F) non-immunogenic system (NIS). (A,D) One-dimensional free-energy profiles along PC1 and PC2. (B,E) Two-dimensional free-energy landscapes projected onto PC1 and PC2, with major conformational minima numbered. (C,F) Representative structures corresponding to these minima, colored according to their occurrence over simulation time, with blue representing later time points. IS shows a pronounced change in the orientation of the TCR$\alpha$ connecting peptide (CP), consistent with a bending motion, whereas NIS retains similar CP orientations. (G,H) Porcupine plots of the dominant motions along PC1 and PC2 in IS-1, with CP bending angles of 127.6$^\circ$ and 96.37$^\circ$, respectively. (I,J) Corresponding motions in NIS-1, showing no pronounced CP bending. (K) Time evolution of the TCR$\alpha$ CP angle for representative IS-1 and NIS-1 trajectories together with their class-averaged profiles. (L) Residue-wise RMSF of the CP in IS and NIS, shown as class averages (solid lines) with corresponding standard deviations (shaded region). (M) Structural representation of the CP highlighting residues with the largest differences in flexibility between IS and NIS. (N) Dynamic cross-correlation matrix (DCCM) of TCR$\alpha$, showing correlated motions between the CP and transmembrane (TM) region.}
\label{fig:pca_bend}
\end{figure}

 Among all TCR domains, the C$\alpha$ domain has shown a distinguished PC1-PC2 conformational landscape projection between immunogenic and non-immunogenic systems (Supplementary Fig.~5A,E), while this separation is weak or absent in the corresponding PC1-PC2 projections of the remaining TCR domains (Supplementary Fig.~5B-D,F-H). To characterize the energetic basis of these conformational differences, we calculated one-dimensional free-energy surfaces (FES) along PC1 and PC2, and also two-dimensional FES combining both, for few IS and NIS (Section~\ref{sec:methods}). FES of representataive IS-1 TCR C$\alpha$ domain exhibited a rugged conformational landscape with deep energy minima corresponding to distinct conformational states (States~1-3; \cref{fig:pca_bend}A,B). Along the primary PC1 axis, which accounts for the largest portion of variance ($\sim 45\%$), energy barriers are relatively low (\cref{fig:pca_bend}A), allowing the C$\alpha$ domain to bend in a hinge-like manner and transition between functionally relevant conformations with ease. In contrast, the representative NIS-1 displayed a comparatively smoother landscape with poorly resolved minima and substantially higher barriers along PC1 (approximately 4~kJ/mol; \cref{fig:pca_bend}D), restricting transitions between conformational states. Structural snapshots corresponding to the three principal free-energy minima (\cref{fig:pca_bend}C,F) further support these observations. IS-1 (\cref{fig:pca_bend}C and F, with state transitions color-coded from white to blue based on the timeframe of the structure and aligned to the original white structure) reveals a coordinated hinge-like movements in the connecting peptide (CP) region that links the C$\alpha$ domain to the transmembrane (TM) region, suggesting a mechanically coupled motion that may facilitate propagation of extracellular conformational changes toward the membrane. On the other hand in NIS-1, the structural ensembles shown in \cref{fig:pca_bend}F, largely overlap and display local motions imposed by the higher PC1 energy barriers. 
 
 Importantly, this pattern of well-defined, rugged minima and guided conformational switching in IS systems, versus shallow and disorganized landscapes in NIS systems, is consistent across multiple IS and NIS trajectories (Supplementary Figure~6). In Supplementary Figure~6, IS uniformly show lower barriers along PC1 when compared to NIS with similar PC1 and PC2 projection profiles, as discussed above. To further assess structural similarities across systems, analysis of additional representative systems (IS-2 and NIS-2; Supplementary Fig.~7) shows the same pattern of connecting-peptide bending in IS-2 and restricted conformational changes in NIS-2. These observations propose that immunogenic peptides guide the C$\alpha$ domain along energetically favorable pathways, involving coordinated motion of the C$\alpha$ CP region. The dominant collective motions underlying these transitions were further visualized using porcupine analysis (\cref{fig:pca_bend}G-J). In the IS, the TCR C$\alpha$ CP exhibits pronounced bending and relaxation motions (\cref{fig:pca_bend}G,H), whereas these dynamics are largely suppressed in the NIS(\cref{fig:pca_bend}I,J), consistent with our C$\alpha$ FES analysis. To quantify these structural rearrangements, we measured the bending angle of the TCR$\alpha$ CP using three reference residues spanning the linker between the C$\alpha$ domain and the transmembrane helix (\cref{fig:pca_bend}G-K; Section~\ref{sec:methods}). In the IS-1, this angle varies substantially, ranging from approximately $96^\circ$ to $127^\circ$ (\cref{fig:pca_bend}K), reflecting significant conformational flexibility. By comparison, the NIS-1 remains nearly rigid, with the angle confined to a narrow range of $147^\circ$-$148^\circ$ (\cref{fig:pca_bend}K). These observations are consistent across the Immunogenic and Non-immunogenic classes (dotted lines in (\cref{fig:pca_bend}K)) indicating that immunogenic systems undergoes differential changes in the connecting-peptide conformations, that can occasionally bring it close to nearby chains or even surrounding lipids.

Residue-level analysis using RMSF profiles further localizes these collective motions to a discrete segment within the TCR$\alpha$ CP (\cref{fig:pca_bend}L). Residues E216-D218 have elevated fluctuations in the IS compared with the NIS, suggesting that it functions as a local hinge facilitating bending motions. The three-dimensional representation of the CP region (blue) and the highly fluctuating  regions-SFETDTN (grey) are shown in \cref{fig:pca_bend}M. The functional importance of this region is supported by previous experimental studies demonstrating that disruption of the conserved FETDxNLN motif within the TCR$\alpha$ CP abolishes productive TCR signalling, implicating its role in transmitting conformational changes from the TCR$\alpha\beta$ heterodimer to the CD3/$\zeta$ signalling complex \cite{backstrom_tcr_motif_1996}. To examine whether these CP hinge motions are correlated with the TCR TM region bending, we calculated the dynamic cross-correlation matrix (DCCM) for the TCR$\alpha$  (\cref{fig:pca_bend}N) and $\beta$ chains together (Supplementary Figure.~8). The DCCM reveals strong positive correlations within the CP itself (residues 206-222), indicating that this segment moves as a coordinated structural unit. In contrast, the TCR$\alpha$  CP exhibits pronounced negative correlations with the adjacent TCR$\alpha$ transmembrane helix, suggesting coordinated anti-correlated motions between these two regions. A similar dynamic coupling pattern is observed for the TCR$\alpha$ CP region and the TCR$\beta$ TM region (Supplementary Fig.~8). These correlation patterns suggests that the TRC$\alpha$ CP acts as a dynamic mechanical coupler linking TCR extracellular conformational rearrangements to the TCR transmembrane region.

Altogether, these results suggest that immune response is associated with a flexible, multi-state conformational landscape that enables dynamic coupling between extracellular ligand recognition and the TM region with TRC$\alpha$ CP as a central linker of this process, especially E216-D218 residues contributing to its dynamic adaptability.

\begin{figure}[!htbp]
  \centering
    \thispagestyle{empty}
  \includegraphics[width=1\linewidth]{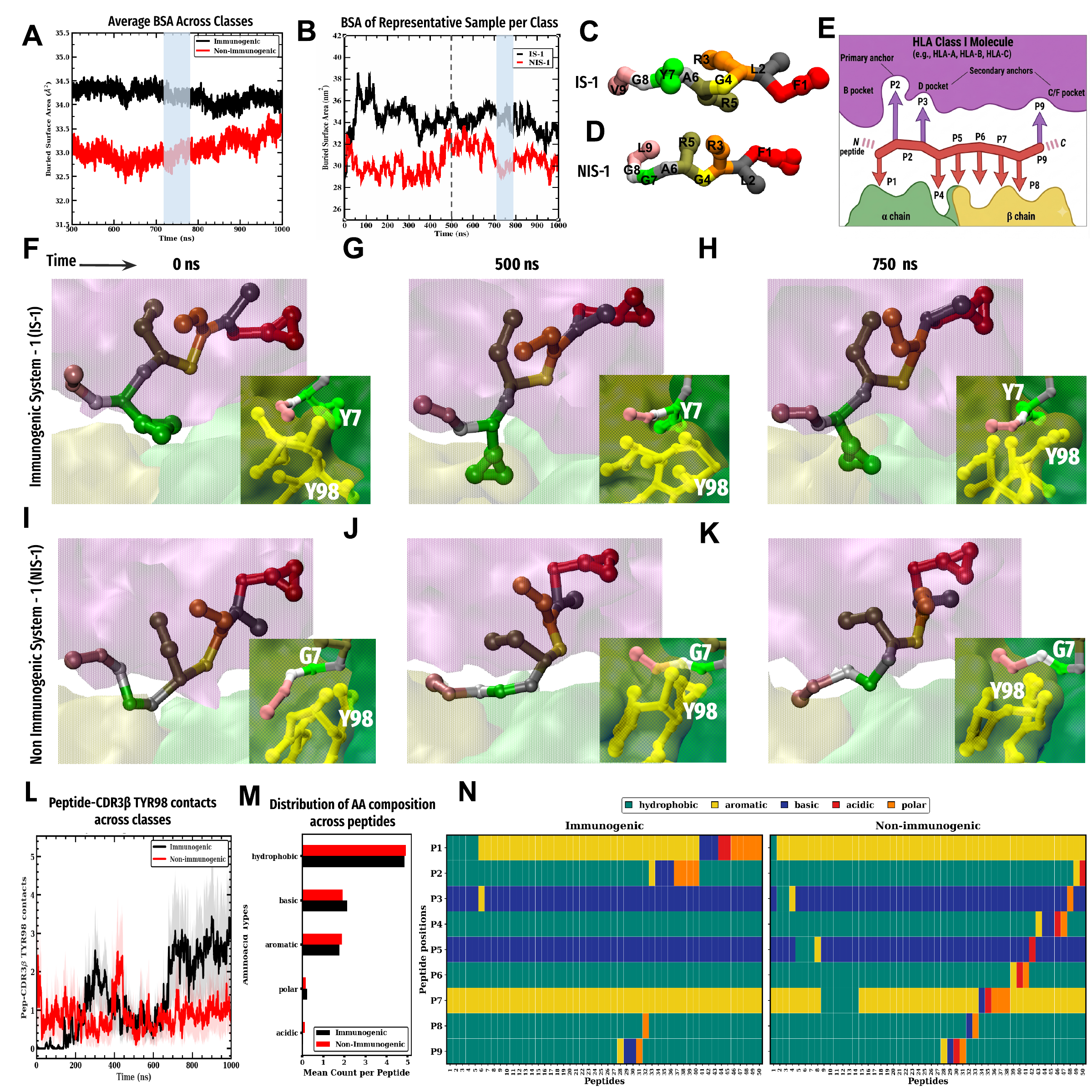}
\caption{\textbf{Time-dependent Buried surface area (BSA) variation determining peptide--CDR3$\beta$ orientations and peptide characteristics in immunogenic (IS) and non-immunogenic systems (NIS).} (A) Average pMHC--TCR interfacial BSA for all IS and NIS over the equilibrated simulation window (500~ns--1~$\mu\text{s}$), with the critical window identified by the interpretability analysis highlighted in blue. (B) Time evolution of BSA for representative IS-1 and NIS-1 systems. (C,D) Peptide structures and sequences of IS-1 and NIS-1, carrying L9V and Y7G mutations, respectively. (E) Schematic representation of the peptide-binding interface,  showing position-specific interactions with HLA-I and TCR in the reference pMHC--TCR structure (PDB ID: 1MI5). Representative conformations of IS-1 (F--H) and NIS-1 (I--K) at 0, 500, and 750~ns, with 750~ns corresponding to the critical time window. MHC, TCR$\alpha$, and TCR$\beta$ are shown in mauve, green, and yellow, respectively, with the peptide shown in licorice representation and colored by residue. In IS-1, peptide residue Y7 progressively reorients at the TCR interface, accompanied by movement of CDR3$\beta$ residue Y98 toward hydrophobic peptide residues (highlighted in the insets), supporting tighter peptide--TCR packing and increased BSA; these rearrangements are not observed for the corresponding G7 residue and CDR3$\beta$ loop in NIS-1. (L) Time evolution of peptide--Y98 contacts in IS and NIS, shown as class averages with shaded regions indicating standard deviations. (M) Average occurrence of hydrophobic, aromatic, basic, acidic, and polar residues in IS and NIS peptides. (N) Position-wise physicochemical characteristics of 50 IS and 50 NIS peptides across positions P1--P9, showing a higher occurrence of polar or acidic residues at central positions in NIS.}

  \label{fig:bsa}
\end{figure}
\subsubsection{Interfacial Buried Surface Area: A constant player}
The fourth most important feature identified by our model interpretability is the interfacial buried surface area (BSA) between the pMHC complex and the TCR. As shown in \cref{fig:bsa}A, IS consistently exhibit higher BSA than NIS, with the time window between 720-780 ns being particularly informative for classifying immunogenicity. While averaging across all IS and NIS smooths out differences, the overall trend remains observable. To explore this behavior in detail, we analyzed representative systems IS-1 and NIS-1. Despite starting from similar BSA values (\cref{fig:bsa}B), reflecting from the structural perturbations by the point mutations, the immunogenic system progressively increases its BSA by approximately 3 Å², corresponding to~10\% rise. This increase reflects the conformational changes the peptides undergo to better complement the TCR binding pocket. While in NIS, the mutations are either disrupting the hydrophobic core interactions being formed or some certain residues that could potentially engage the TCR remain restricted due to stabilizing interactions with the MHC. Although BSA iscorrelated with binding affinity, it is not sufficient to explain immunogenicity, as affinity-based approaches alone often do not accurately predict T-cell activation \cite{beyond_mhc_binding}. 

To uncover the structural basis underlying these differences, we next examined how the peptide mutations alter residue positioning within the pMHC-TCR interface. In a pMHC-TCR complex, previous studies on the presentation of HLA class I peptides have stated that the positions of peptides P2, P3 and P9 primarily anchor the peptide within the MHC binding groove, whereas the central residues remain solvent exposed and actively interact with TCRs ~\cite{calis2013properties} (illustrated schematically in \cref{fig:bsa}E). Structural analysis throughout the simulation (0 ns, 500 ns, 750 ns, corresponding to the center of the critical window, and 1~$\mu$s; \cref{fig:bsa}F-H) reveals that, in IS, peptide residue Y7 progressively reorients and inserts into the hydrophobic groove of the TCR, increasing peptide-TCR contacts and consequently the buried surface area. In contrast, this insertion is absent in the NIS-1, where the Y7G mutation reduces hydrophobicity of the peptide, leading to a further decrease in solvent-accessible surface area (SASA) and limiting peptide-TCR interactions (\cref{fig:bsa}I-K). In TCR, the CDR3$\beta$ loop forms the major peptide interacting region and contributes the most for TCR-specific peptide recognition~\cite{rossjohn2015tcr}. Interestingly, the CDR3$\beta$ loop is enriched in aromatic residues, particularly tyrosines, which frequently participate in hydrophobic packing at the center of the peptide--TCR interface~\cite{stadinski_hydrophobic_2016,rossjohn2015tcr}. In the LC13 TCR, residue Y98 forms part of the CDR3$\beta$ loop (ASSLGQAYEQY). \cref{fig:bsa}F-H, inset panels show that CDR3$\beta$ residue Y98 reorients itself during the simulation to account for more tight packing. At 500 ns, Y98 remains relatively distant from the peptide residue Y7, whereas by the center of the critical window (750 ns), it directly interacts with the inserted aromatic side chain, strengthening the peptide-TCR interface.

To quantify this, we calculated the number of contacts formed between peptide residue Y7 and CDR3$\beta$ residue Y98 over time. As shown in \cref{fig:bsa}L, the number of contacts increases by approximately two contacts in the IS, whereas in NIS it remains the same. This increase in local aromatic interactions coincides with the rise in buried surface area, directly linking peptide rearrangement to the enhanced peptide-TCR interface identified by our model. Further, to confirm this hypothesis, we analyzed additional systems, IS-2 and NIS-2 (Supplementary Figure.~9). IS-2 contains a F1W mutation, while NIS-2 carries an A6I mutation (Supplementary Figure.~9C,D). Consistent with our previous observations, IS-2 shows a $\sim18\%$ increase in BSA over time (Supplementary Figure.~9B), accompanied by enhanced movements of aromatic side chains within the TCR binding pocket. Interestingly, as phenylalanine is mutated to tryptophan in IS-2, the bulkier tryptophan engages more effectively with the TCR interface, enhancing contacts (Supplementary Figure.~9E,F,G). In contrast, the BSA of NIS-2 (A6I) remains the same, likely due to stabilizing interactions with MHC, particularly due to the hydrophobic mutation at position 6, which restricts the flexibility of the peptide (Supplementary Figure.~9H,I,J). These results suggest that specific point mutations can modulate peptide-TCR complementarity, contributing to the observed differences in BSA between IS and NIS. Although this orientation of peptide residues may vary depending on the MHC allele, this behavior is consistent with class I viral peptides, which are typically 9 amino acids in length. 

We next investigated sequence-level contributions to understand their influence on the BSA, amino acids were grouped into five physicochemical classes: hydrophobic, aromatic, polar, acidic, and basic. Global composition analysis (\cref{fig:bsa}M) shows similar distributions across immunogenic and non-immunogenic peptides, indicating that overall amino acid frequency is not a distinguishing factor. However, position-specific analysis (\cref{fig:bsa}N) reveals clear patterns: hydrophobic residues are enriched at positions P2, P4, and P6, while aromatic residues frequently occupy P1 and P7 in both classes. A key distinction emerges at intermediate positions (P2--P7), which primarily interact with the TCR (\cref{fig:bsa}N). Immunogenic peptides largely lack polar, acidic, or basic residues at these positions, whereas non-immunogenic peptides frequently contain such residues. These polar or charged residues likely reduce favorable insertion into the hydrophobic TCR interface, thereby weakening TCR engagement. These observations are consistent with prior studies demonstrating that central peptide positions (P4--P6) are critical for TCR recognition and that aromatic residues enhance immunogenic potential \cite{calis2013properties}. Furthermore, TCR recognition is driven primarily by peptide-specific interactions and conformational sampling rather than overall binding affinity \cite{cole2014tcr}. Structural studies further support that peptide-induced conformational changes are transmitted across the TCR-CD3 complex to initiate signaling \cite{dong2019structural, brazin_t_2018}.

Overall, our results demonstrate that while BSA contributes to immunogenicity, it is not the primary determinant. Instead, dynamic structural rearrangements, residue positioning, and local physicochemical compatibility at the TCR interface govern immune activation, underscoring the need for dynamic, structure-based predictive models. (\cref{fig:bsa}H).

\subsection{TCR$\alpha$ TM Bending to Immunogenicity \textit{via} CD3 Activation}
As discussed in the Introduction and in \cref{sch:overflow}, the CD3 complex surrounding the TCR
plays a key role in transmitting conformational changes of the TCR
into downstream intracellular signaling events. Since our previous analyses highlighted
membrane bending of the TCR$\alpha$ chain as an important dynamic
feature associated with immunogenicity, we extended our simulations
to include the CD3 complex in order to investigate how these
TCR$\alpha$ conformational changes propagate further into the
signaling machinery. 
To investigate structural coupling between antigen recognition and CD3-mediated signaling, we constructed a complete membrane-bound TCR-CD3 complex using the cryo-EM structure of the human TCR--CD3 assembly (PDB: 6JXR)~\cite{dong2019structural}. The TCR--CD3 complex  was structurally aligned with the pMHC--TCR complex (represented in black) used throughout this study to generate a membrane-embedded pMHC--TCR--CD3 system (Supplementary Fig.~10A). The resulting model comprises the pMHC--TCR complex together with the six associated CD3 subunits (CD3$\delta$, CD3$\gamma$, two CD3$\epsilon$, and two CD3$\zeta$ chains) surrounding the TCR within the membrane (Supplementary Fig.~10B). Five immunogenic (IS) and non-immunogenic (NIS) systems were subsequently simulated for 3~$\mu$s using the same coarse-grained molecular dynamics protocol described in Section~\ref{sec:methods}, enabling direct comparison of CD3 rearrangements during immune activation.

\begin{figure}[!htbp]
  \centering  %
  \includegraphics[width=1\textwidth]{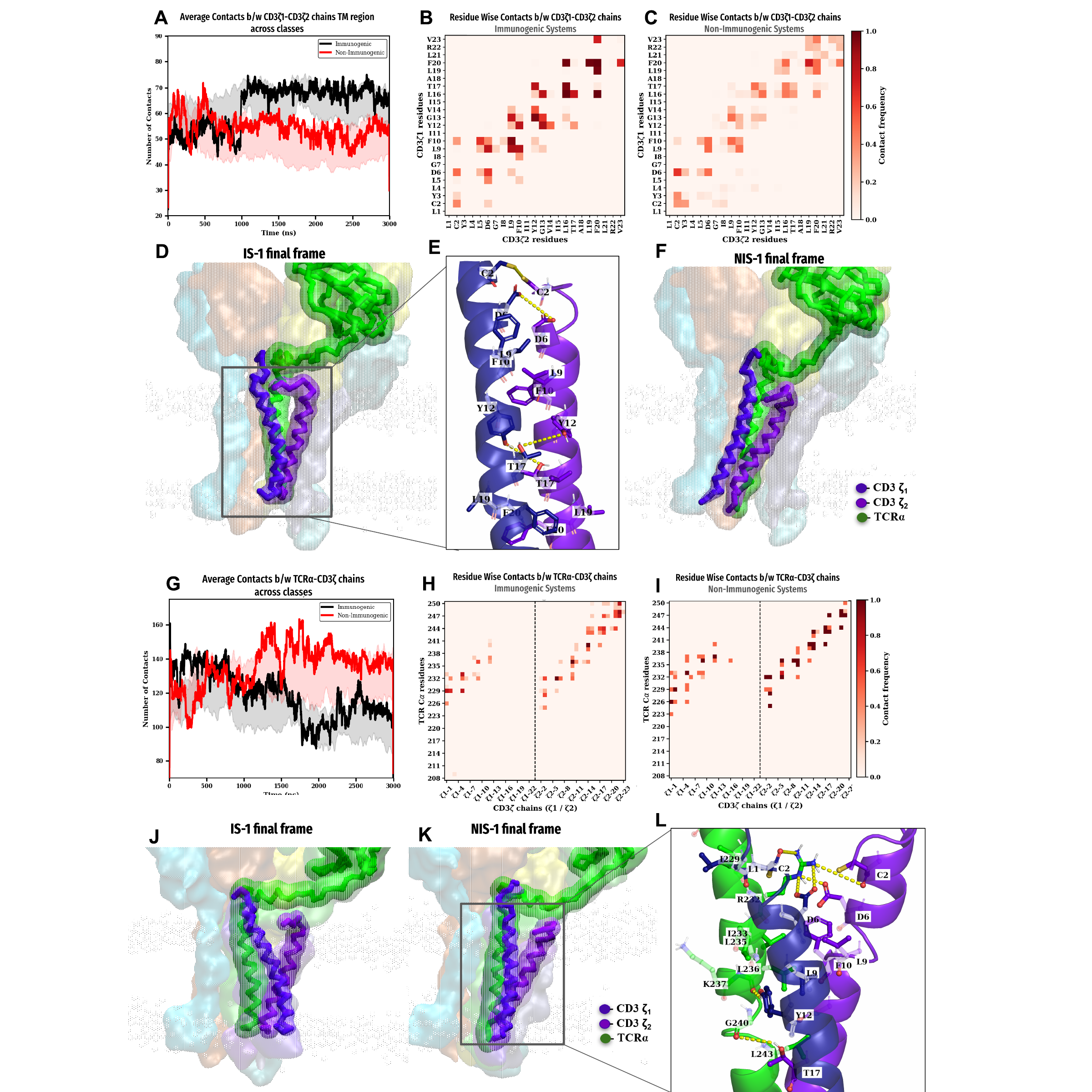}
  \thispagestyle{empty}

  \caption{\textbf{Structural basis of CD3$\zeta$ chain association and their dissociation from TCR$\alpha$ in immunogenic and non-immunogenic systems.} (A) Time evolution of the total number of contacts between the two CD3$\zeta$ chains in immunogenic (IS) and non-immunogenic systems (NIS). Thick lines represent class-averaged profiles and shaded regions indicate the corresponding standard deviations. (B,C) Residue-wise CD3$\zeta$--CD3$\zeta$ contact-frequency maps for IS and NIS, respectively, with darker red indicating higher contact frequencies. IS exhibits broader and more persistent inter-chain contacts than NIS. (D,F) Final conformations of representative systems IS-1 and NIS-1, respectively, showing closer association of the CD3$\zeta$ chains in IS-1. (E) Interfacial interactions stabilizing the CD3$\zeta$--CD3$\zeta$ association in IS-1, with hydrogen bonds indicated by yellow dashed lines. (G) Time evolution of TCR$\alpha$--CD3$\zeta$ contacts in IS and NIS, shown as class averages with corresponding standard deviations. (H,I) Residue-wise TCR$\alpha$--CD3$\zeta$ contact-frequency maps for IS and NIS, showing reduced contacts in IS compared with NIS. (J,K) Final conformations of representative IS-1 and NIS-1 systems, respectively, highlighting the separation of CD3$\zeta$ from TCR$\alpha$ in IS-1 and their retained association in NIS-1. (L) Interfacial interactions stabilizing the TCR$\alpha$--CD3$\zeta$ association in NIS-1, with hydrogen bonds indicated by yellow dashed lines. In the structural representations, TCR$\alpha$, CD3$\zeta$1, and CD3$\zeta$2 are shown in green, navy blue, and purple, respectively, with the remaining TCR--CD3 subunits shown as transparent surfaces.}
  \label{fig:cd3_zeta}  %
\end{figure}

Consistent with previous studies, TCR activation is associated with coordinated conformational rearrangements in both TCR and CD3 subunits~\cite{brazin_t_2018,al-aghbar_interplay_2022}. In our simulations, IS exhibit distinct CD3 reorganization compared to NIS. Most notably, in IS the two CD3$\zeta$ chains move close to each other throughout the simulation, undergoing a clear transition after approximately 1~$\mu$s leading to an increase of 20 contacts(\cref{fig:cd3_zeta}A). Following this transition, the number of inter-chain contacts reaches a stable plateau, whereas NIS remains constant. This increased $\zeta$--$\zeta$ association is complemented by a reduction in interactions between the CD3$\zeta$ dimer and the TCR$\alpha$ transmembrane helix (\cref{fig:cd3_zeta}G), suggesting a coordinated rearrangement of the CD3 assembly driven by bending of the TCR$\alpha$ TM region. To determine which regions drive this structural reorganization, we calculated residue-level contact frequencies between the two CD3$\zeta$ chains (\cref{fig:cd3_zeta}B,C). In IS, the interaction interface is broader with stable contacts distributed across a continuous stretch of neighbouring residues. In particular, residues within the F10--V23 region show consistently higher contact occupancies, suggesting that closer association of the two CD3$\zeta$ transmembrane helices is initiated near the C-terminal end and propagates toward the N-terminal (extracellular) side, resulting in a more stronger interaction interface. In contrast, NIS exhibit fewer interactions, which remain confined to a smaller set of residues. This localized interaction pattern is consistent with the reduced number of inter-chain contacts observed in \cref{fig:cd3_zeta}A.

Structural snapshots provide direct visualization of this transition. Representative IS-1 and NIS-1 systems begin from nearly identical CD3$\zeta$ orientations; however, by the end of the simulation the two CD3$\zeta$ TM regions in IS-1 move closer, forming a compact dimeric arrangement (\cref{fig:cd3_zeta}D), whereas little structural reorganization is observed in NIS-1 (\cref{fig:cd3_zeta}F). To characterize the molecular basis of this reorganization, the final coarse-grained structure of IS-1 was converted to an atomistic representation (\cref{fig:cd3_zeta}E). The reconstructed structure reveals the formation of a well-defined interaction partners between the two CD3$\zeta$ chains. Specifically, reciprocal hydrogen-bond interactions between Y12 of one chain and T17 of the opposing chain, together with hydrogen bonds involving the conserved D6 and C2 residues (highlighted in yellow in \cref{fig:cd3_zeta}E), stabilize the interface. Additional hydrophobic contacts formed by residues L9, F10, L19, and F20 further stabilize this association. Overall, the immunogenic system forms a much tighter interface between the two CD3$\zeta$ helices than observed at the beginning of the simulation, in agreement with the conserved transmembrane packing reported from experimentally determined TCR--CD3 complexes~\cite{dong2019structural,call2006conserved}.

As discussed above, the increased association between the two CD3$\zeta$ chains is complemented by loss of interactions between TCR$\alpha$ and the CD3$\zeta$ dimer by ~29\% (\cref{fig:cd3_zeta}G). Residue-resolved contact maps further localize these changes (\cref{fig:cd3_zeta}H,I). In NIS, interactions between TCR$\alpha$ and CD3$\zeta$ are distributed across a broad region of the TM interface remain largely preserved throughout the simulation. In contrast, IS display a reduced contact occupancies rather than complete loss of interactions. Instead, specific interaction hotspots are selectively weakened or disappear, indicating localized remodeling of the TCR$\alpha$--CD3$\zeta$ interface during immunogenic activation. Structural comparison of the IS-1, NIS-1 final simulation frames illustrates the molecular basis of this contact redistribution (\cref{fig:cd3_zeta}J,K,L). The bending of TCR$\alpha$ TM helix  is sufficient to increase its separation from the adjacent CD3$\zeta$ dimer, leading to disruption of several stabilizing interactions present in the initial structure. The atomistic reconstruction of the final NIS-1 state (\cref{fig:cd3_zeta}L) highlights the interaction network that is maintained at the TCR$\alpha$-CD3$\zeta$ interface but becomes weakened in IS. Among these interactions, TCR$\alpha$ R232 emerges as a central interaction hub, remaining in close proximity to the conserved D6 and C2 residues of both CD3$\zeta$ chains and forming a dynamic network of electrostatic and hydrogen-bonding interactions (hydrogen bonds are highlighted in yellow in \cref{fig:cd3_zeta}L) followed by K237 interactions with Y12. TCR$\alpha$ has a nice stretch of hydrophobic residues I229, I233, L235, L236, G240, and L243 involving in more hydrophobic interactions with the residues L1, L9, F10, and Y12 of the CD3$\zeta$ chains. In immunogenic systems, bending of the TCR$\alpha$ transmembrane helix weakens both the electrostatic interactions centered around R232 and the surrounding hydrophobic packing, resulting in reduction in TCR$\alpha$--CD3$\zeta$ contacts. Many of these residues have previously been identified as important contributors to TCR--CD3 transmembrane assembly and stabilization~\cite{call2006conserved,call2010architecture,dong2019structural}.

Taken Together, these observations are consistent with our initial models proposing that TCR activation is driven by rearrangements in TM helix orientation rather than large-scale conformational changes of the receptor ectodomain. The weakening of TCR$\alpha$--CD3$\zeta_1$ interactions, together with enhanced CD3$\zeta$--CD3$\zeta$ association, reveals a coordinated redistribution of transmembrane contacts within the TCR--CD3 complex. This structural reorganization provides a plausible mechanism by which extracellular ligand binding may be coupled to the membrane-proximal events that initiate T-cell activation.

\FloatBarrier

\subsection{Case study: Proposing potential vaccine candidates}
As DynamiT captures the structural changes associated with T-cell activation, we next explored its potential for vaccine discovery. Given a pMHC--TCR complex, DynamiT requires only a short molecular dynamics simulation to predict whether the complex is likely to elicit an immunogenic or non-immunogenic response. Although the model was trained using Epstein--Barr virus (EBV) peptide systems, it generalized well to independent viral datasets, achieving an accuracy of 74\% on 30 out-of-class pMHC--TCR complexes spanning multiple viruses. This suggests that although peptide sequences, TCR repertoires, and peptide recognition modes differ substantially across pathogens, the structural mechanisms that propagate extracellular antigen recognition into downstream T-cell activation are largely conserved. To assess its practical utility, we applied DynamiT to identify potential vaccine candidates for hepatitis C virus (HCV), hantavirus, and adenovirus. We first used HCV as a proof-of-concept to evaluate whether DynamiT could serve as a computational alternative to conventional experimental screening, enabling rapid and cost-effective prioritization of candidate peptides for subsequent in vitro and in vivo validation. We then extended this approach to hantavirus and adenovirus, two clinically important pathogens for which broadly effective vaccines remain unavailable or limited.
\begin{table}[htbp]

\caption{Potential peptide vaccines for HCV, adenovirus, and hantavirus with already experimentally known peptides highlighted in purple}
\centering
\renewcommand{\arraystretch}{1.55}
\setlength{\tabcolsep}{10pt}

\begin{tabularx}{\textwidth}{
    >{\centering\arraybackslash}X
    >{\centering\arraybackslash}X
    >{\centering\arraybackslash}X
}
\hline
\textbf{HCV}
&
\textbf{Adenovirus}
&
\textbf{Hantavirus}
\\
\hline

\textcolor{peptidepurple}{\textbf{CINGVCWTV}}
&
\textcolor{peptidepurple}{\textbf{TDLGQNLLY}}
&
\textcolor{peptidepurple}{\textbf{ISNQEPLKL}}
\\
\hline

\textcolor{peptidepurple}{\textbf{CVNGVCWTV}}
&
\textbf{LTDLGQNLLY}
&
\textbf{QLGQRKIVL}
\\
\hline

\textcolor{peptidepurple}{\textbf{TISGILWTV}}
&
\textcolor{peptidepurple}{\textbf{LLDQLIEEV}}
&
\textcolor{peptidepurple}{\textbf{QLGQRIIVL}}
\\
\hline

\textcolor{peptidepurple}{\textbf{TVNGVLWTV}}
&
\textbf{LLDQLIIEV}
&
\textbf{QLGQRIKVL}
\\
\hline

\textcolor{peptidepurple}{\textbf{TINGVLWTV}}
&
\textbf{LLDQWIEEV}
&
\textbf{QLGWRIKVL}
\\
\hline

&
\textbf{LLDELIEEV}
&
\textbf{QLGYRIKVL}
\\
\hline

&
\textbf{LLDILIEEV}
&
\\
\hline

&
\textbf{LLDWLIEEV}
&
\\
\hline
\end{tabularx}

\label{tab:viral_peptide_sequences}
\end{table}

\subsubsection{\textit{HCV epitopes}}
To evaluate DynamiT as a practical screening framework, we selected hepatitis C virus (HCV) as a proof-of-concept because a comprehensive study had previously characterized the immunogenicity of several NS3$_{1073}$ peptide variants across multiple viral genotypes using T-cell activation assays, cytokine measurements, and MHC binding experiments~\cite{fytili_cross_genotype_2008}. The reported NS3$_{1073}$ peptide variants CINGVCWTV and CVNGVCWTV, which are known to elicit strong T-cell responses, were included together with genotype variants 11 (SISGVLWTV) and 13 (TISGILWTV) from genotype 2, and variants 27 (TINGVLWTV) and 28 (TVNGVLWTV) from genotype 6~\cite{fytili_cross_genotype_2008}. In addition, we evaluated the Influenza NA$_{231}$ peptide, which has previously been investigated alongside HCV because of structural similarities in peptide presentation and potential cross-reactive immune recognition. This dataset provided an ideal benchmark to assess whether DynamiT could reproduce the reported immunogenic responses without performing the underlying experimental workflow.

We extracted the peptide sequences together with the corresponding HLA and TCR information (TRAV4, TRAJ43, TRBV4, and TRBJ2-7 reported by Grant \textit{et al.}~\cite{quigley_lack_2016}), constructed the corresponding pMHC--TCR complexes (pMHC-TCR initial structure and docking through TCRModel \cite{yin_tcrmodel2_2023}), performed MD simulations, and predicted their immunogenicity using DynamiT (\cref{sch:overflow}E). The overall computational workflow for the HCV case study is illustrated in Supplementary Fig.~S11. DynamiT successfully reproduced the reported immunogenicity trends with an overall agreement of 85.7\%. Both immunogenic and non-immunogenic peptide variants were classified consistently with the published observations (Immunogenic peptides are presented in \cref{tab:viral_peptide_sequences}), demonstrating that the dynamic signatures learned by DynamiT are sufficient to recover experimentally measured T-cell responses directly from molecular simulations. Although the current workflow still requires structural modeling followed by a short molecular dynamics simulation, it substantially reduces the number of peptide candidates requiring downstream experimental evaluation. These results demonstrate the utility of DynamiT as a computational prioritization framework for narrowing large peptide libraries to a smaller set of high-confidence immunogenic epitopes for subsequent \textit{in vitro} and \textit{in vivo} validation.

\subsubsection{\textit{Adenovirus epitopes}}
Human adenoviruses are common pathogens responsible for respiratory, ocular, and gastrointestinal infections, with severe disease occurring primarily in immunocompromised individuals~\cite{lion2014adenovirus}. Although adenoviral vectors have been widely exploited for vaccine and gene-delivery applications, effective vaccines against wild-type adenoviral infections remain limited, highlighting the need to identify immunogenic T-cell epitopes that could support future vaccine development~\cite{lion2014adenovirus}. We therefore evaluated whether DynamiT could be used to prioritize potential adenoviral vaccine epitopes.

We selected two experimentally reported adenoviral peptides, TDLGQNLLY from the hexon protein and LLDQLIEEV from the E1A protein, from the VDJdb database~\cite{shugay2018vdjdb}, together with their associated HLA alleles and TCR information obtained from the Huisman group dataset~\cite{huisman_public_tcr_2022}. When complete TCR chain annotations were unavailable, the missing TCR sequences were reconstructed using TCRdesign~\cite{li_tcrdesign_2025} together with our in-house TCR pairing strategy (Section~\ref{sec:methods}). To explore sequence space beyond the reported epitopes, we generated all possible single amino acid variants for each peptide. These candidate peptides were initially screened using the benchmark immunogenicity prediction models employed in this study \cite{vita2025immune},\cite{schmidt2023prime2},\cite{morrone2024tscape},\cite{bigmhc2023},\cite{immunostruct2025}, and a consensus of the top-ranked immunogenic candidates across these models was selected for further evaluation, as existing sequence-based predictors often produce a considerable number of false-positive predictions. For each shortlisted peptide, a membrane-bound pMHC--TCR complex was generated using TCRmodel~\cite{yin_tcrmodel2_2023}, followed by MD simulations and subsequently evaluation using DynamiT for immunogenicity prediction (\cref{sch:overflow}E). The peptide sequences, reconstructed TCRs, and final immunogenicity predictions are summarized in Supplementary Table~S3 with the final list of potential Adenovirus vaccine candidates are presented in \cref{tab:viral_peptide_sequences}.

\subsubsection{\textit{Hantavirus epitopes}}
Hantaviruses are zoonotic pathogens that cause hemorrhagic fever with renal syndrome (HFRS) and hantavirus cardiopulmonary syndrome (HCPS) in severe conditions, with severe cases reaching a mortality rate of 50\% ~\cite{hanta_who}. Despite their significant public health burden, there are currently no widely licensed antiviral therapies or vaccines for hantavirus infections, highlighting the need to identify effective T-cell epitopes for vaccine development. Therefore, we evaluated whether DynamiT could be used to propose potential hantavirus vaccine candidates. Among hantavirus proteins, the nucleocapsid protein (NP5) is one of the most extensively investigated vaccine targets ~\cite{hanta_np5}. We selected the NP5 peptide (QLGQRIIVL), presented by HLA-E*01:03, as the starting template for our computational screening workflow ~\cite{hanta_np5}. Unlike adenovirus, no experimentally characterized TCR repertoire was available for this peptide. A comprehensive survey of the literature, immune databases, and publicly available single-cell TCR sequencing datasets yielded no experimentally resolved TCRs for this epitope. Therefore, following the workflow described for adenovirus (\cref{sch:overflow}E), we first predicted potential TCR binders computationally.

TCRdesign \cite{li_tcrdesign_2025} was first used to identify candidate CDR3$\beta$ sequences together with their corresponding TCR$\beta$ chains, with TRBV10-3*04 emerging as the highest-ranked TCR$\beta$ candidate. The corresponding TCR$\alpha$ chains were subsequently reconstructed using our in-house TCR pairing strategy (Section~\ref{sec:methods}). Since the reconstructed TCR$\alpha$ chain introduces additional uncertainty, we evaluated the two highest-ranked TCR$\alpha\beta$ receptor pairs, namely TRAV8-4*05/TRBV10-3*04 and TRAV12-3*01/TRBV10-3*04 (Supplementary Table~S4). Next as discussed in case of Adenovirus, all possible single amino acid variants of the NP5 peptide were generated and initially screened using the benchmark immunogenicity prediction models. We then evaluated a set of consensus-positive and consensus-negative peptide variants using DynamiT. Benchmark-negative peptides served as controls to identify the TCR pair that best distinguished immunogenic from non-immunogenic variants (Supplementary Table~S4). TRAV12-3*01/TRBV10-3*04 pair correctly identified the wt and the negative peptides correctly and was used for all subsequent analyses. Then additional positive variants of the NP5 peptide were simulated and evaluated using DynamiT, resulting in a final set of predicted immunogenic and non-immunogenic candidates, summarized in the Supplementary Table~S5. The final shortlisted hantavirus peptide candidates are presented in \cref{tab:viral_peptide_sequences}.

\section{Methods}\label{sec:methods}
In this section, we describe the methodologies used to develop a viral vaccine prediction model based on TCR dynamics, integrating molecular dynamics (MD) simulations and machine learning (ML). We begin with the generation of multiple transmembrane (TM)embedded MHC-epitope-TCR systems, followed by the optimization of both all-atom and coarse-grained MD simulation protocols. We then outline the analyses performed to validate the simulation protocols and to extract key features from the simulation data that characterize TCR dynamics. Further, we describe the development of a time series based ML classification model built on these features, along with its tuning and performance evaluation. Finally, we present the details of the interpretability analysis of the ML model to identify the key features driving immunogenicity classification.

\subsection{System generation}
MHC-peptide-TCR complexes were retrieved from the Protein Data Bank (PDB) and inspected for missing residues, which were reconstructed using PDBFixer \cite{eastman2013openmm}. Amino acid substitutions within the peptide were introduced using the Mutagenesis plugin in PyMOL Molecular Graphics System \cite{schrodinger2015pymol} to generate starting structures for multiple peptide variants. Transmembrane (TM) regions corresponding to MHC and TCR were obtained from UniProt, and full-length structures were modeled using AlphaFold2, resulting in complete MHC-peptide-TCR complexes including TM segments \cite{uniprot2023}\cite{jumper2021alphafold}. The TM-extended complexes were subsequently embedded into lipid bilayers representing cell membrane compositions of dendritic cell \cite{luhr_maturation_2020} and T-cell \cite{zech_accumulation_2009} and arranged appropriately using the INSANE script for coarse-grained simulation setups or CHARMM-GUI for all-atom system preparation \cite{wassenaar2015insane} \cite{jo2008charmmgui}.

\subsection{All-atom Molecular Dynamics Simulations}
Protein-lipid bilayer systems were embedded in rectangular simulation boxes and solvated using the SPC/E water model, followed by neutralization and addition of 150 mM NaCl. Energy minimization was performed using the steepest descent algorithm for 10,000 steps. Systems were gradually heated from 0 to 310 K over 15 ns under the NVT ensemble using annealing. Subsequently, NPT equilibration was carried out with gradually reduced positional restraints on heavy atoms. Production runs were performed for 100 ns using a 2 fs integration time step under periodic boundary conditions, with three independent trajectories to ensure reproducibility. All simulations were carried out using GROMACS 2024.5 ~\cite{abraham2015gromacs} with the CHARMM36m all-atom force field~\cite{huang2017charmm36m}. Temperature was maintained at 310 K using the Nosé-Hoover thermostat, and pressure was controlled at 1 bar using the Parrinello-Rahman semi-isotropic barostat. Long-range electrostatic interactions were treated using the Particle Mesh Ewald (PME) method~\cite{darden1993pme}, with a cutoff of 1.0 nm for short-range interactions. Bond constraints involving hydrogen atoms were applied using the LINCS algorithm.

\subsection{Coarse-Grained Molecular Dynamics Simulations}
Atomistic structures were converted to coarse-grained representations using the \texttt{martinize} script~\cite{monticelli2008martini} within the Martini 2.2 force field~\cite{marrink2007martini} with DSSP-derived secondary structure assignment and backbone position restraints. Protein-membrane systems were constructed using the \texttt{insane} tool~\cite{wassenaar2015insane}, embedding proteins into lipid bilayers representative of dendritic and T-cell membranes \cite{luhr_maturation_2020,zech_accumulation_2009}. Systems were solvated with Martini water, neutralized, and ionized to a physiological concentration of 150 mM NaCl. Energy minimization was performed using the steepest descent algorithm, followed by a step-wise equilibration process. During equilibration, position restraints were applied to protein backbone beads and gradually released. An initial time step of 5 fs was used and gradually increased to 20 fs. Temperature was raised to 310 K using annealing and was maintained with the velocity-rescale thermostat, while pressure was controlled at 1 bar using the Parrinello-Rahman barostat with semi-isotropic coupling, allowing independent scaling of the membrane plane and bilayer normal. Extended equilibration (up to 50 ns) ensured proper membrane relaxation and lipid mixing. Production runs were carried out under the NPT ensemble using GROMACS 2022.5~\cite{abraham2015gromacs} with a 20 fs integration time step, in three replicates. Electrostatic interactions were treated using a reaction-field approach with a cutoff of 1.2 nm and a relative dielectric constant of 15, while van der Waals interactions were truncated between 0.9 and 1.2 nm according to standard Martini parameters. The neighbors lists were updated every 10 steps using a Verlet scheme. The Center-of-mass motion was removed separately for the protein-membrane and solvent components. For visualization and structural interpretation, the coarse-grained trajectories were converted into bond-represented models using MartiniGlass~\cite{martiniglass2022}, allowing a clearer representation of molecular connectivity and transmembrane organization.

\subsection{MD trajectory analysis and visualization}
\subsubsection{Cell Membrane Property Analysis}

To characterize the structural and dynamical properties of the lipid bilayer under different simulation conditions, we evaluated key biophysical parameters including interdigitation fraction, membrane thickness, lipid tail order parameters, lipid density distribution, membrane curvature, and root mean square deviation (RMSD). Together, these metrics provide a comprehensive description of bilayer packing, ordering, and stability.

\paragraph{Interdigitation Fraction ($f_{\text{inter}}$):}
Interdigitation quantifies the extent of overlap between lipid tails from opposing leaflets, influencing membrane permeability and hydrophobic mismatch. It was computed based on the volumetric overlap of hydrophobic chains:
\begin{equation}
f_{\text{inter}} = \frac{V_{\text{overlap}}}{V_{\text{leaflet1}} + V_{\text{leaflet2}}}
\end{equation}
where $V_{\text{overlap}}$ represents the volume of overlapping tail regions, and $V_{\text{leaflet1}}$ and $V_{\text{leaflet2}}$ denote the total tail volumes of the respective leaflets. A spatial binning scheme along the bilayer normal ($z$-axis) was employed to resolve interdigitation profiles over time.

\paragraph{Lipid Order Parameters ($S_{CD}$):}
Acyl chain ordering was quantified using the deuterium order parameter:
\begin{equation}
S_{CD} = \left\langle \frac{3 \cos^2 \theta - 1}{2} \right\rangle
\end{equation}
where $\theta$ is the angle between the segmental vector (e.g., C-H bond or pseudo-bond) and the bilayer normal. Higher $S_{CD}$ values indicate increased chain alignment and reduced flexibility. Order parameters were computed per carbon segment and averaged across all lipids and trajectory frames.

\paragraph{Membrane curvature analysis}
For each trajectory, headgroup beads per lipid species was selected and upper and lower membrane leaflets were identified using a connectivity-based leaflet detection method. Leaflet surfaces were projected onto a regular \(12 \times 12\) nm lateral grid, and local leaflet heights were obtained by averaging headgroup \(z\)-coordinates within each grid cell. The bilayer midplane was defined as

\begin{equation}
h(x,y,t)=\frac{z_{\mathrm{upper}}(x,y,t)+z_{\mathrm{lower}}(x,y,t)}{2},
\end{equation}

where \(h(x,y,t)\) is the membrane midplane height at lateral position \((x,y)\) and time \(t\). Time-averaged midplanes were calculated over the selected trajectory window. Membrane undulations were quantified as height fluctuations relative to the mean membrane plane,

\begin{equation}
\Delta h(x,y)=h(x,y)-\langle h(x,y)\rangle_{x,y},
\end{equation}

where \(\langle h(x,y)\rangle_{x,y}\) denotes the spatial average over the membrane surface. The resulting height maps were Gaussian-smoothed to suppress grid-scale noise while preserving mesoscale bending modes and visualized as two-dimensional contour maps. The same reconstructed midplane was additionally mean-centered and plane-corrected and rendered as a three-dimensional surface to visualize large-scale membrane curvature, depressions, and protrusions.

\paragraph{Root Mean Square Deviation (RMSD):}
RMSD was used to monitor structural deviations of the membrane relative to the initial configuration:
\begin{equation}
\text{RMSD}(t) = \sqrt{ \frac{1}{N} \sum_{i=1}^{N} \left[ (x_i(t) - x_i^{\text{ref}})^2 + (y_i(t) - y_i^{\text{ref}})^2 + (z_i(t) - z_i^{\text{ref}})^2 \right] }
\end{equation}
where $(x_i(t), y_i(t), z_i(t))$ are the coordinates of atom $i$ at time $t$, and $(x_i^{\text{ref}}, y_i^{\text{ref}}, z_i^{\text{ref}})$ correspond to the reference structure. RMSD reflects lipid mobility, membrane relaxation, and overall structural stability.

\subsubsection{Buried Surface Area (BSA)}
Buried surface area was computed to quantify the extent of intermolecular contact formed between interacting biomolecular components upon complex formation. BSA was calculated as the difference between the solvent-accessible surface area (SASA) of the isolated components and that of the corresponding complex:
\begin{equation}
\text{BSA} = \text{SASA}_{\text{A}} + \text{SASA}_{\text{B}} - \text{SASA}_{\text{AB}}
\end{equation}
where $\text{SASA}_{\text{A}}$ and $\text{SASA}_{\text{B}}$ denote the solvent-accessible surface areas of the individual molecules (e.g., TCR and pMHC), and $\text{SASA}_{\text{AB}}$ represents the SASA of the bound complex. SASA values were calculated using a rolling probe algorithm with a probe radius of 1.4~\AA. Higher BSA values indicate increased interfacial contact and stronger molecular association.

\subsubsection{Angular analysis}

Angular metrics were computed to characterize the relative orientation of molecular components. The angle $\theta$ between three points was defined as

\begin{equation}
\theta = \cos^{-1}
\left(
\frac{\mathbf{u}\cdot\mathbf{v}}
{|\mathbf{u}|\,|\mathbf{v}|}
\right),
\end{equation}

where $\mathbf{u}$ and $\mathbf{v}$ are vectors constructed from three selected coordinates.

Transmembrane (TM) bending was quantified using three centers of mass (COMs): (i) the COM of the TCR constant domain, (ii) the COM of the first four residues of the TM helix (TCR$\alpha$: VIGF; TCR$\beta$: ILLG), and (iii) the COM of the last four TM residues (TCR$\alpha$: LRLW; TCR$\beta$: LMAM). The angle formed by these three reference points describes the orientation of the TM helix relative to the extracellular constant domain. Similarly, bending of the TCR$\alpha$ connecting peptide (CP) was quantified using the C$\alpha$ backbone beads of three representative residues spanning the linker between the C$\alpha$ domain and the transmembrane helix (S201, L210, and N222). Time-dependent angular profiles and their corresponding averages were used to compare conformational flexibility between immunogenic and non-immunogenic systems.

\subsubsection{Inter-domain Contacts and Contact Maps}
Interactions between distinct regions were quantified using minimum distance (mindist) analysis and contact maps. A contact was defined when the minimum distance between any pair of atoms from two regions was less than a cutoff of 0.6~nm:
\begin{equation}
C_{ij} =
\begin{cases}
1, & d_{ij}^{\mathrm{min}} < 0.6~\text{nm} \\
0, & \text{otherwise}
\end{cases}
\end{equation}
where $d_{ij}^{\mathrm{min}}$ is the minimum interatomic distance between regions $i$ and $j$. Contact maps were constructed by averaging $C_{ij}$ over simulation time to identify persistent interactions.

\subsubsection{Radius of Gyration ($R_g$)}
The radius of gyration was used to evaluate the compactness of the system:
\begin{equation}
R_g = \sqrt{ \frac{1}{N} \sum_{i=1}^{N} \left| \mathbf{r}_i - \mathbf{r}_{\mathrm{COM}} \right|^2 }
\end{equation}
where $\mathbf{r}_i$ is the position of atom $i$ and $\mathbf{r}_{\mathrm{COM}}$ is the center of mass. Temporal evolution of $R_g$ reflects structural expansion, compaction, and stability.

\subsubsection{Principal Component Analysis (PCA)}
Principal component analysis (PCA) was employed to extract dominant collective motions from the MD trajectories and to identify large-scale conformational changes. PCA was performed by constructing the covariance matrix of atomic positional fluctuations:
\begin{equation}
C_{ij} = \left\langle (x_i - \langle x_i \rangle)(x_j - \langle x_j \rangle) \right\rangle
\end{equation}
where $x_i$ and $x_j$ represent the Cartesian coordinates of atoms $i$ and $j$, respectively, and $\langle \cdot \rangle$ denotes time averaging over the trajectory. Diagonalization of the covariance matrix yields eigenvectors (principal components) and corresponding eigenvalues, which describe the direction and magnitude of collective motions. The first few principal components capture the most significant conformational variability and were used to compare dynamic behavior between different systems and functional states.

\subsection{Free Energy Landscape}
The free energy surface (FES) was constructed using the first two principal components (CV1 and CV2) obtained from principal component analysis (PCA) of the molecular dynamics (MD) trajectory, representing the dominant conformational motions of the system.

For each trajectory frame, projections $(CV1_i, CV2_i)$ were extracted and used to estimate the probability density. The joint probability distribution $P(CV1, CV2)$ was computed using either a two-dimensional histogram or kernel density estimation (KDE). The marginal distributions were obtained as:
\begin{equation}
P(CV1) = \int P(CV1, CV2)\, dCV2
\end{equation}
\begin{equation}
P(CV2) = \int P(CV1, CV2)\, dCV1
\end{equation}
The free energy surface was calculated via Boltzmann inversion:
\begin{equation}
G(CV1, CV2) = -k_B T \ln \left( \frac{P(CV1, CV2)}{P_{\max}} \right)
\end{equation}
One-dimensional free energy profiles were similarly obtained:
\begin{equation}
G(CV1) = -k_B T \ln \left( \frac{P(CV1)}{P_{\max}} \right)
\end{equation}
\begin{equation}
G(CV2) = -k_B T \ln \left( \frac{P(CV2)}{P_{\max}} \right)
\end{equation}
where $k_B$ is the Boltzmann constant, $T$ is the simulation temperature, and $P_{\max}$ is the maximum observed probability used for normalization. All free energy values were shifted such that the global minimum is zero:
\begin{equation}
G_{\text{shifted}} = G - \min(G)
\end{equation}

\subsection{Machine Learning-Based Epitope Immunogenicity Prediction}
\label{sec:ml_immunogenicity_prediction}

We developed a machine learning framework to predict epitope immunogenicity from molecular dynamics (MD) trajectories of pMHC--TCR complexes. The framework uses structural and dynamical descriptors instead of antigen-processing or MHC-binding signals alone. It therefore models conformational changes in the complete pMHC--TCR complex that may be associated with T-cell activation. We formulated the task as supervised binary classification of multivariate time series. For system $i$, the MD trajectory is represented as $\mathbf{X}_i={\mathbf{x}_i^{(1)},\mathbf{x}_i^{(2)},\ldots,\mathbf{x}_i^{(T)}}$, where $\mathbf{x}_i^{(t)}\in\mathbb{R}^{P}$ is a $P$-dimensional feature vector at time point $t$. Each system has a binary label $y_i\in{0,1}$, where $y_i=1$ denotes an immunogenic system and $y_i=0$ denotes a weakly or non-immunogenic system.

The dataset contained $N=100$ pMHC--TCR systems, with 50 systems in each class. Each system was represented by $P=25$ biophysical features extracted at $T=2496$ time points from the final 500~ns of a $1~\mu\mathrm{s}$ simulation. Although the number of systems was modest, each trajectory required substantial computational resources. Experimental immunogenicity measurements are also available for only a limited number of structurally characterized pMHC--TCR complexes. The selected features described both local peptide recognition and global receptor dynamics. They included TCR membrane-bending angles, buried surface area at the TCR--pMHC interface, distances between peptide residues and TCR complementarity-determining regions, RMSD and RMSF of receptor domains, and principal components of TCR motion.

Our proposed model, DynamiT, combines Piecewise Aggregate Approximation (PAA), MANTIS time-series embeddings, and an XGBoost classifier. PAA standardized the temporal dimension from 2500 to 2496 frames while retaining the complete simulation interval~\cite{keogh2001piecewise}. MANTIS, a pretrained time-series foundation model, then encoded each processed trajectory as a fixed-dimensional representation~\cite{feofanov2025mantis}. XGBoost used these representations to produce the final immunogenicity probabilities~\cite{chen2016xgboost}.

We compared DynamiT with two groups of baseline methods. The first group comprised existing immunogenicity predictors: T-SCAPE~\cite{morrone2024tscape}, PRIME2.1~\cite{schmidt2023prime2}, the IEDB immunogenicity model~\cite{vita2025immune}, BigMHC~\cite{bigmhc2023}, and ImmunoStruct~\cite{immunostruct2025}. For BigMHC and ImmunoStruct, we used only zero-shot scores. Neither model was trained nor fine-tuned on the present dataset. The second group comprised time-series classifiers applied to the MD-derived trajectories. These classifiers included Arsenal~\cite{middlehurst2021hivecote}, ROCKET~\cite{dempster2020rocket}, MUSE~\cite{schaefer2018muse}, the Shapelet Transform~\cite{bostrom2017shapelet}, $k$-nearest neighbours~\cite{cover1967nearest}, and a classifier trained on summary statistics.

We estimated model performance using random 10-fold cross-validation repeated five times. During each repeat, the dataset was divided into ten folds. Nine folds were used for training, and the remaining fold was used for evaluation. This procedure produced 50 held-out fold evaluations for each model. We assessed performance using accuracy, precision, recall, $F_1$ score and Mathews Correlation Coefficient (MCC).

We evaluated model generalization on an external set of experimentally annotated pMHC--TCR complexes. This set contained distinct viral epitopes, TCRs, and MHC backgrounds, including systems associated with cytomegalovirus, influenza A virus, hepatitis C virus, and SARS-CoV-2. We processed these systems with the same MD feature-extraction and preprocessing pipeline. They were not used for model fitting, hyperparameter selection, or cross-validation. This analysis tested whether the learned trajectory-level patterns transferred to independent pMHC--TCR recognition systems.

\subsection{Model Interpretability Analysis}
\label{sec:methods_interpretability}

We applied post hoc interpretability methods to determine which molecular features and trajectory intervals influenced the final model. We wrapped the trained pipeline as a probability-output function that mapped each multivariate trajectory to its predicted class probabilities. This interface enabled perturbation-based attribution with Time Interpret and Captum~\cite{enguehard2023timeinterpret,kokhlikyan2020captum}.

For global interpretation, we used grouped feature ablation with a zero-valued reference in the preprocessed input space~\cite{kokhlikyan2020captum}. Channel-level ablation replaced all time points of one MD feature with the reference value. This analysis produced one overall importance score for each molecular feature. Temporal-window ablation replaced contiguous time intervals across all features and identified trajectory regions that influenced the model output. Channel-by-time ablation replaced each feature within coarse temporal windows. It therefore produced a two-dimensional attribution map across molecular features and simulation time. We computed all attributions for the experimentally annotated class and averaged them across samples. Training and held-out test samples were analyzed separately to obtain their global importance profiles.

For local interpretation, we used AugmentedOcclusion, BayesLime, and BayesKernelShap to explain individual predictions. AugmentedOcclusion applied sliding-window perturbations and used reference segments sampled from the training distribution~\cite{enguehard2023timeinterpret}. BayesLime fitted a Bayesian local surrogate model and reported feature attributions with credible intervals~\cite{enguehard2023timeinterpret,ribeiro2016why}. BayesKernelShap estimated local Shapley-value attributions together with their uncertainty~\cite{enguehard2023timeinterpret,lundberg2017unified}. Each method targeted the experimentally annotated class. The resulting explanations therefore measured how each feature or temporal region increased or decreased the probability assigned to that class. These perturbation-based scores describe model sensitivity and do not establish causal molecular mechanisms.

\subsection{Reconstruction of Paired TCR$\alpha\beta$ Receptors}

For peptide candidates lacking experimentally determined TCR pairs, TCR$\beta$ sequences together with their corresponding TRBV gene assignments were first predicted using TCRdesign \cite{li_tcrdesign_2025}. Since no computational framework currently predicts paired TCR$\alpha$ chains directly from peptide sequences, the corresponding TCR$\alpha$ chain was assigned using a nearest-neighbour sequence matching strategy. Given a predicted TCR$\beta$ sequence ($B^{*}$), the closest experimentally observed TCR$\beta$ sequence ($B_i$) in the paired TCR$\alpha\beta$ repertoire of the TCRdesign training dataset was identified by minimizing the Levenshtein edit distance:

\begin{equation}
i^{*}=\arg\min_{i} d_{L}(B^{*},B_{i}),
\end{equation}

where $d_{L}$ denotes the Levenshtein edit distance between two TCR$\beta$ amino acid sequences. The TCR$\alpha$ chain paired with the identified TCR$\beta$ sequence ($B_{i^{*}}$) was subsequently assigned to the predicted TCR$\beta$ sequence, and the reconstructed TCR$\alpha\beta$ receptor was used for downstream pMHC--TCR structural modelling and molecular dynamics simulations.

\section{Conclusion}

In this study, we developed DynamiT, a computational framework that combines molecular dynamics simulations with an interpretable time-series classifier to investigate T-cell activation and predict peptide immunogenicity. Our results show that immune activation is not solely determined by pMHC--TCR binding or static structural properties, but is strongly linked to specific conformational changes in the TCR. Model attribution ranked the TCR$\beta$ transmembrane bending angle as the leading predictive feature. The dominant collective motion of the TCR$\alpha$ constant domain ranked second, followed by the TCR$\alpha$ transmembrane bending angle. An extended in-depth analysis of TCR--CD3 complexes revealed that TCR TM bending further loosens CD3 $\zeta$ chains and exposes phosphorylation sites that trigger downstream signaling. Although buried surface area remains an important descriptor of peptide--TCR interactions, our analyses show that dynamic structural features contribute more strongly to immunogenicity prediction.

DynamiT achieved an accuracy of 73.3\% on an independent external cohort of 30 viral pMHC--TCR systems. Among the 12 evaluated methods, it achieved the highest accuracy, $F_1$ score, precision, and Matthews correlation coefficient, and obtained the best mean rank across the five evaluation metrics. These results show that DynamiT provided the most balanced performance on this cohort. They also indicate that the learned dynamic signatures transferred across distinct viral epitopes, TCRs, and MHC backgrounds. However, the small external cohort limits conclusions about broader generalizability or a conserved activation mechanism.

We further evaluated the practical use of DynamiT through three case studies. For hepatitis C virus, DynamiT reproduced the reported immunogenicity trends with an agreement of 85.7\%. This result supports its use for prioritizing peptide candidates before experimental evaluation. We then applied the same workflow to adenovirus and hantavirus and shortlisted candidate epitopes for subsequent validation. These adenovirus and hantavirus predictions remain computational proposals and require experimental confirmation.

Overall, our study establishes a direct link between TCR structural dynamics and peptide immunogenicity while demonstrating how molecular dynamics simulations and interpretable machine learning can be integrated for mechanistic immune response prediction. DynamiT is intended to complement existing screening pipelines by evaluating candidates shortlisted by faster sequence-based methods and prioritizing a smaller set for experimental validation. Although the current framework requires molecular dynamics simulations as an initial step, ongoing advances in enhanced sampling methods, accelerated molecular simulations, and AI-based structural dynamics prediction are expected to substantially improve its computational efficiency. We anticipate that this work will provide a foundation for future studies exploring the role of structural dynamics in immunogenicity prediction, while also advancing our understanding of pMHC--TCR and pMHC--TCR--CD3 signaling and facilitating better peptide vaccine designs.

\bibliography{references}%

@article{brazin_t_2018,
    title = {The {T} {Cell} {Antigen} {Receptor} $\alpha$ {Transmembrane} {Domain} {Coordinates} {Triggering} through {Regulation} of {Bilayer} {Immersion} and {CD3} {Subunit} {Associations}},
    volume = {49},
    issn = {1074-7613},
    doi = {10.1016/j.immuni.2018.09.007},
    language = {English},
    number = {5},
    journal = {Immunity},
    author = {Brazin, Kristine N. and Mallis, Robert J. and Boeszoermenyi, Andras and Feng, Yinnian and Yoshizawa, Akihiro and Reche, Pedro A. and Kaur, Pavanjeet and Bi, Kevin and Hussey, Rebecca E. and Duke-Cohan, Jonathan S. and Song, Likai and Wagner, Gerhard and Arthanari, Haribabu and Lang, Matthew J. and Reinherz, Ellis L.},
    month = nov,
    year = {2018},
    pmid = {30389415},
    pages = {829--841.e6},
}

@article{tung2011popisk,
  title={POPISK: T-cell reactivity prediction using support vector machines and string kernels},
  author={Tung, Chun-Wei and Ziehm, Matthias and K{\"a}mper, Andreas and Kohlbacher, Oliver and Ho, Shinn-Ying},
  journal={BMC bioinformatics},
  volume={12},
  number={1},
  pages={446},
  year={2011},
  publisher={Springer}
}

@article{jensen2024elife93934,
  title={Enhancing TCR specificity predictions by combined pan- and peptide-specific training, loss-scaling, and sequence similarity integration},
  author={Jensen, Mathias Fynbo and Nielsen, Morten},
  journal={eLife},
  year={2024},
  volume={12},
  pages={RP93934},
  doi={10.7554/eLife.93934.3},
}

@article{stranzl2010netctlpan,
  title={NetCTLpan: pan-specific MHC class I pathway epitope predictions},
  author={Stranzl, Thomas and Larsen, Mette Voldby and Lundegaard, Claus and Nielsen, Morten},
  journal={Immunogenetics},
  volume={62},
  number={6},
  pages={357--368},
  year={2010},
  publisher={Springer}
}

@article{nilsson2025netmhcpan,
  title={NetMHCpan-4.2: improved prediction of CD8+ epitopes by use of transfer learning and structural features},
  author={Nilsson, Jonas Birkelund and Greenbaum, Jason and Peters, Bjoern and Nielsen, Morten},
  journal={Frontiers in Immunology},
  volume={16},
  pages={1616113},
  year={2025},
  publisher={Frontiers Media SA}
}

@article{vita2025immune,
  title={The immune epitope database (IEDB): 2024 update},
  author={Vita, Randi and Blazeska, Nina and Marrama, Daniel and IEDB Curation Team Members Shackelford Deborah Zalman Leora Foos Gabriele Zarebski Laura Chan Kenneth Reardon Brian Fitzpatrick Sidne Busse Matthew Coleman Sara Sedwick Caitlin Edwards Lindy MacFarlane Catriona Ennis Marcus and Duesing, Sebastian and Bennett, Jason and Greenbaum, Jason and De Almeida Mendes, Marcus and Mahita, Jarjapu and Wheeler, Daniel K and others},
  journal={Nucleic Acids Research},
  volume={53},
  number={D1},
  pages={D436--D443},
  year={2025},
  publisher={Oxford University Press}
}

@article{rammensee1999syfpeithi,
  title={SYFPEITHI: database for MHC ligands and peptide motifs},
  author={Rammensee, H-G and Bachmann, Jutta and Emmerich, Niels Philipp Nikolaus and Bachor, Oskar Alexander and Stevanovi{\'c}, SSYFPEITHI},
  journal={Immunogenetics},
  volume={50},
  number={3},
  pages={213--219},
  year={1999},
  publisher={Springer}
}

@article{shugay2018vdjdb,
  title={VDJdb: a curated database of T-cell receptor sequences with known antigen specificity},
  author={Shugay, Mikhail and Bagaev, Dmitriy V and Zvyagin, Ivan V and Vroomans, Renske M and Crawford, Jeremy Chase and Dolton, Garry and Komech, Ekaterina A and Sycheva, Anastasiya L and Koneva, Anna E and Egorov, Evgeniy S and others},
  journal={Nucleic acids research},
  volume={46},
  number={D1},
  pages={D419--D427},
  year={2018},
  publisher={Oxford University Press}
}

@article{knapp2014large,
  title={Large scale characterization of the LC13 TCR and HLA-B8 structural landscape in reaction to 172 altered peptide ligands: a molecular dynamics simulation study},
  author={Knapp, Bernhard and Dunbar, James and Deane, Charlotte M},
  journal={PLoS computational biology},
  volume={10},
  number={8},
  pages={e1003748},
  year={2014},
  publisher={Public Library of Science San Francisco, USA}
}

@article{knapp2019mhc,
  title={MHC binding affects the dynamics of different T-cell receptors in different ways},
  author={Knapp, Bernhard and van der Merwe, P Anton and Dushek, Omer and Deane, Charlotte M},
  journal={PLOS Computational Biology},
  volume={15},
  number={9},
  pages={e1007338},
  year={2019},
  publisher={Public Library of Science San Francisco, CA USA}
}

@article{alba2020molecular,
  title={Molecular dynamics simulations reveal canonical conformations in different pMHC/TCR interactions},
  author={Alba, Josephine and Di Rienzo, Lorenzo and Milanetti, Edoardo and Acuto, Oreste and D’Abramo, Marco},
  journal={Cells},
  volume={9},
  number={4},
  pages={942},
  year={2020},
  publisher={MDPI}
}

@article{van2023tcr,
  title={TCR-pMHC complex formation triggers CD3 dynamics},
  author={van Eerden, Floris J and Sherif, Aalaa Alrahman and Llamas-Covarrubias, Mara Anais and Millius, Arthur and Lu, Xiuyuan and Ishizuka, Shigenari and Yamasaki, Sho and Standley, Daron M},
  journal={eLife},
  volume={12},
  year={2023},
  publisher={eLife Sciences Publications Limited}
}

@article{ochoa1997ability,
  title={The ability of peptides to induce cytotoxic T cells in vitro does not strongly correlate with their affinity for the H-2Ld molecule: implications for vaccine design and immunotherapy},
  author={Ochoa-Garay, Jorge and McKinney, Denise M and Kochounian, Harold H and McMillan, Minnie},
  journal={Molecular immunology},
  volume={34},
  number={3},
  pages={273--281},
  year={1997},
  publisher={Elsevier}
}

@article{al-aghbar_interplay_2022,
    title = {The interplay between membrane topology and mechanical forces in regulating {T} cell receptor activity},
    volume = {5},
    copyright = {2022 The Author(s)},
    issn = {2399-3642},
    doi = {10.1038/s42003-021-02995-1},
    language = {en},
    number = {1},
    journal = {Communications Biology},
    author = {Al-Aghbar, Mohammad Ameen and Jainarayanan, Ashwin K. and Dustin, Michael L. and Roffler, Steve R.},
    month = jan,
    year = {2022},
    pages = {40},
}

@article{kjer-nielsen_structural_2003,
    title = {A {Structural} {Basis} for the {Selection} of {Dominant} $\alpha\beta$ {T} {Cell} {Receptors} in {Antiviral} {Immunity}},
    volume = {18},
    issn = {1074-7613},
    doi = {10.1016/S1074-7613(02)00513-7},
    language = {English},
    number = {1},
    journal = {Immunity},
    author = {Kjer-Nielsen, Lars and Clements, Craig S. and Purcell, Anthony W. and Brooks, Andrew G. and Whisstock, James C. and Burrows, Scott R. and McCluskey, James and Rossjohn, Jamie},
    month = jan,
    year = {2003},
    pmid = {12530975},
    pages = {53--64},
}

@article{croft_most_2019,
    title = {Most viral peptides displayed by class {I} {MHC} on infected cells are immunogenic},
    volume = {116},
    doi = {10.1073/pnas.1815239116},
    number = {8},
    journal = {Proceedings of the National Academy of Sciences},
    author = {Croft, Nathan P. and Smith, Stewart A. and Pickering, Jana and Sidney, John and Peters, Bjoern and Faridi, Pouya and Witney, Matthew J. and Sebastian, Prince and Flesch, Inge E. A. and Heading, Sally L. and Sette, Alessandro and La Gruta, Nicole L. and Purcell, Anthony W. and Tscharke, David C.},
    month = feb,
    year = {2019},
    pages = {3112--3117},
}

@article{firbas_immunogenicity_2006,
    title = {Immunogenicity and safety of a novel therapeutic hepatitis {C} virus ({HCV}) peptide vaccine: {A} randomized, placebo controlled trial for dose optimization in 128 healthy subjects},
    volume = {24},
    issn = {0264-410X},
    shorttitle = {Immunogenicity and safety of a novel therapeutic hepatitis {C} virus ({HCV}) peptide vaccine},
    doi = {10.1016/j.vaccine.2006.03.009},
    number = {20},
    journal = {Vaccine},
    author = {Firbas, Christa and Jilma, Bernd and Tauber, Erich and Buerger, Vera and Jelovcan, Sandra and Lingnau, Karen and Buschle, Michael and Frisch, Jürgen and Klade, Christoph S.},
    month = may,
    year = {2006},
    pages = {4343--4353},
}

@article{heitmann_covid_19_2022,
    title = {A {COVID}-19 peptide vaccine for the induction of {SARS}-{CoV}-2 {T} cell immunity},
    volume = {601},
    copyright = {2022 The Author(s)},
    issn = {1476-4687},
    doi = {10.1038/s41586-021-04232-5},
    language = {en},
    number = {7894},
    journal = {Nature},
    author = {Heitmann, Jonas S. and Bilich, Tatjana and Tandler, Claudia and Nelde, Annika and Maringer, Yacine and Marconato, Maddalena and Reusch, Julia and Jäger, Simon and Denk, Monika and Richter, Marion and Anton, Leonard and Weber, Lisa Marie and Roerden, Malte and Bauer, Jens and Rieth, Jonas and Wacker, Marcel and Hörber, Sebastian and Peter, Andreas and Meisner, Christoph and Fischer, Imma and Löffler, Markus W. and Karbach, Julia and Jäger, Elke and Klein, Reinhild and Rammensee, Hans-Georg and Salih, Helmut R. and Walz, Juliane S.},
    month = jan,
    year = {2022},
    pages = {617--622},
}

@article{li_global_2024,
    title = {Global burden of viral infectious diseases of poverty based on {Global} {Burden} of {Diseases} {Study} 2021},
    volume = {13},
    issn = {2049-9957},
    doi = {10.1186/s40249-024-01234-z},
    language = {en},
    number = {1},
    journal = {Infectious Diseases of Poverty},
    author = {Li, Xin-Chen and Zhang, Yan-Yan and Zhang, Qi-Yu and Liu, Jing-Shu and Ran, Jin-Jun and Han, Le-Fei and Zhang, Xiao-Xi},
    month = oct,
    year = {2024},
    pages = {71},
}

@article{haider_global_2025,
    title = {Global dengue epidemic worsens with record 14 million cases and 9000 deaths reported in 2024},
    volume = {158},
    issn = {1201-9712},
    doi = {10.1016/j.ijid.2025.107940},
    language = {English},
    journal = {International Journal of Infectious Diseases},
    author = {Haider, Najmul and Hasan, Mohammad Nayeem and Onyango, Joshua and Billah, Masum and Khan, Sakirul and Papakonstantinou, Danai and Paudyal, Priyamvada and Asaduzzaman, Md},
    month = sep,
    year = {2025},
    pmid = {40449873},
}

@article{kuroki_broad_spectrum_2021,
    title = {Broad-{Spectrum} {Antiviral} {Peptides} and {Polymers}},
    volume = {10},
    copyright = {© 2021 Wiley-VCH GmbH},
    issn = {2192-2659},
    doi = {10.1002/adhm.202101113},
    language = {en},
    number = {23},
    journal = {Advanced Healthcare Materials},
    author = {Kuroki, Agnès and Tay, Joyce and Lee, Guan Huei and Yang, Yi Yan},
    year = {2021},
    pages = {2101113},
}

@article{keogh2001piecewise,
  title={Dimensionality Reduction for Fast Similarity Search in Large Time Series Databases},
  author={Keogh, Eamonn and Chakrabarti, Kaushik and Pazzani, Michael and Mehrotra, Sharad},
  journal={Knowledge and Information Systems},
  volume={3},
  number={3},
  pages={263--286},
  year={2001},
  publisher={Springer},
  doi={10.1007/PL00011669},
}

@article{ding_four_1999,
  title   = {Four {A6}-{TCR}/peptide/{HLA}-{A2} structures that generate very different {T} cell signals are nearly identical},
  author  = {Ding, Ya-Han and Baker, Brian M. and Garboczi, David N. and Biddison, William E. and Wiley, Don C.},
  journal = {Immunity},
  year    = {1999},
  volume  = {11},
  number  = {1},
  pages   = {45--56},
  doi     = {10.1016/S1074-7613(00)80080-1}
}

@article{feofanov2025mantis,
  title={Mantis: Lightweight Calibrated Foundation Model for User-Friendly Time Series Classification},
  author={Feofanov, Vasilii and Cao, Yushu and Chen, Gaozhong and Sefti, Mahyar and Lin, Guangji and Cheng, Mingzhao and Xu, Chongqing and Vazirgiannis, Michalis and Hartvigsen, Thomas},
  journal={arXiv preprint arXiv:2502.15637},
  year={2025},
}

@article{chen2016xgboost,
  title={{XGBoost}: A Scalable Tree Boosting System},
  author={Chen, Tianqi and Guestrin, Carlos},
  journal={Proceedings of the 22nd ACM SIGKDD International Conference on Knowledge Discovery and Data Mining},
  pages={785--794},
  year={2016},
  publisher={ACM},
  doi={10.1145/2939672.2939785}
}

@article{quigley_lack_2016,
  title = {Lack of heterologous cross-reactivity toward {HLA}-{A}*02:01-restricted viral epitopes is underpinned by distinct T cell receptor signatures},
  author = {Quigley, Michael F. and Greenaway, Helen Y. and Venturi, Vanessa and Lindsay, Robert and Quinn, Kerrie M. and Seder, Robert A. and Douek, Daniel C. and Davenport, Miles P. and Price, David A. and Gruta, Nicole L. La and Turner, S. J.},
  journal = {Journal of Biological Chemistry},
  year = {2016},
  volume = {291},
  number = {47},
  pages = {24398--24413},
  month = nov,
  doi = {10.1074/jbc.M116.752121},
  pmid = {27645996}
}

@article{fytili_cross_genotype_2008,
  title = {Cross-genotype-reactivity of the immunodominant {HCV} {CD8} {T}-cell epitope {NS3}-1073},
  author = {Fytili, P. and Dalekos, G. N. and Schlaphoff, V. and Suneetha, P. V. and Sarrazin, C. and Zauner, W. and Zachou, K. and Berg, T. and Manns, M. P. and Klade, C. S. and Cornberg, M. and Wedemeyer, H.},
  journal = {Vaccine},
  year = {2008},
  month = jul,
  volume = {26},
  number = {31},
  pages = {3818--3826},
  doi = {10.1016/j.vaccine.2008.05.045},
  pmid = {18582999}
}

@article{richard_staggered_2021,
  title   = {Staggered starts in the race to {T} cell activation},
  author  = {Richard, Arianne C. and Frazer, Gordon L. and Ma, Claire Y. and Griffiths, Gillian M.},
  journal = {Trends in Immunology},
  year    = {2021},
  month   = nov,
  volume  = {42},
  number  = {11},
  pages   = {994--1008},
  doi     = {10.1016/j.it.2021.09.004},
  pmid    = {34649777},
  pmcid   = {PMC7612485}
}

@article{dong2019structural,
  title={Structural basis of assembly of the human T cell receptor-CD3 complex},
  author={Dong, Dapeng and Zheng, Lin and Lin, Jingjing and Zhang, Baoyu and Zhu, Yufei and Li, Ningning and Xie, Sheng and Wang, Yu and Gao, Ning and Huang, Zhenhua},
  journal={Nature},
  volume={573},
  number={7775},
  pages={546--552},
  year={2019},
  publisher={Nature Publishing Group}
}

@article{call2006conserved,
  title   = {The Structure of the {$\zeta\zeta$} Transmembrane Dimer Reveals Features Essential for Its Assembly with the {T} Cell Receptor},
  author  = {Call, Matthew E. and Schnell, Jason R. and Xu, Chenqi and Lutz, Regina A. and Chou, James J. and Wucherpfennig, Kai W.},
  journal = {Cell},
  volume  = {127},
  number  = {2},
  pages   = {355--368},
  year    = {2006},
  pmid    = {17055436},
  pmcid   = {PMC3466601}
}

@article{call2010architecture,
  title   = {A conserved {$\alpha\beta$} transmembrane interface forms the core of a compact {T}-cell receptor--{CD}3 structure within the membrane},
  author  = {Krshnan, Logesvaran and Park, Soohyung and Im, Wonpil and Call, Melissa J. and Call, Matthew E.},
  journal = {Proceedings of the National Academy of Sciences of the United States of America},
  volume  = {113},
  number  = {43},
  pages   = {E6649--E6658},
  year    = {2016},
  doi     = {10.1073/pnas.1611445113},
  pmcid   = {PMC5086997},
  pmid    = {27791034}
}

@article{eastman2013openmm,
  title={OpenMM 4: a reusable, extensible, hardware independent library for high performance molecular simulation},
  author={Eastman, Peter and Friedrichs, Mark S. and Chodera, John D. and others},
  journal={Journal of Chemical Theory and Computation},
  volume={9},
  number={1},
  pages={461--469},
  year={2013}
}

@misc{schrodinger2015pymol,
  title={The PyMOL Molecular Graphics System, Version 1.8},
  author={{Schr{\"o}dinger, LLC}},
  year={2015}
}

@article{jumper2021alphafold,
  title={Highly accurate protein structure prediction with AlphaFold},
  author={Jumper, John and Evans, Richard and Pritzel, Alexander and others},
  journal={Nature},
  volume={596},
  number={7873},
  pages={583--589},
  year={2021}
}

@article{wassenaar2015insane,
  title={Computational lipidomics with insane: a versatile tool for generating custom membranes for molecular simulations},
  author={Wassenaar, Tsjerk A. and Ing{\'o}lfsson, Helgi I. and others},
  journal={Journal of Chemical Theory and Computation},
  volume={11},
  number={5},
  pages={2144--2155},
  year={2015}
}

@article{jo2008charmmgui,
  title={CHARMM-GUI: a web-based graphical user interface for CHARMM},
  author={Jo, Sunhwan and Kim, Taeyong and others},
  journal={Journal of Computational Chemistry},
  volume={29},
  number={11},
  pages={1859--1865},
  year={2008}
}

@article{uniprot2023,
  title={UniProt: the Universal Protein Knowledgebase in 2023},
  author={{The UniProt Consortium}},
  journal={Nucleic Acids Research},
  volume={51},
  number={D1},
  pages={D523--D531},
  year={2023},
  doi={10.1093/nar/gkac1052}
}

@article{abraham2015gromacs,
  title={GROMACS: High performance molecular simulations through multi-level parallelism from laptops to supercomputers},
  author={Abraham, Mark J and Murtola, Teemu and Schulz, Roland and Páll, Szilárd and Smith, Jeremy C and Hess, Berk and Lindahl, Erik},
  journal={SoftwareX},
  volume={1},
  pages={19--25},
  year={2015}
}

@article{darden1993pme,
  title={Particle mesh Ewald: An N log(N) method for Ewald sums in large systems},
  author={Darden, Tom and York, Darrin and Pedersen, Lee},
  journal={Journal of Chemical Physics},
  volume={98},
  number={12},
  pages={10089--10092},
  year={1993}
}

@article{marrink2007martini,
  title={The MARTINI force field: Coarse grained model for biomolecular simulations},
  author={Marrink, Siewert J and Risselada, H Jelger and Yefimov, Sergei and Tieleman, D Peter and de Vries, Alex H},
  journal={Journal of Physical Chemistry B},
  volume={111},
  number={27},
  pages={7812--7824},
  year={2007}
}

@article{monticelli2008martini,
  title={The MARTINI coarse-grained force field: Extension to proteins},
  author={Monticelli, Luca and Kandasamy, Senthil and Periole, Xavier and Larson, Ronald G and Tieleman, D Peter and Marrink, Siewert J},
  journal={Journal of Chemical Theory and Computation},
  volume={4},
  number={5},
  pages={819--834},
  year={2008}
}

@article{plesnar2011cholesterol,
  title={Saturation with cholesterol increases vertical order and smoothes the surface of the phosphatidylcholine bilayer},
  author={Plesnar, Edyta and others},
  journal={Langmuir},
  volume={27},
  number={6},
  pages={2737--2747},
  year={2011}
}

@article{venturoli2008mesoscopic,
  title={Mesoscopic models of biological membranes},
  author={Venturoli, Maddalena and others},
  journal={Physics Reports},
  volume={437},
  number={1--2},
  pages={1--54},
  year={2008}
}

@article{martiniglass2022,
  title={MartiniGlass: visualizing Martini coarse-grained molecular dynamics simulations},
  author={Schott-Verdugo, Stefan and Gohlke, Holger},
  journal={Journal of Chemical Information and Modeling},
  volume={62},
  number={2},
  pages={353--359},
  year={2022},
  doi={10.1021/acs.jcim.1c01005}
}

@article{call2002organizing,
  title={The organizing principle in the formation of the T cell receptor-CD3 complex},
  author={Call, Matthew E and Wucherpfennig, Kai W},
  journal={Cell},
  volume={111},
  number={7},
  pages={967--979},
  year={2002}
}

@article{backstrom_tcr_motif_1996,
  title={A motif within the T cell receptor $\alpha$ chain constant region connecting peptide domain controls antigen responsiveness},
  author={Backstrom, Britt-Marie and Milia, E. and Peter, A. and Jaureguiberry, Beatrice and Baldari, Cosima T. and Palmer, Ed},
  journal={Immunity},
  volume={5},
  number={5},
  pages={437--447},
  year={1996},
  month={nov},
  publisher={Elsevier},
  doi={10.1016/S1074-7613(00)80500-2}
}

@article{beyond_mhc_binding,
  title={Beyond MHC binding: immunogenicity prediction tools to refine neoantigen selection in cancer patients},
  journal={Frontiers in Immunology},
  year={2017},
  author={Wells, D. K. and others}
}

@article{cole2014tcr,
  title={T-cell receptor (TCR)-peptide specificity overrides affinity-enhancing TCR-MHC interactions},
  journal={Journal of Biological Chemistry},
  year={2014},
  author={Cole, D. K. and others}
}

@article{luhr_maturation_2020,
  title   = {Maturation of Monocyte-Derived {DCs} Leads to Increased Cellular Stiffness, Higher Membrane Fluidity, and Changed Lipid Composition},
  author  = {L{"u}hr, Jennifer J. and Alex, Nils and Amon, Lukas and Kr{"a}ter, Martin and Kub{\'a}nkov{\'a}, Mark{\'e}ta and Sezgin, Erdinc and Lehmann, Christian H. K. and Heger, Lukas and Heidkamp, Gordon F. and Smith, Ana-Sun{\v{c}}ana and Zaburdaev, Vasily and B{"o}ckmann, Rainer A. and Levental, Ilya and Dustin, Michael L. and Eggeling, Christian and Guck, Jochen and Dudziak, Diana},
  journal = {Frontiers in Immunology},
  year    = {2020},
  volume  = {11},
  pages   = {590121},
  month   = nov,
  doi     = {10.3389/fimmu.2020.590121},
}

@article{zech_accumulation_2009,
  title   = {Accumulation of raft lipids in {T}-cell plasma membrane domains engaged in {TCR} signalling},
  author  = {Zech, Tobias and Ejsing, Christer S. and Gaus, Katharina and de Wet, Ben and Shevchenko, Andrej and Simons, Kai and Harder, Thomas},
  journal = {The EMBO Journal},
  year    = {2009},
  volume  = {28},
  number  = {5},
  pages   = {466--476},
  month   = jan,
  doi     = {10.1038/emboj.2009.6},
  pmid    = {19177148},
  pmcid   = {PMC2657588}
}

@article{piggot_acyl_chain_order_2017,
  title   = {On the Calculation of Acyl Chain Order Parameters from Lipid Simulations},
  author  = {Piggot, Thomas J. and Allison, Jane R. and Sessions, Richard B. and Essex, Jonathan W.},
  journal = {Journal of Chemical Theory and Computation},
  year    = {2017},
  month   = sep,
  volume  = {13},
  number  = {11},
  pages   = {5683--5696},
  doi     = {10.1021/acs.jctc.7b00643},
}

@article{ferreira_cholesterol_popc_order_2013,
  title   = {Cholesterol and {POPC} segmental order parameters in lipid membranes: solid state $^1$H--$^{13}$C {NMR} and {MD} simulation studies},
  author  = {Mendes Ferreira, Tiago and Coreta-Gomes, Filipe and Ollila, O. H. Samuli and Moreno, Maria Jo{\~a}o and Vaz, Winchil L. C. and Topgaard, Daniel},
  journal = {Physical Chemistry Chemical Physics},
  year    = {2013},
  volume  = {15},
  number  = {6},
  pages   = {1976--1989},
  doi     = {10.1039/C2CP42738A},
}

@article{jiang_membrane_curvature_2025,
  title   = {Membrane curvature: a key regulator of membrane protein structure and function},
  author  = {Jiang, Pu and Zhang, Cong and Lin, Cong and Wang, Yibo and Wang, Xiaohui},
  journal = {Drug Discovery Today},
  year    = {2025},
  doi     = {10.1016/j.drudis.2025.104550},
}

@article{lion2014adenovirus,
  title   = {Adenovirus infections in immunocompetent and immunocompromised patients},
  author  = {Lion, T.},
  journal = {Clinical Microbiology Reviews},
  year    = {2014},
  volume  = {27},
  number  = {3},
  pages   = {441--462}
}

@article{morrone2024tscape,
  title={T-SCAPE: T cell immunogenicity scoring via cross-domain aided predictive engine},
  journal={Science Advances},
  year={2024}
}

@article{bigmhc2023,
  title={BigMHC improves prediction of peptide presentation and immunogenicity with deep learning},
  journal={Nature Machine Intelligence},
  year={2023}
}

@article{immunostruct2025,
  title={ImmunoStruct: Multimodal deep learning for peptide–MHC immunogenicity prediction},
  journal={Nature Machine Intelligence},
  year={2025}
}

@article{schmidt2023prime2,
  title={Improved prediction of antigen presentation and TCR recognition with MixMHCpred2.2 and PRIME2.0},
  journal={Nature Biotechnology},
  year={2023}
}

@article{cover1967nearest,
  title = {Nearest Neighbor Pattern Classification},
  author = {Cover, Thomas M. and Hart, Peter E.},
  journal = {IEEE Transactions on Information Theory},
  volume = {13},
  number = {1},
  pages = {21--27},
  year = {1967},
  doi = {10.1109/TIT.1967.1053964}
}

@misc{hanta_who,
  author       = {{World Health Organization}},
  title        = {Hantavirus infections},
  howpublished = {Fact sheet},
}

@misc{who_health_stats,
  author       = {{World Health Organization}},
  title        = {World Health Statistics 2025: Monitoring Health for the SDGs},
  year         = {2025},
  howpublished = {\url{https://www.who.int/data/gho/publications/world-health-statistics}}
}

@article{huang2017charmm36m,
  author = {Huang, Jing and Rauscher, Sarah and Nawrocki, Grzegorz and others},
  title = {CHARMM36m: an improved force field for folded and intrinsically disordered proteins},
  journal = {Nature Methods},
  volume = {14},
  pages = {71--73},
  year = {2017}
}

@article{hanta_np5,
  title   = {Hantaan virus-derived peptides that stabilize {HLA}-{E} could abrogate inhibition of {CD56}\textsuperscript{dim}{NKG2A}+ {NK} cells},
  author  = {Xue, Manling and Tang, Kang and Zhang, Yusi and Xu, Xiaoyue and Zhang, Chunmei and Zuo, Jiajia and Wang, Fenglan and Zhang, Xiyue and Zheng, Xuyang and Zhuang, Ran and Zhang, Yun and Jin, Boquan and Ma, Ying},
  journal = {PLoS Pathogens},
  year    = {2025},
  month   = jul,
  day     = {18},
  volume  = {21},
  number  = {7},
  pages   = {e1012717},
  doi     = {10.1371/journal.ppat.1012717},
  pmid    = {40680069},
  pmcid   = {PMC12303380}
}

@article{rossjohn2015tcr,
  title={T cell antigen receptor recognition of antigen-presenting molecules},
  author={Rossjohn, Jamie and Gras, Stephanie and Miles, John J and Turner, Stephen J and Godfrey, Dale I and McCluskey, James},
  journal={Annual review of immunology},
  volume={33},
  number={1},
  pages={169--200},
  year={2015},
  publisher={Annual Reviews}
}

@article{calis2013properties,
  title={Properties of MHC class I presented peptides that enhance immunogenicity},
  author={Calis, Jorg JA and Maybeno, Matt and Greenbaum, Jason A and Weiskopf, Daniela and De Silva, Aruna D and Sette, Alessandro and Ke{\c{s}}mir, Can and Peters, Bjoern},
  journal={PLoS computational biology},
  volume={9},
  number={10},
  pages={e1003266},
  year={2013},
  publisher={Public Library of Science San Francisco, USA}
}

@article{enguehard2023timeinterpret,
  title   = {Time Interpret: a Unified Model Interpretability Library for Time Series},
  author  = {Enguehard, Joseph},
  journal = {arXiv preprint arXiv:2306.02968},
  year    = {2023}
}

@article{kokhlikyan2020captum,
  title   = {Captum: A unified and generic model interpretability library for PyTorch},
  author  = {Kokhlikyan, Narine and Miglani, Vivek and Martin, Miguel and Wang, Edward and Alsallakh, Bilal and Reynolds, Jonathan and Melnikov, Alexander and Kliushkina, Natalia and Araya, Carlos and Yan, Siqi and Reblitz-Richardson, Orion},
  journal = {arXiv preprint arXiv:2009.07896},
  year    = {2020}
}

@misc{ribeiro2016why,
  title        = {``Why Should I Trust You?'': Explaining the Predictions of Any Classifier},
  author       = {Ribeiro, Marco Tulio and Singh, Sameer and Guestrin, Carlos},
  year         = {2016},
  howpublished = {In: \emph{Proceedings of the 22nd ACM SIGKDD International Conference on Knowledge Discovery and Data Mining}, pp.~1135--1144},
  doi          = {10.1145/2939672.2939778}
}

@misc{lundberg2017unified,
  title        = {A Unified Approach to Interpreting Model Predictions},
  author       = {Lundberg, Scott M. and Lee, Su-In},
  year         = {2017},
  howpublished = {In: \emph{Advances in Neural Information Processing Systems} (NeurIPS), vol.~30},
  note         = {Proceedings paper}
}

@article{middlehurst2021hivecote,
  title   = {{HIVE-COTE} 2.0: A New Meta Ensemble for Time Series
             Classification},
  author  = {Middlehurst, Matthew and Large, James and Flynn, Michael and
             Lines, Jason and Bostrom, Aaron and Bagnall, Anthony},
  journal = {Machine Learning},
  volume  = {110},
  pages   = {3211--3243},
  year    = {2021},
  doi     = {10.1007/s10994-021-06057-9}
}

@article{dempster2020rocket,
  title   = {{ROCKET}: Exceptionally Fast and Accurate Time Series
             Classification Using Random Convolutional Kernels},
  author  = {Dempster, Angus and Petitjean, Fran{\c{c}}ois and Webb, Geoffrey I.},
  journal = {Data Mining and Knowledge Discovery},
  volume  = {34},
  pages   = {1454--1495},
  year    = {2020},
  doi     = {10.1007/s10618-020-00701-z}
}

@article{schaefer2018muse,
  title   = {Multivariate Time Series Classification with {WEASEL+MUSE}},
  author  = {Sch{\"a}fer, Patrick and Leser, Ulf},
  journal = {arXiv preprint arXiv:1711.11343},
  year    = {2018},
  doi     = {10.48550/arXiv.1711.11343}
}

@misc{bostrom2017shapelet,
  title        = {Binary Shapelet Transform for Multiclass Time Series Classification},
  author       = {Bostrom, Aaron and Bagnall, Anthony},
  year         = {2017},
  howpublished = {In: \emph{Transactions on Large-Scale Data- and Knowledge-Centered Systems XXXII} (LNCS 10420), Springer, pp.~24--46},
  doi          = {10.1007/978-3-662-55608-5_2}
}

@article{farrell_tcell_2020,
  title   = {T Cell Membrane Heterogeneity Aids Antigen Recognition and T Cell Activation},
  author  = {Farrell, Megan V. and Webster, Samantha and Gaus, Katharina and Goyette, Jesse},
  journal = {Frontiers in Cell and Developmental Biology},
  volume  = {8},
  pages   = {609},
  year    = {2020},
  month   = jul,
  doi     = {10.3389/fcell.2020.00609},
}

@article{stadinski_hydrophobic_2016,
  title   = {Hydrophobic {CDR3} residues promote the development of self-reactive {T} cells},
  author  = {Stadinski, Brian D. and Shekhar, Karthik and G{\'o}mez-Touri{\~n}o, Iria and Jung, Jonathan and Sasaki, Katsuhiro and Sewell, Andrew K. and Peakman, Mark and Chakraborty, Arup K. and Huseby, Eric S.},
  journal = {Nature Immunology},
  year    = {2016},
  pmid    = {27348411},
  pmcid   = {PMC4955740}
}

@article{huisman_public_tcr_2022,
  title   = {Public T-Cell Receptors (TCRs) Revisited by Analysis of the Magnitude of Identical and Highly-Similar TCRs in Virus-Specific T-Cell Repertoires of Healthy Individuals},
  author  = {Huisman, Wesley and Hageman, Lois and Leboux, Didier A. T. and Khmelevskaya, Alexandra and Efimov, Grigory A. and Roex, Marthe C. J. and Amsen, Derk and Falkenburg, J. H. Frederik and Jedema, Inge},
  journal = {Frontiers in Immunology},
  year    = {2022},
  volume  = {13},
  pages   = {851868},
  doi     = {10.3389/fimmu.2022.851868},
  pmid    = {35401538},
  pmcid   = {PMC8987591}
}

@article{li_tcrdesign_2025,
  title   = {{TCRdesign}: an antigen-specific generative language model for de novo design of {T}-cell receptors},
  author  = {Li, Xiaokun and Yang, Qiang and Xu, Long and Dong, Weihe and Wang, Kuanquan and Dong, Suyu and Wang, Wei and Luo, Gongning and Zhang, Xianyu and Yang, Tiansong},
  journal = {Briefings in Bioinformatics},
  year    = {2025},
  volume  = {26},
  number  = {6},
  month   = nov,
  pages   = {bbaf691},
  doi     = {10.1093/bib/bbaf691},
}

@article{yin_tcrmodel2_2023,
  title   = {{TCRmodel2}: high-resolution modeling of {T} cell receptor recognition using deep learning},
  author  = {Yin, R. and Ribeiro-Filho, H. V. and Lin, V. and Gowthaman, R. and Cheung, M. and Pierce, B. G.},
  journal = {Nucleic Acids Research},
  year    = {2023},
  doi     = {10.1093/nar/gkad356},
}

\clearpage

\end{document}


\title[Supplementary Information]{Supplementary Information for: An
  Integrative Computational Approach to Predict Viral Epitopes by
Targeting the MHC-TCR Complexation}

\maketitle
\tableofcontents
\clearpage %

\FloatBarrier
\section{Supplementary Figures}

\begin{figure}[!htbp]
  \centering
  \includegraphics[width=0.9\linewidth]{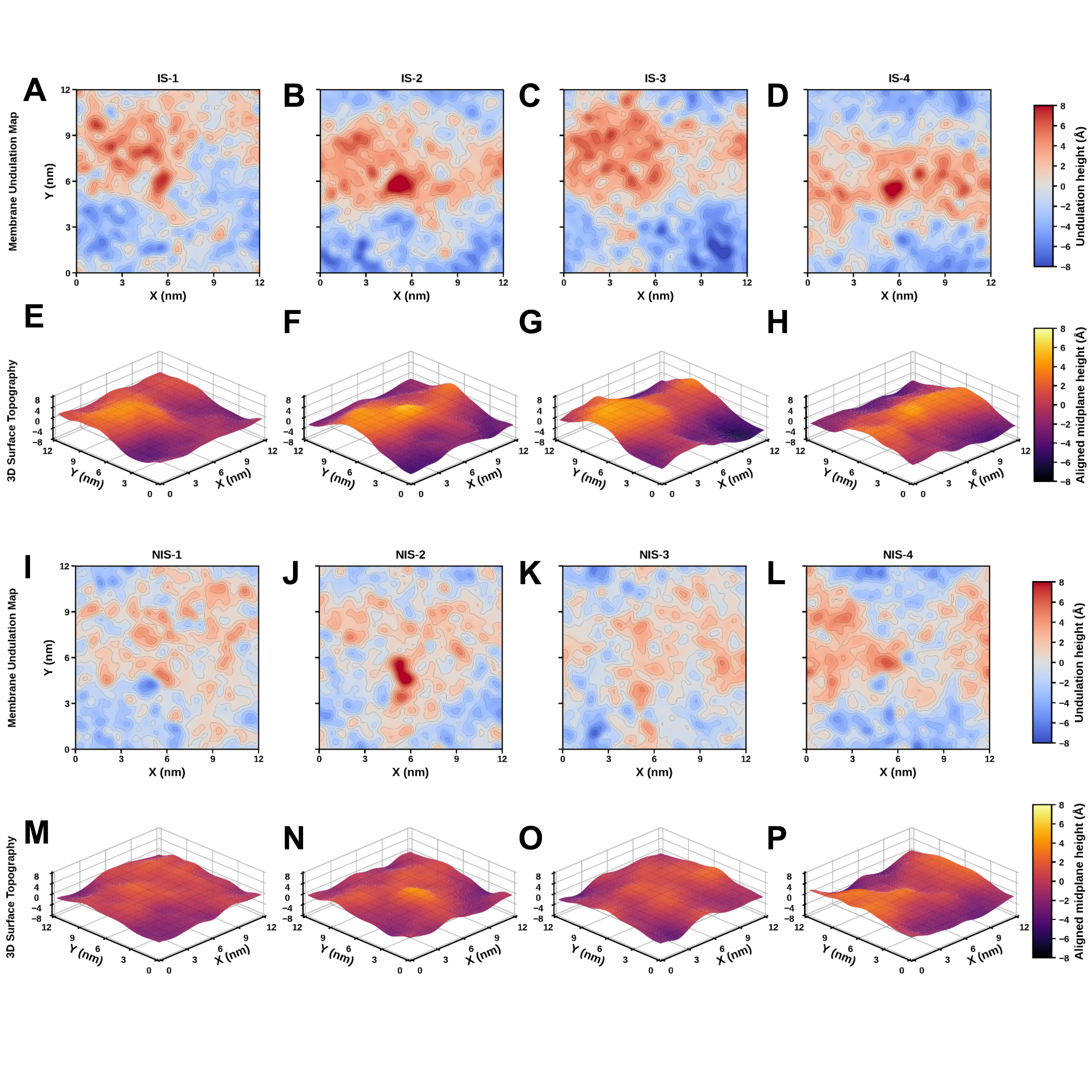}
\caption*{\textbf{Supplementary Figure S1.} \textbf{Membrane curvature analysis of representative immunogenic (IS1--IS4) and non-immunogenic (NIS1--NIS4) systems.} The top two rows (A--H) correspond to immunogenic systems, whereas the bottom two rows (I--P) correspond to non-immunogenic systems. The first and third rows show two-dimensional membrane undulation maps illustrating local height deviations of the membrane midplane relative to the mean plane. The corresponding second and fourth rows present three-dimensional surface topographies, providing spatial reconstructions of membrane curvature. Immunogenic systems consistently exhibit larger membrane deformations and more pronounced local curvature than non-immunogenic systems, indicating enhanced membrane remodeling during immunogenic activation.}

  \label{fig:S1}
\end{figure}

\begin{figure}[!htbp]
  \centering
  \includegraphics[width=0.9\linewidth]{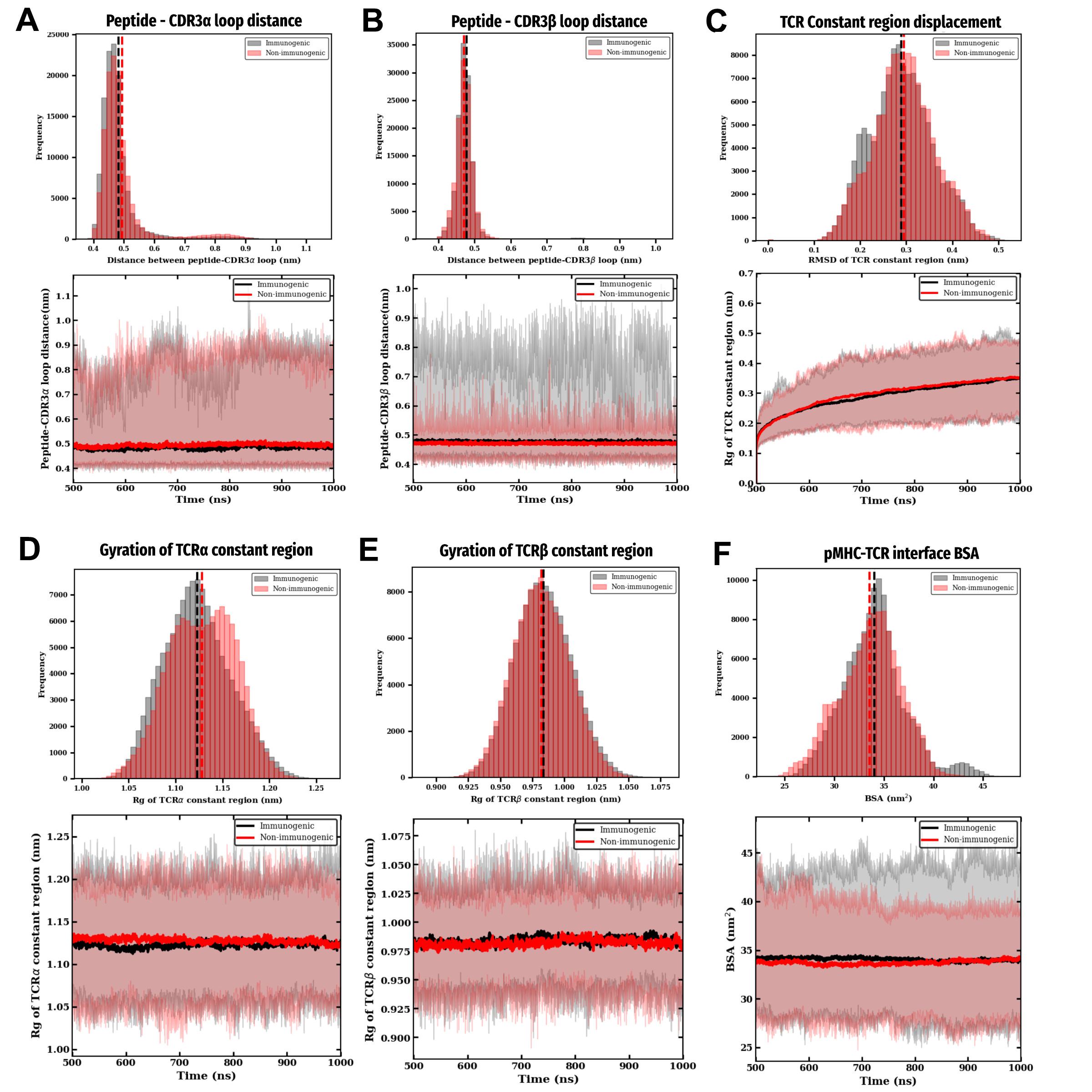}
\caption*{\textbf{Supplementary Figure S2.} \textbf{Distribution and temporal evolution of different structural features across immunogenic (IS, black) and non-immunogenic (NIS, red) systems.} (A) Peptide--CDR3$\alpha$ loop distance. (B) Peptide--CDR3$\beta$ loop distance. (C) Root-mean-square displacement (RMSD) of the TCR constant region. (D) Radius of gyration ($R_g$) of the TCR$\alpha$ constant domain. (E) Radius of gyration ($R_g$) of the TCR$\beta$ constant domain. (F) Buried surface area (BSA) at the pMHC--TCR interface. For each feature, the upper panel shows the distribution across all sampled frames, with dashed lines indicating the class mean, while the lower panel shows the corresponding class-averaged time evolution over the final 500~ns of the simulations (500~ns--1~$\mu$s); shaded regions represent the standard deviation across systems.}
  \label{fig:S2}
\end{figure}

\begin{figure}[!t]
  \centering
  \includegraphics[width=0.9\linewidth]{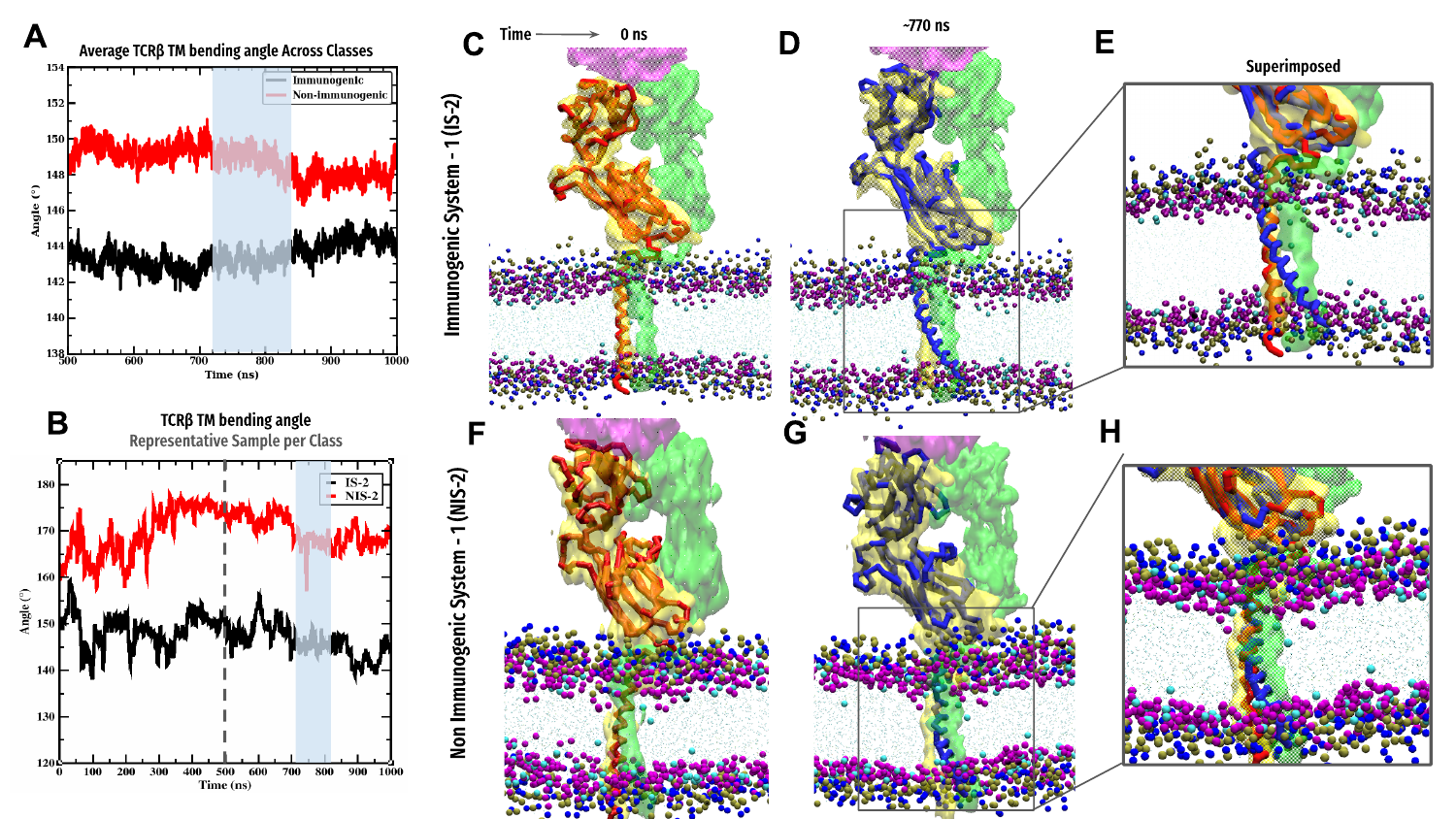}
  \caption*{\textbf{Supplementary Figure S3.}\textbf{ TCR$\beta$ transmembrane (TM) bending dynamics in immunogenic and non-immunogenic systems.} (A) Average TCR$\beta$ TM bending angle for all immunogenic (IS) and non-immunogenic systems (NIS) during the final 500~ns of the simulations (500~ns--1~$\mu$s). (B) Time evolution of TCR$\beta$ TM bending angle for one representative IS (IS-2) and NIS system (NIS-2). The blue shaded region indicates the critical window identified by the interpretability analysis as contributing most strongly to the predictive importance of this feature. (C--H) Structural comparison of the TCR$\beta$ TM region in IS-2 and NIS-2, showing the initial structure (red), the average structure from the critical window (blue), and their superimposed orientations. Panels C--E correspond to IS-2 and illustrate a pronounced bending motion, whereas panels F--H correspond to NIS-2 and show comparatively limited bending. In all structural panels, the equilibriated MHC, TCR$\alpha$, and TCR$\beta$ are shown in magenta, green, and yellow, respectively.}
  \label{fig:S3}
\end{figure}
\begin{figure}[!t]
  \centering
  \includegraphics[width=0.9\linewidth]{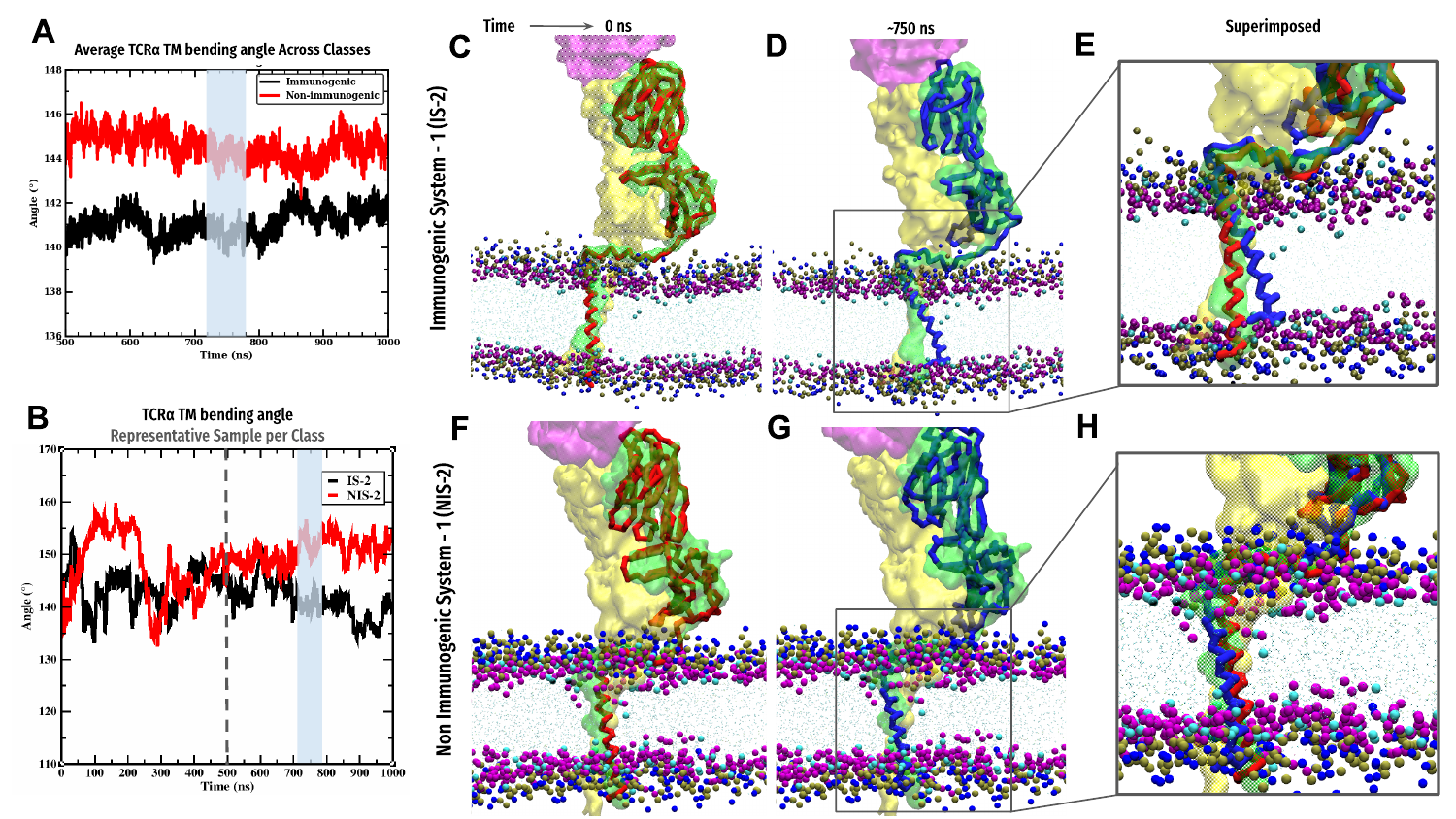}
  \caption*{\textbf{Supplementary Figure S4.}\textbf{ TCR$\alpha$ transmembrane (TM) bending dynamics in immunogenic and non-immunogenic systems.} (A) Average TCR$\alpha$ TM bending angle for all immunogenic (IS) and non-immunogenic systems (NIS) during the final 500~ns of the simulations (500~ns--1~$\mu$s). (B) Time evolution of TCR$\alpha$ TM bending angle for one representative IS (IS-2) and NIS system (NIS-2). The blue shaded region indicates the critical window identified by the interpretability analysis as contributing most strongly to the predictive importance of this feature. (C--H) Structural comparison of the TCR$\alpha$ TM region in IS-2 and NIS-2, showing the initial structure (red), the average structure from the critical window (blue), and their superimposed orientations. Panels C--E correspond to IS-2 and illustrate a pronounced bending motion, whereas panels F--H correspond to NIS-2 and show comparatively limited bending. In all structural panels, the equilibriated MHC, TCR$\alpha$, and TCR$\beta$ are shown in magenta, green, and yellow, respectively.}
  \label{fig:S4}
\end{figure}

\begin{figure}[!t]
  \centering
  \IfFileExists{figures/all_pca.pdf}{%
    \includegraphics[width=\textwidth,height=0.48\textheight,keepaspectratio]{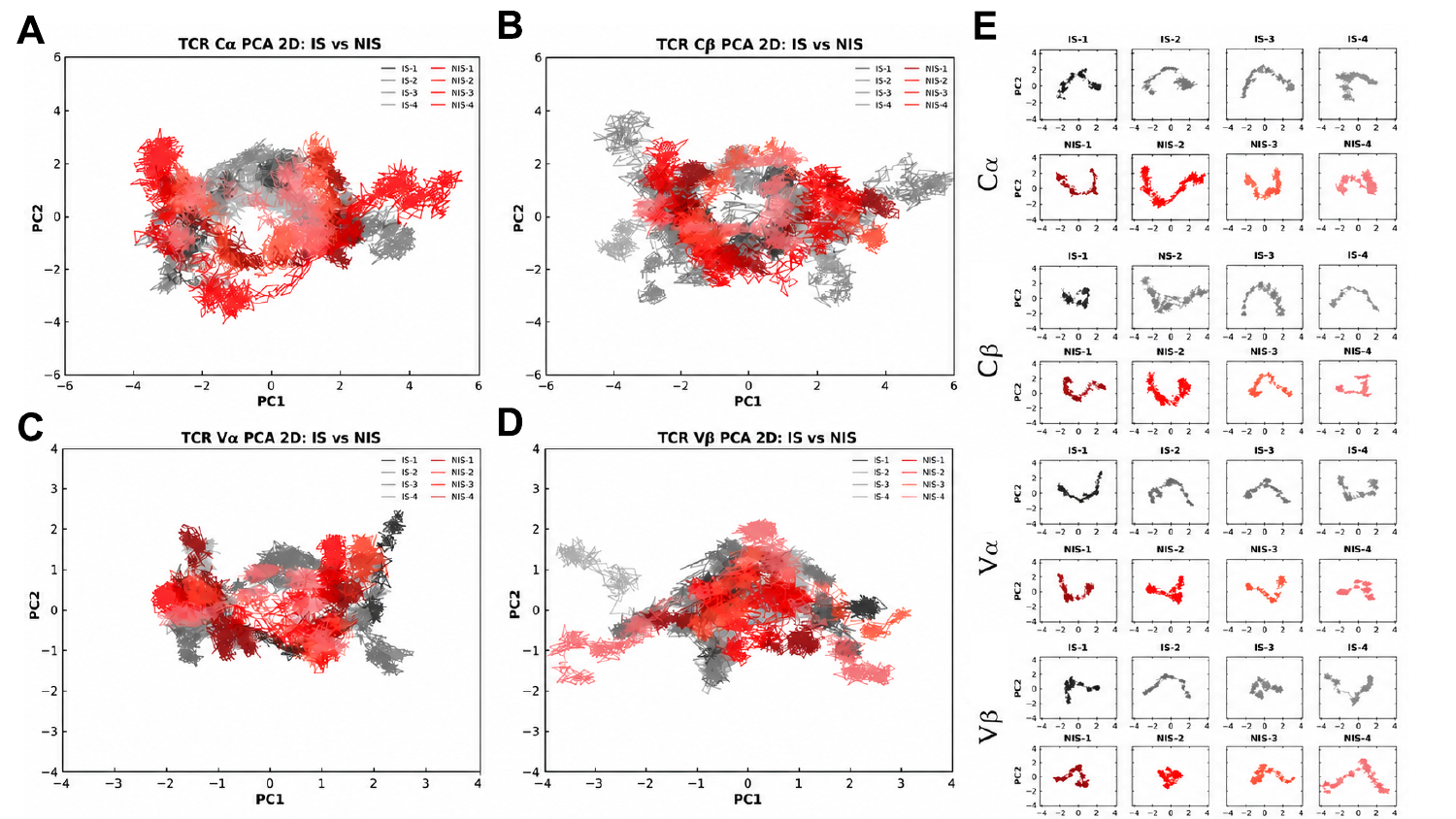}%
  }{%
    \fbox{\parbox{0.9\textwidth}{\centering Upload \texttt{images/all\_pca\_supp.svg}.}}%
  }
  \caption*{\textbf{Supplementary Figure S5. Principal component analysis (PCA) of TCR domain motions in immunogenic and non-immunogenic systems.} (A--D) Two-dimensional PCA projections (PC1 vs.\ PC2) for the TCR$\alpha$ constant (C$\alpha$), TCR$\beta$ constant (C$\beta$), TCR$\alpha$ variable (V$\alpha$), and TCR$\beta$ variable (V$\beta$) domains, respectively, across four immunogenic (IS) and four non-immunogenic (NIS) systems. Gray projections represent IS, whereas red projections represent NIS. The left panels show the overlaid projections of all IS and NIS trajectories, while the right panels show the corresponding PCA conformational landscapes for each individual system, illustrating the conformational space sampled by each TCR domain during the simulations. Among the four domains, the C$\alpha$ domain shows the most distinct separation in conformational sampling between the IS and NIS systems.}
  \label{fig:S5}
\end{figure}

\begin{figure}[!t]
  \centering
  \includegraphics[width=0.95\linewidth]{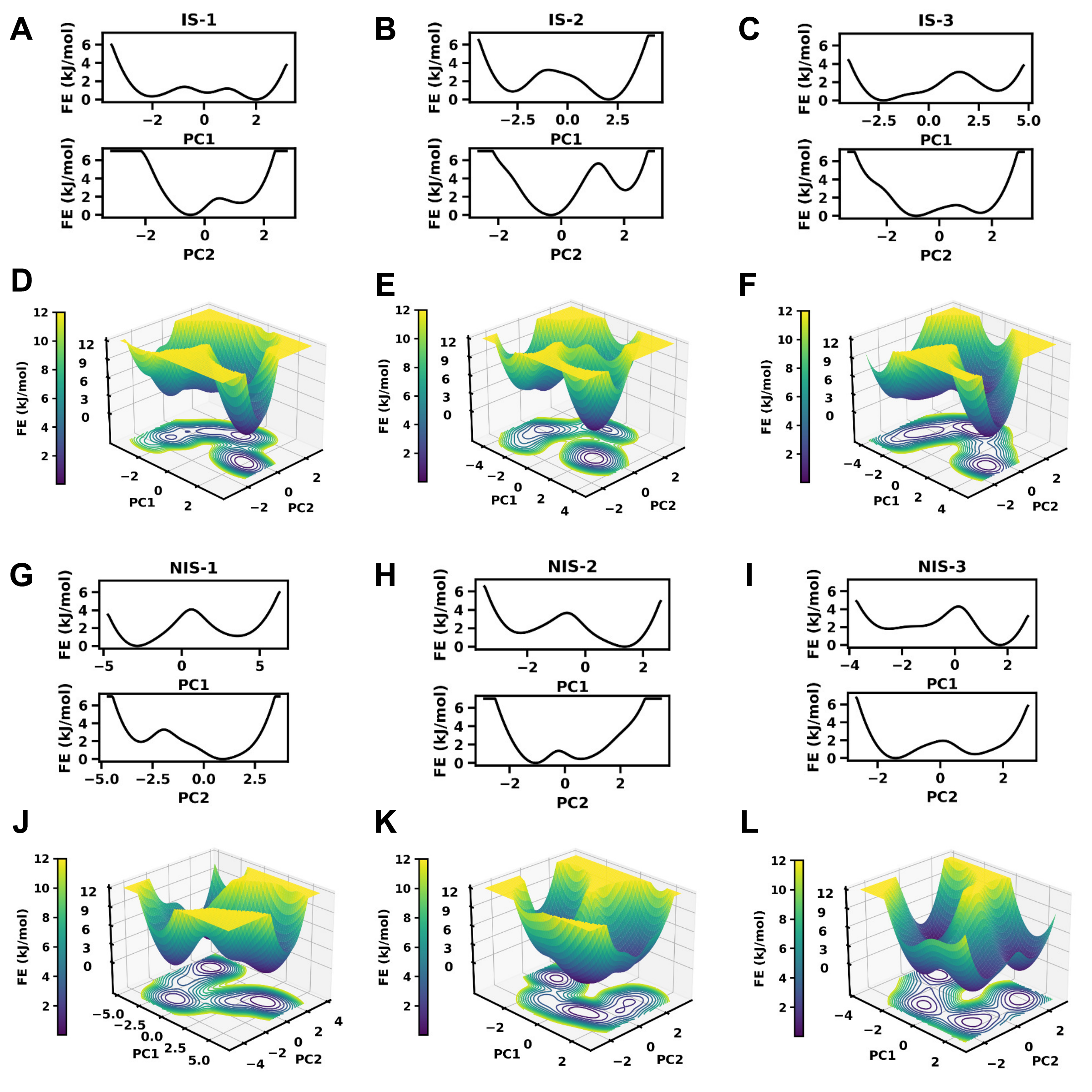}
  \caption*{\textbf{Supplementary Figure S6.} \textbf{Principal component analysis (PCA) and free-energy surfaces (FES) of the TCR$\alpha$ constant (C$\alpha$) domain for immunogenic (IS) and non-immunogenic (NIS) systems.} (A--C) One-dimensional free-energy profiles projected along the first (PC1) and second (PC2) principal components for representative immunogenic systems (IS-1, IS-2, and IS-3). (D--F) Corresponding two-dimensional free-energy surfaces projected onto PC1 and PC2 for the same immunogenic systems. (G--I) One-dimensional free-energy profiles along PC1 and PC2 for representative non-immunogenic systems (NIS-1, NIS-2, and NIS-3). (J--L) Corresponding two-dimensional free-energy surfaces projected onto PC1 and PC2 for the non-immunogenic systems. The overall free-energy landscapes are consistent across systems within each class, with immunogenic systems exhibiting multiple well-defined energy minima separated by relatively low barriers along PC1.}
  \label{fig:S6}
\end{figure}

\begin{figure}[!t]
  \centering
  \includegraphics[width=0.9\linewidth]{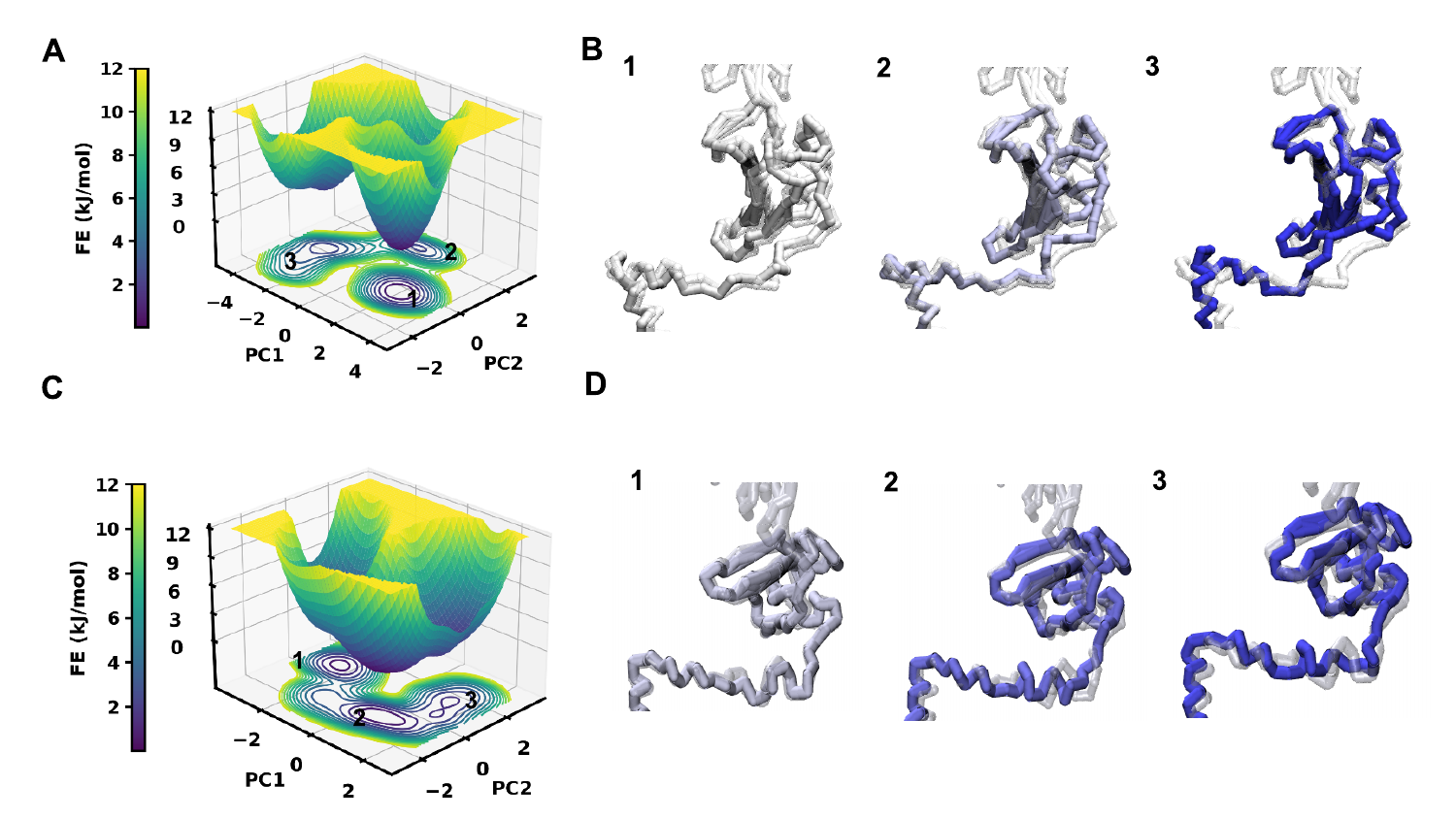}
   \caption*{\textbf{Supplementary Figure S7.} \textbf{Principal component analysis of TCR$\alpha$ constant-region dynamics.} (A,B) Representative immunogenic system (IS-2) and (C,D) representative non-immunogenic system (NIS-2). (A,C) Three-dimensional free-energy landscapes (FES) of the TCR$\alpha$ constant region projected along PC1 and PC2, with free energy represented on the Z-axis, highlighting the conformational space sampled by IS-2 and NIS-2. (B,D) Representative minimum-energy conformations extracted from the corresponding FES, showing a pronounced bending of the TCR$\alpha$ connecting peptide region in IS-2, which is not observed in NIS-2.}

  \label{fig:S7}
\end{figure}

\begin{figure}[!t]
  \centering
  \includegraphics[width=0.7\linewidth]{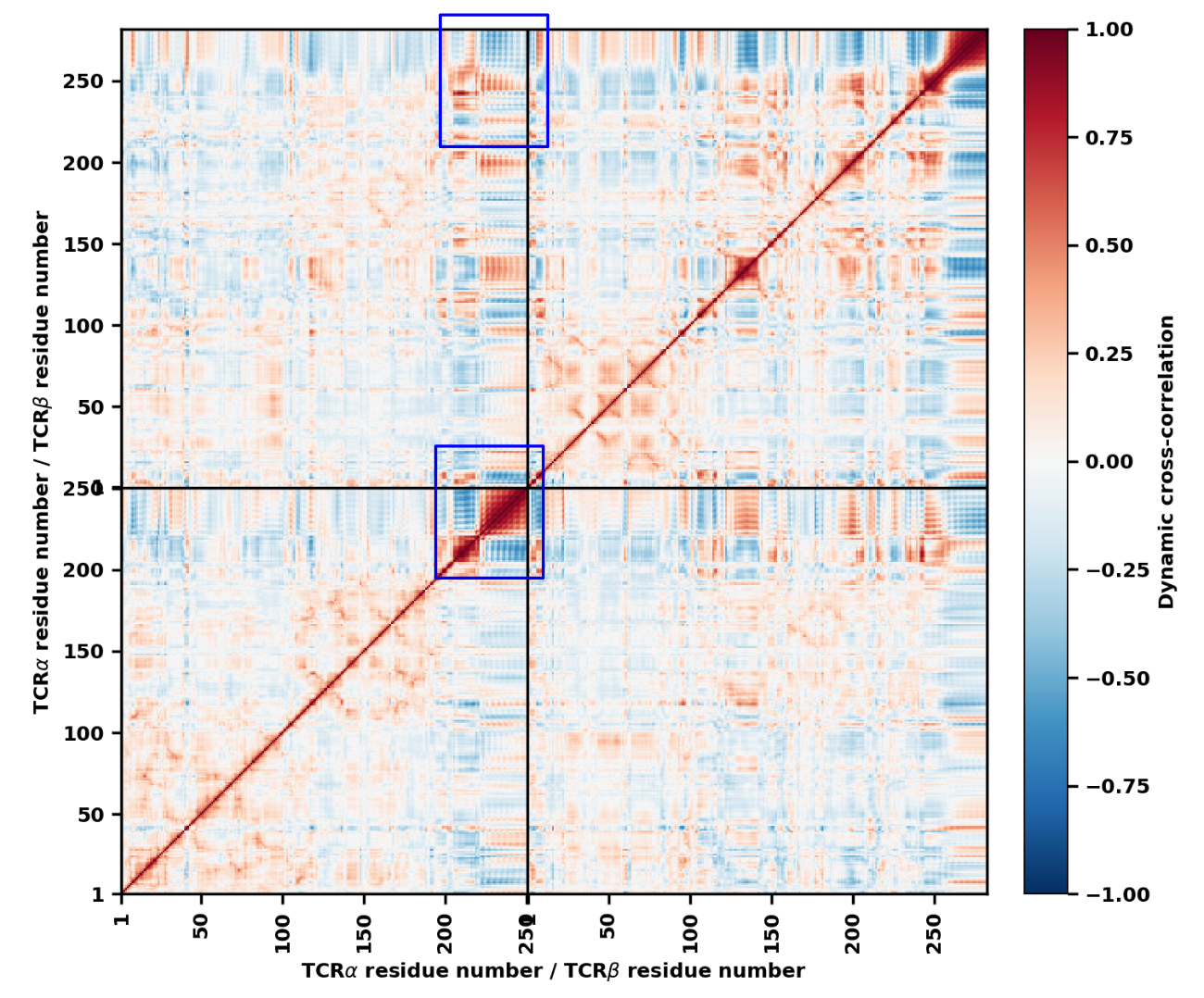}
\caption*{\textbf{Supplementary Figure S8.} \textbf{Dynamic cross-correlation matrix (DCCM) of the TCR$\alpha$ and TCR$\beta$ chains calculated over the simulation time. }Positive correlations (red) indicate residues moving in the same direction, whereas negative correlations (blue) indicate anti-correlated motions. Blue boxes highlight the connecting-peptide (CP) regions of TCR$\alpha$ and TCR$\beta$, which exhibit strong intra-segment positive correlations and pronounced anti-correlated motions with their respective transmembrane helices.}
  \label{fig:S8}
\end{figure}

\begin{figure}[!t]
  \centering
  \adjustbox{trim=0 30mm 0 0,clip}{\includegraphics[width=1\linewidth]{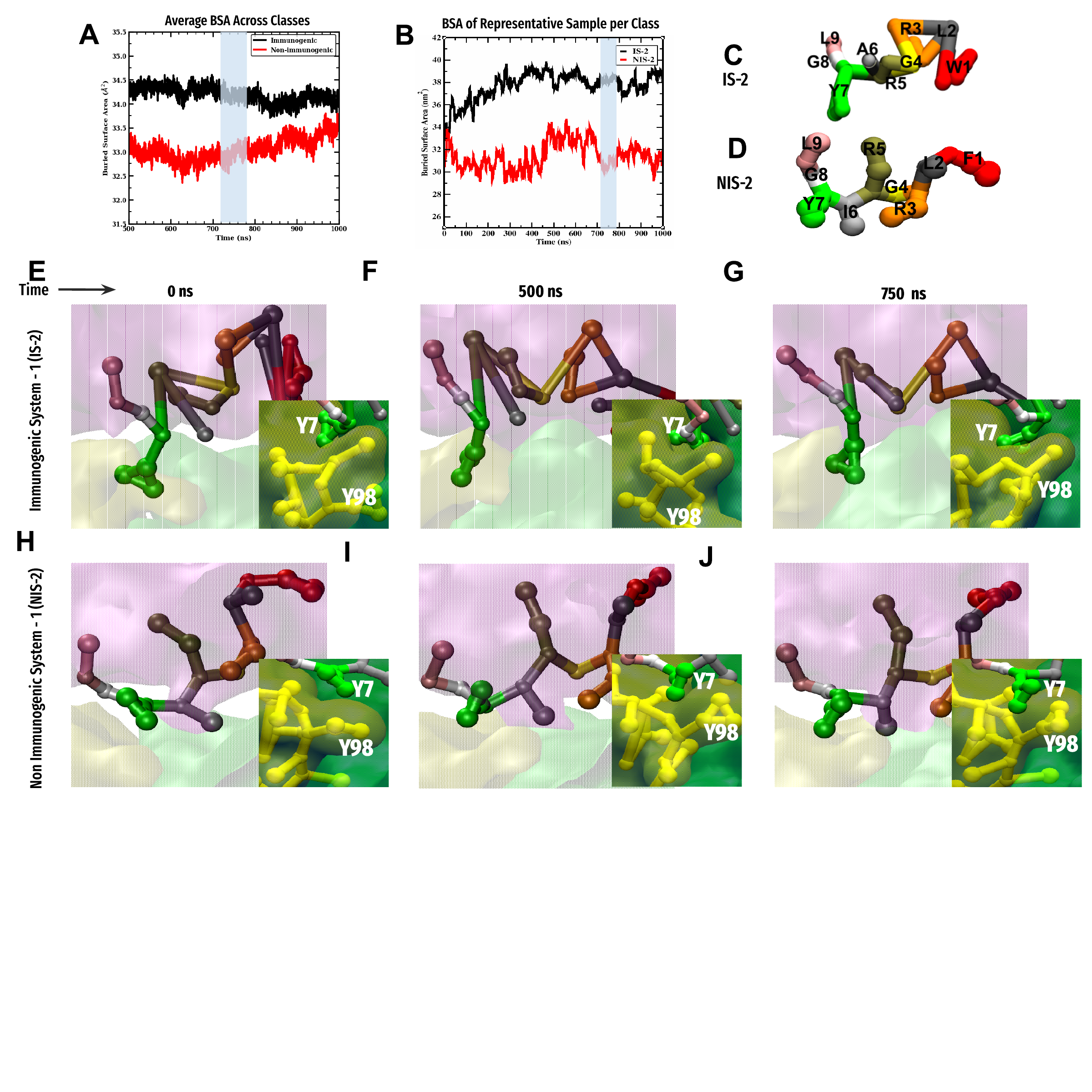}}
  \caption*{\textbf{Supplementary Figure S9}:\textbf{ Buried surface area (BSA) variation and peptide characteristics in immunogenic (IS-2) and non-immunogenic systems (NIS-2).} (A) Average BSA at the pMHC--TCR interface for all IS and NIS, computed over the equilibrated simulation window (500~ns--1~$\mu\text{s}$).with the critical window identified by the interpretability analysis
highlighted in blue.  (B) Time evolution of BSA for one representative IS-2 and one
representative NIS-2, highlighting differences in interfacial buried surface area (C,D) Peptide structures and corresponding amino acid sequences for representative IS-2 and NIS-2 complexes having F1W and A6I mutations respectively.(E--J) Representative conformations of IS-2 and NIS-2 at 0, 500, and 750~ns showing reorientation of peptide and CDR3$\beta$ loop regions at the pMHC-TCR interface. MHC, TCR$\alpha$, and TCR$\beta$ are shown in mauve, green, and yellow, respectively, with the peptide shown in licorice representation and colored by residue. (E--G) In IS-2 peptide residue Y7 reorients and goes deeper into the TCR groove. Insets highlight the corresponding reorientation of CDR3$\beta$ residue Y98, bringing it into closer proximity to Y7. (H--J) Corresponding conformations of NIS-2, where Y7 does not exhibit a similar reorientation over the simulation.}
  
  \label{fig:S9}
\end{figure}

\begin{figure}[!t]
  \centering
  \includegraphics[width=1\linewidth]{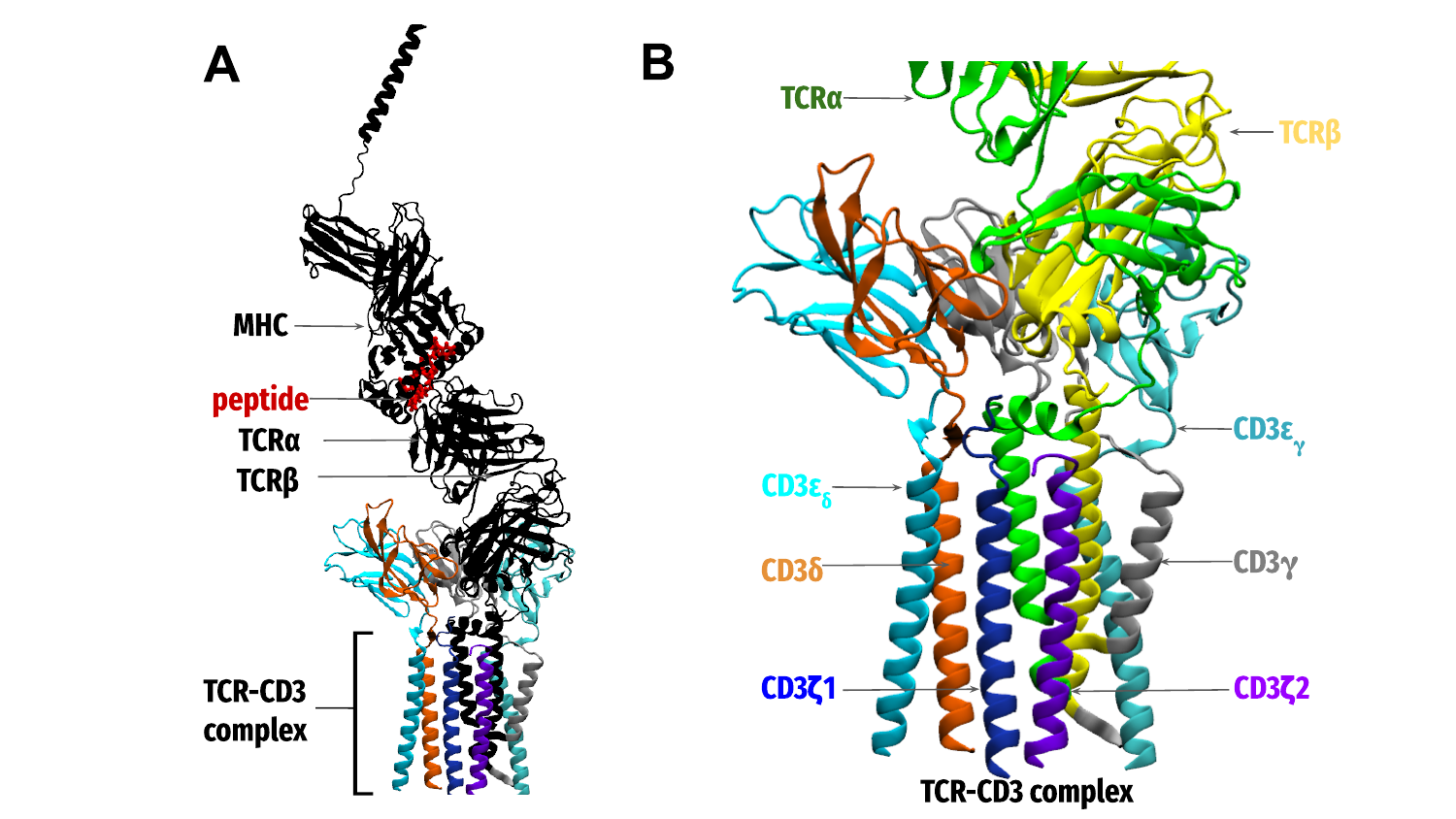}
  \caption*{\textbf{Supplementary Figure S10.} \textbf{Construction of the membrane-embedded pMHC--TCR--CD3 system.} (A) Structural alignment of the TCR--CD3 cryo-EM assembly (PDB: 6JXR) to the reference pMHC--TCR complex used throughout this study. (B) Resulting membrane-bound pMHC--TCR--CD3 model showing the pMHC--TCR complex together with the six CD3 subunits (CD3$\delta$, CD3$\gamma$, two CD3$\epsilon$, and two CD3$\zeta$ chains) surrounding the TCR within the membrane.}

  \label{fig:S10}
\end{figure}

\begin{figure}[!t]
  \centering
  \includegraphics[width=1\textwidth]{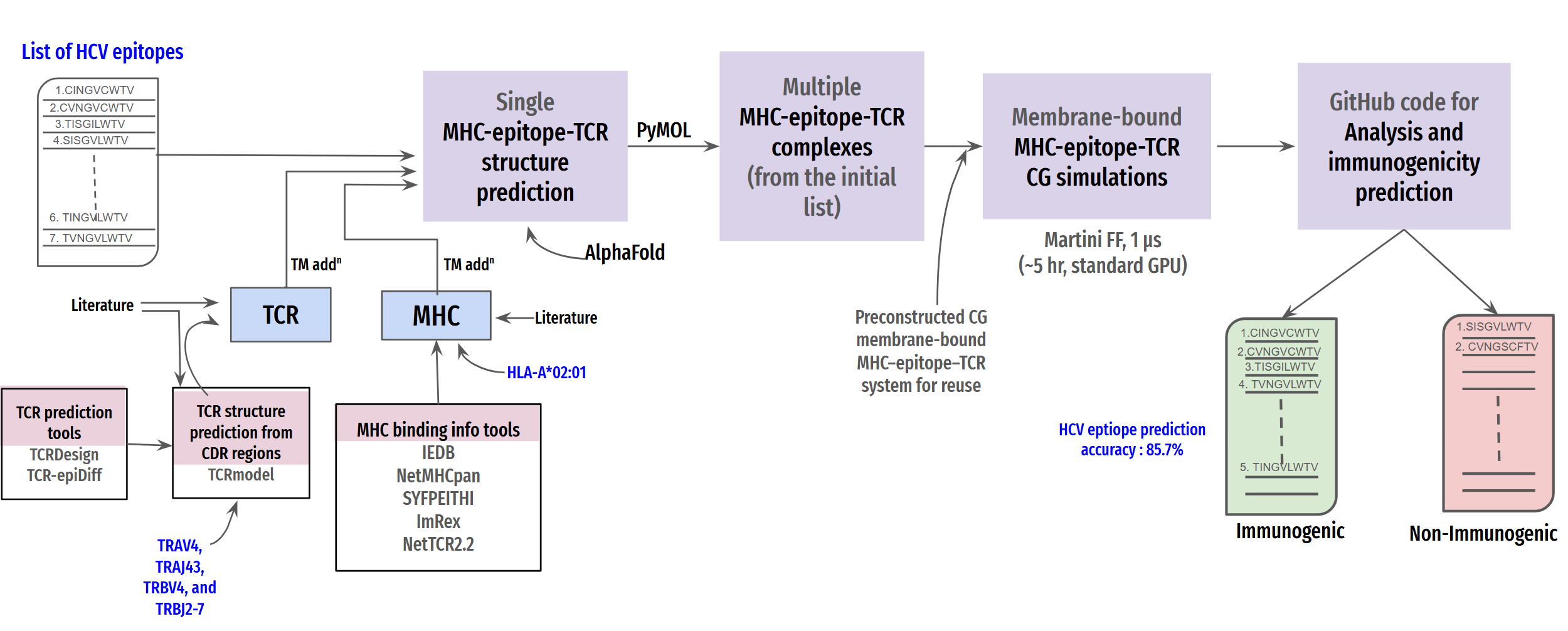}
  \caption*{\textbf{Supplementary Figure S11: Schematic workflow of the DynamiT immunogenicity prediction pipeline.} Starting from a set of candidate peptide sequences, pMHC--TCR complex structures are obtained from available structural information or generated using the modeling protocol described in this study. The resulting complexes are subjected to coarse-grained (CG) simulations, and trajectory-derived dynamic features are used by DynamiT to predict peptide immunogenicity. The application of the complete workflow is demonstrated using an HCV case study, with HCV-specific steps and information highlighted in blue.}
  
  \label{fig:S11}
\end{figure}

\raggedbottom

\FloatBarrier
\clearpage

\section{Supplementary Tables}

\begin{table}[!htbp]
\caption{Complete list of structural and dynamic features extracted from the molecular dynamics trajectories and used as input variables for machine learning-based immunogenicity prediction. The feature set includes intermolecular distances, RMSD, radius of gyration, principal component (PC) projections, buried surface area (BSA), and angular descriptors characterizing TCR conformational dynamics.}
\centering
\renewcommand{\arraystretch}{1.2}
\rowcolors{2}{blue!5}{gray!5}
\begin{tabular}{>{\centering\arraybackslash}p{0.6\linewidth}}
\rowcolor{blue!30}\textbf{Features}\\
\toprule
Pair distance between MHC and CDR$\alpha$1 loop\\
Pair distance between MHC and CDR$\alpha$2 loop\\
Pair distance between peptide and CDR$\alpha$3 loop\\
Pair distance between MHC and CDR$\beta$1 loop\\
Pair distance between MHC and CDR$\beta$2 loop\\
Pair distance between peptide and CDR$\beta$3 loop\\
RMSD of TCR$\alpha$ constant region\\
RMSD of TCR$\beta$ constant region\\
RMSD of TCR constant region\\
RMSD of TCR$\alpha$ variable region\\
RMSD of TCR$\beta$ variable region\\
RMSD of TCR variable region\\
pMHC--TCR buried surface area\\
Radius of gyration of TCR$\alpha$ constant region\\
Radius of gyration of TCR$\beta$ constant region\\
PC1 of TCR$\alpha$ constant region\\
PC1 of TCR$\beta$ constant region\\
PC1 of TCR$\alpha$ variable region\\
PC1 of TCR$\beta$ variable region\\
PC2 of TCR$\alpha$ constant region\\
PC2 of TCR$\beta$ constant region\\
PC2 of TCR$\alpha$ variable region\\
PC2 of TCR$\beta$ variable region\\
Angle between TCR$\alpha$ constant and membrane\\
Angle between TCR$\beta$ constant and membrane\\
\bottomrule
\end{tabular}

\label{tab:S1}
\end{table}

{\scriptsize
\setlength{\LTleft}{0pt}
\setlength{\LTright}{0pt}
\setlength{\tabcolsep}{4pt}
\renewcommand{\arraystretch}{1.15}
\begin{longtable}{p{0.7cm} p{1.0cm} p{1.2cm} p{1.5cm} p{3.2cm} p{2.3cm} p{1.0cm} p{2.0cm}}
\captionsetup{
    margin={0pt,-0.07\textwidth}
}
  \caption{ List of different virus pMHC--TCR complexes used in the test set along with their immunogenicity information, type of mutations, corresponding viral sources, MHC type, and TCR type.}\label{tab:S2}\\
  \toprule
  {\textbf{S.\,No}} & {\textbf{PDB ID}} & \shortstack{\textbf{Immuno-}\\\textbf{genicity}} &
  {\textbf{Mutation}} & {\textbf{Virus}} & \shortstack{\textbf{MHC}\\\textbf{Type}} & \shortstack{\textbf{TCR}\\\textbf{Type}} & {\textbf{Reference}} \\
  \midrule
  \endfirsthead

  \toprule
  {\textbf{S.\,No}} & {\textbf{PDB ID}} & \shortstack{\textbf{Immuno-}\\\textbf{genicity}} &
  {\textbf{Mutation}} & {\textbf{Virus}} & \shortstack{\textbf{MHC}\\\textbf{Type}} & \shortstack{\textbf{TCR}\\\textbf{Type}} & {\textbf{Reference}} \\
  \midrule
  \endhead

  1  & 1ao7 & High & WT  & Human T-cell Lymphotropic Virus & HLA-A*0201 & A6 &
  \href{https://pubmed.ncbi.nlm.nih.gov/9586631}{9586631} \\
  2  & 1qrn & Low  & P6A & Human T-cell Lymphotropic Virus & HLA-A*0201 & A6 &
  \href{https://pubmed.ncbi.nlm.nih.gov/9586631}{9586631} \\
  3  & 1qse & High & V7R & Human T-cell Lymphotropic Virus & HLA-A*0201 & A6 &
  \href{https://pubmed.ncbi.nlm.nih.gov/9586631}{9586631} \\
  4  & 1qsf & Low  & Y8A & Human T-cell Lymphotropic Virus & HLA-A*0201 & A6 &
  \href{https://pubmed.ncbi.nlm.nih.gov/9586631}{9586631} \\

  5  & 5jzi & High & WT   & Hepatitis C virus & HLA-A*0201 & 1406 &
  \href{https://pubmed.ncbi.nlm.nih.gov/28572406}{28572406} \\
  6  & 5jzi & High & V10A & Hepatitis C virus & HLA-A*0201 & 1406 &
  \href{https://pubmed.ncbi.nlm.nih.gov/28572406}{28572406} \\
  7  & 5jzi & Low  & G6A  & Hepatitis C virus & HLA-A*0201 & 1406 &
  \href{https://pubmed.ncbi.nlm.nih.gov/28572406}{28572406} \\
  8  & 5jzi & Low  & K1A  & Hepatitis C virus & HLA-A*0201 & 1406 &
  \href{https://pubmed.ncbi.nlm.nih.gov/28572406}{28572406} \\

  9  & 3sjv & High & WT  & Epstein--Barr virus & HLA-A*0201 & AS01 &
  \href{https://pubmed.ncbi.nlm.nih.gov/24759101}{24759101} \\
  10 & 3sjv & Low  & G8W & Epstein--Barr virus & HLA-A*0201 & AS01  &
  \href{https://pubmed.ncbi.nlm.nih.gov/24759101}{24759101} \\
  11 & 4prh & High & E5D & Epstein--Barr virus & HLA-B*0801 & LC13 &
  \href{https://pubmed.ncbi.nlm.nih.gov/24759101}{24759101} \\
  12 & 4prh & Low  & E5Q & Epstein--Barr virus & HLA-B*0801 & LC13 &
  \href{https://pubmed.ncbi.nlm.nih.gov/24759101}{24759101} \\

  13 & 5w1v & High & WT  & Cytomegalovirus & HLA-A*0201 & GF4 &
  \href{https://doi.org/10.1074/jbc.M117.807719}{28972140} \\
  14 & 2esv & High & WT  & Cytomegalovirus & HLA-A*0201 & RA14 &
  \href{https://www.nature.com/articles/ni1312}{16474394} \\

  15 & 7n5p & High & WT  & Influenza A virus & HLA-A*0201 & 6218 &
  \href{https://pubmed.ncbi.nlm.nih.gov/35999236}{35999236} \\
  16 & 7n5p & Low  & R7K & Influenza A virus & HLA-A*0201 & 6218 &
  \href{https://pubmed.ncbi.nlm.nih.gov/35999236}{35999236} \\

  17 & 5tez & High & WT  & Influenza A virus & HLA-A*0201 & F50 &
  \href{https://doi.org/10.1074/jbc.M117.810382}{28931605} \\

  18 & 7n1f & High & WT  & SARS-CoV-2 & HLA-A*0201 & YLQ7 &
  \href{https://www.nature.com/articles/s41467-021-27669-8}{35013235} \\
  19 & 7n1f & Low  & P4L & SARS-CoV-2 & HLA-A*0201 & YLQ7 &
  \href{https://www.nature.com/articles/s41467-021-27669-8}{35013235} \\

  20 & 3mv7 & High & WT  & Epstein--Barr virus & HLA-B*0801 & LC13 &
  \href{http://www.jem.org/cgi/doi/10.1084/jem.20100603}{20566715} \\
  21 & 3mv7 & Low  & E5D & Epstein--Barr virus & HLA-B*0801 & LC13 &
  \href{http://www.jem.org/cgi/doi/10.1084/jem.20100603}{20566715} \\
  22 & 3mv8 & Low  & WT  & Epstein--Barr virus & HLA-B*0801 & LC13 &
  \href{http://www.jem.org/cgi/doi/10.1084/jem.20100603}{20566715} \\
  23 & 3mv8 & Low  & E5D & Epstein--Barr virus & HLA-B*0801 & LC13 &
  \href{http://www.jem.org/cgi/doi/10.1084/jem.20100603}{20566715} \\
  24 & 3mv9 & High & WT  & Epstein--Barr virus & HLA-B*0801 & LC13 &
  \href{http://www.jem.org/cgi/doi/10.1084/jem.20100603}{20566715} \\
  25 & 3mv9 & Low  & E5D & Epstein--Barr virus & HLA-B*0801 & LC13 &
  \href{http://www.jem.org/cgi/doi/10.1084/jem.20100603}{20566715} \\

  26 & 1bd2 & High & WT  & Human T-cell Lymphotropic Virus & HLA-A*0201 & A6 &
  \href{https://pubmed.ncbi.nlm.nih.gov/9586631}{9586631} \\
  27 & 1bd2 & Low  & Y5A & Human T-cell Lymphotropic Virus & HLA-A*0201 & A6 &
  \href{https://pubmed.ncbi.nlm.nih.gov/9586631}{9586631} \\
  28 & 1bd2 & Low  & Y8A & Human T-cell Lymphotropic Virus & HLA-A*0201 & A6 &
  \href{https://pubmed.ncbi.nlm.nih.gov/9586631}{9586631} \\

  29 & 5wkh & High & WT  & Dengue virus--3 & HLA-A*0201 & D30 &
  \href{https://doi.org/10.1038/ni.3850}{28945243} \\
  30 & 5wkh & Low  & N9D\_R10K & Dengue virus--3 & HLA-A*0201 & D30 &
  \href{https://doi.org/10.1038/ni.3850}{28945243} \\

  \bottomrule
\end{longtable}
}

\begin{table*}[ht]
\centering
\captionsetup{
    margin={0.11pt,0\textwidth}
}
\caption{ Different possible adenovirus peptide candidates evaluated using DynamiT (predicted immunogenic response (IR): 0--No, 1--Yes). Wild-type peptides are highlighted in purple and single amino acid variants are shown in black along with their corresponding MHC, TCR$\alpha$ and TCR$\beta$ gene details. TCR gene annotations experimentally known are in black and the ones reconstructed using our in-house TCR$\alpha\beta$ pairing strategy are highlighted in blue.}

\label{tab:S3}

\small
\setlength{\tabcolsep}{2pt}
\renewcommand{\arraystretch}{1}
\begin{tabular}{cllllllc}
\toprule
\textbf{S.No }& \textbf{Viral Protein }& \textbf{Peptide} & \textbf{MHC} & \textbf{TCR$\alpha$ genes} & \textbf{TCR$\beta$ genes} & \textbf{IR} \\
\midrule
1 & Hexon & \textcolor{peptidepurple}{TDLGQNLLY} & HLA-A*01:01 & \textcolor{blue}{TRAV38-2/DV8*01} & TRBV20-1, TRBJ2-3 & 1 \\
2 & Hexon & LTDLGQNLLY & HLA-A*01:01 & \textcolor{blue}{TRAV38-2/DV8*01} & TRBV20-1, TRBJ2-3 & 1 \\
3 & E1A & \textcolor{peptidepurple}{LLDQLIEEV} & HLA-A*02:01 & TRAV19, TRAJ4 & \textcolor{blue}{TRBV5-6*01} & 1 \\
4 & E1A (E7I) & LLDQLIIEEV & HLA-A*02:01 & TRAV19, TRAJ4 & \textcolor{blue}{TRBV5-6*01} & 1 \\
5 & E1A (L5W) & LLDQWIEEV & HLA-A*02:01 & TRAV19, TRAJ4 & \textcolor{blue}{TRBV5-6*01} & 1 \\
6 & E1A (Q4E) & LLDELIEEV & HLA-A*02:01 & TRAV19, TRAJ4 & \textcolor{blue}{TRBV5-6*01} & 1 \\
7 & E1A (Q4I) & LLDILIEEV & HLA-A*02:01 & TRAV19, TRAJ4 & \textcolor{blue}{TRBV5-6*01} & 1 \\
8 & E1A (Q4W) & LLDWLIEEV & HLA-A*02:01 & TRAV19, TRAJ4 & \textcolor{blue}{TRBV5-6*01} & 1 \\
9 & E1A (Q4F) & LLDFLIEEV & HLA-A*02:01 & TRAV19, TRAJ4 & \textcolor{blue}{TRBV5-6*01} & 0 \\
10 & E1A (Q4D) & LLDDLIEEV & HLA-A*02:01 & TRAV19, TRAJ4 & \textcolor{blue}{TRBV5-6*01} & 0 \\
11 & E1A (I6K) & LLDQLKEEV & HLA-A*02:01 & TRAV19, TRAJ4 & \textcolor{blue}{TRBV5-6*01} & 0 \\
\bottomrule
\end{tabular}
\end{table*}

\begin{table*}[ht]
\centering
\captionsetup{
    margin={0.1pt,0\textwidth}
}
\caption{Selection of the reconstructed TCR$\alpha\beta$ receptor for hantavirus peptide screening. Two TCR$\alpha\beta$ receptors predicted using TCRdesign together with our in-house TCR pairing strategy were evaluated using the wild-type NP5 peptide and representative benchmark-negative peptide variants. The TCR pair showing the best agreement with the benchmark consensus was selected for subsequent screening of additional peptide variants. Wild-type peptide is shown in purple. Predicted immunogenic response (IR): 0--No, 1--Yes)}

\label{tab:hanta_tcr_selection}
\small
\setlength{\tabcolsep}{3pt}
\renewcommand{\arraystretch}{1}
\begin{tabular}{clclllllc}
\toprule
\textbf{S.No }&
\textbf{Viral Protein} &
\textbf{Consensus} &
\textbf{Peptide} &
\textbf{MHC} &
\textbf{TCR$\alpha$ genes} &
\textbf{TCR$\beta$ genes} &
\textbf{IR} \\
 &
 &
\textbf{Benchmark }&
 &
 &
 &
 &
&\\
\midrule

1 &
NP5 &
1 &
\textcolor{peptidepurple}{QLGQRIIVL} &
HLA-E*01:03 &
TRAV8-4*05 &
TRBV10-3*04 &
0 \\

2 &
NP5 &
1 &
\textcolor{peptidepurple}{QLGQRIIVL} &
HLA-E*01:03 &
TRAV12-3*01 &
TRBV10-3*04 &
1 \\

3 &
NP5 (I6K) &
0 &
QLGQRKIVL &
HLA-E*01:03 &
TRAV8-4*05 &
TRBV10-3*04 &
1 \\

4 &
NP5 (I6K) &
0 &
QLGQRKIVL &
HLA-E*01:03 &
TRAV12-3*01 &
TRBV10-3*04 &
0 \\

5 &
NP5 (I7K) &
0 &
QLGQRIKVL &
HLA-E*01:03 &
TRAV8-4*05 &
TRBV10-3*04 &
1 \\

6 &
NP5 (I7K) &
0 &
QLGQRIKVL &
HLA-E*01:03 &
TRAV12-3*01 &
TRBV10-3*04 &
0 \\

\bottomrule
\end{tabular}
\end{table*}

\begin{table*}[ht]
\centering
\captionsetup{
    margin={0pt,-0.05\textwidth}
}
\caption{Predicted immunogenicity of hantavirus NP5 peptide variants using the selected TCR$\alpha\beta$ receptor (TRAV12-3*01/TRBV10-3*04). Wild-type peptide is highlighted in purple. Predicted immunogenic response (IR) is reported as 1 (immunogenic) and 0 (non-immunogenic).}
\label{tab:hanta_candidates}

\small
\begin{tabular}{cllllllc}
\toprule
\textbf{S.No} &
\textbf{Viral Protein} &
\textbf{Peptide} &
\textbf{MHC }&
\textbf{TCR$\alpha$ genes} &
\textbf{TCR$\beta$ genes} &
\textbf{IR} \\
\midrule

1 &
NP5 &
\textcolor{peptidepurple}{QLGQRIIVL} &
HLA-E*01:03 &
TRAV12-3*01 &
TRBV10-3*04 &
1 \\

2 &
NP5 (I6K) &
QLGQRKIVL &
HLA-E*01:03 &
TRAV12-3*01 &
TRBV10-3*04 &
0 \\

3 &
NP5 (I7K) &
QLGQRIKVL &
HLA-E*01:03 &
TRAV12-3*01 &
TRBV10-3*04 &
0 \\

4 &
NP5 (R5F) &
QLGQFIIVL &
HLA-E*01:03 &
TRAV12-3*01 &
TRBV10-3*04 &
1 \\

5 &
NP5 (Q1Y) &
YLGQRIIVL &
HLA-E*01:03 &
TRAV12-3*01 &
TRBV10-3*04 &
1 \\

6 &
NP5 (Q1K) &
KLGQRIIVL &
HLA-E*01:03 &
TRAV12-3*01 &
TRBV10-3*04 &
0 \\

7 &
NP5 (Q4D) &
QLGDRIIVL &
HLA-E*01:03 &
TRAV12-3*01 &
TRBV10-3*04 &
1 \\

8 &
NP5 (G3D) &
QLDQRIIVL &
HLA-E*01:03 &
TRAV12-3*01 &
TRBV10-3*04 &
1 \\

9 &
NP5 (Q1V) &
VLGQRIIVL &
HLA-E*01:03 &
TRAV12-3*01 &
TRBV10-3*04 &
0 \\

10 &
NP5 (R5Y) &
QLGQYIIVL &
HLA-E*01:03 &
TRAV12-3*01 &
TRBV10-3*04 &
1 \\

\bottomrule
\end{tabular}
\end{table*}